\documentclass[%
 reprint,
 amsmath,amssymb,
 aps,
floatfix,
]{revtex4-2}
\usepackage{placeins}
\usepackage{graphicx}
\usepackage{dcolumn}
\usepackage{bm}
\usepackage{gensymb}
\usepackage{makecell}
\usepackage[caption=false]{subfig}
\usepackage{hyperref}
\usepackage[T1]{fontenc}
\usepackage[utf8]{inputenc}
\newcommand{\subfigimg}[3][,]{%
  \setbox1=\hbox{\includegraphics[#1]{#3}}
  \leavevmode\rlap{\usebox1}
  \rlap{\hspace*{0pt}\raisebox{\dimexpr\ht1-2\baselineskip}{#2}}
  \phantom{\usebox1}
}

\begin{document}

\preprint{}

\title{Origin of the 50~Hz harmonics in the transverse beam spectrum of the Large Hadron Collider}
\thanks{Research supported by the HL-LHC project.}

\author{S.~Kostoglou}
\email{sofia.kostoglou@cern.ch} 
\affiliation{%
 CERN, Geneva 1211, Switzerland \\
}%
\affiliation{%
 National Technical University of Athens, Athens 15780, Greece\\
}%

\author{G.~Arduini}%
\affiliation{%
 CERN, Geneva 1211, Switzerland \\
}%

\author{L.~Intelisano}%
\affiliation{%
 CERN, Geneva 1211, Switzerland \\
}%
\affiliation{%
 INFN, Sapienza  Universit\`a di Roma, Rome 00185, Italy\\
}%

\author{Y.~Papaphilippou}%
\affiliation{%
 CERN, Geneva 1211, Switzerland \\
}%
 \author{G.~Sterbini}%
\affiliation{%
 CERN, Geneva 1211, Switzerland \\
}%

\date{\today}

\begin{abstract}

Since the beginning of the Large Hadron Collider (LHC) commissioning, spectral components at harmonics of the mains frequency (50~Hz) have been observed in the transverse beam spectrum. This paper presents an overview of the most important observations, collected during the latest physics operation of the LHC in 2018, which clearly indicates that the harmonics are the result of a real beam excitation rather than an instrumental feature. Based on these findings, potential sources of the perturbation are discussed and a correlation with power supply ripple originating from the magnets' power supplies is presented. 

\end{abstract}

\maketitle

\section{Introduction}
In particle accelerators, studies of the beam spectrum can reveal important information concerning the existence of external noise sources that perturb the motion of the particles. Noise effects such as power supply ripple, ground motion and the noise induced by the transverse feedback system are an important issue for the single-particle beam dynamics in past, present and future accelerators. In the presence of non-linearities such as non-linear magnets and beam-beam effects, depending on the spectral components and the nature of the source, power supply ripple can act as a diffusion mechanism for the particles in the beam distribution, through the excitation of resonances in addition to the ones driven by the lattice non-linearities, an effect that can prove detrimental to the beam lifetime \cite{intro1993_1, intro1993_2, intro1994_1, intro1994_2}. This paper focuses on the investigation of such a mechanism that has been observed in the transverse beam spectrum of the Large Hadron Collider (LHC) \cite{Bruning:782076}, which is contaminated by harmonics of 50~Hz \cite{50Hz_1, 50Hz_2, 50Hz_3, 50Hz_4}.

Observations of harmonics of the mains power frequency in the beam spectrum have been reported in the past from several accelerators including the Super Proton Synchrotron (SPS) \cite{SPS,SPS1_tune, SPS2_tune, SPS3_tune, SPS2}, the Hadron-Electron Ring Accelerator (HERA) \cite{intro1994_1, HERA1_tune, HERA2_tune}, the Relativistic Heavy Ion Collider (RHIC) \cite{RHIC1, RHIC2} and the Tevatron \cite{Tevatron1, Tevatron2}. The studies in the SPS excluded the factor of instrumentation noise as the origin of the perturbation and it was shown that the beam was excited by high-order harmonics, in the form of dipolar excitations, mainly affecting the horizontal plane \cite{SPS}. The source of the perturbation was identified as the main dipoles by injecting an external sinusoidal ripple on their power supply. 

The study conducted at RHIC demonstrated that high-order harmonics (\(h>\)100) were visible in several unrelated instruments as a result of a real beam excitation rather than an artifact of the instrumentation system \cite{RHIC1}. By modifying machine parameters such as the betatron tune and the coupling the source was identified as a dipolar field error. Through a set of experiments, a correlation with power supply ripple was established and specifically, with the 12-pulse line-commutated thyristor power supplies of the main dipoles \cite{RHIC2}. 

A similar observation of 50~Hz high-order harmonics perturbing the beam spectrum in the form of dipolar excitations is also systematically made in the LHC. In this paper, we present the analysis of the experimental data acquired during the 2018 LHC operation aiming to identify the origin of the perturbation. The key observations that lead to the understanding that the harmonics are the result of a real beam excitation are presented in Section~\ref{Sec:experimental_observations}. A correlation with power supply ripple arising from the power supplies of the main dipoles is confirmed with dedicated experiments for the first time in the LHC operation. The analytical formalism of dipolar power supply ripple and the tracking simulations aiming to determine whether the observed power supply ripple leads to a degradation of the beam performance are treated in a second paper \cite{previous}.

\section{Experimental observations in the LHC}
\label{Sec:experimental_observations}
Throughout this paper, the main observable is the representation of the beam signal in frequency domain as computed with the Fast Fourier Transform (FFT). Based on the Fourier analysis, information concerning the origin of the 50~Hz can be extracted. This can be achieved by following the evolution of the lines in frequency domain, both in terms of amplitude and phase, during normal operation, i.e., without any modification in the beam or machine parameters (Sections~\ref{section_normal_operation}). Then, the findings are further extended by observing the response of the harmonics during modifications in the beam or machine configuration (Sections~\ref{section_changing_operation}). These modifications refer to changes, first, in the betatron motion with parameters such as the tune, the phase advance and the beam energy, second, in the power supplies and last, in the settings of the transverse damper.

\subsection{Overview of the LHC beam modes}
\begin{figure}
\includegraphics[width = \columnwidth]{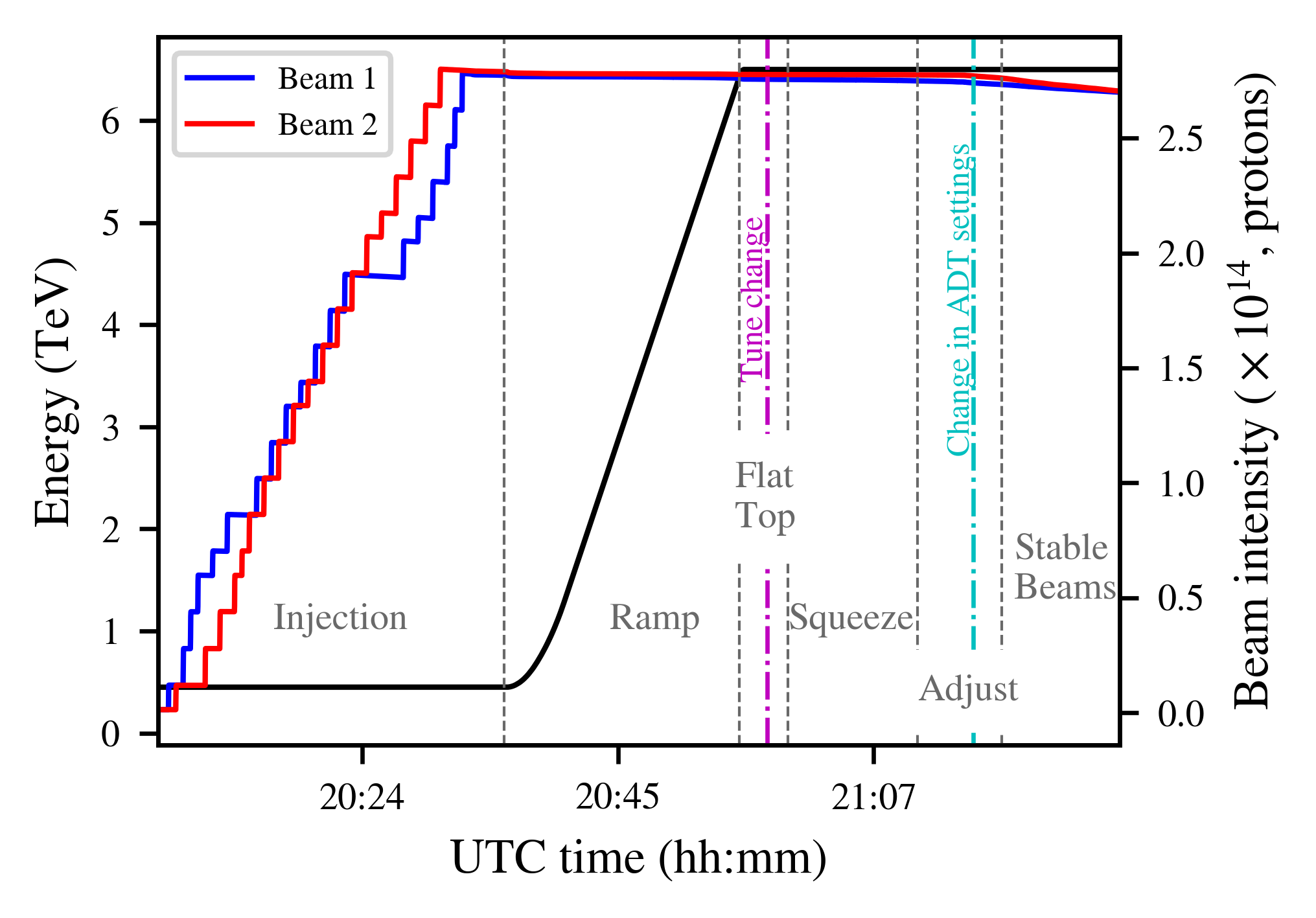} 
\caption{\label{fig:cycle} The LHC operational cycle in 2018 (Fill 7333). The intensity evolution of Beam 1 (blue) and 2 (red) and the beam energy (black) are illustrated. The vertical gray lines denote the beam modes and important modifications during the cycle such as the change of tune (magenta) and the modification in the transverse damper's settings (cyan).}
\end{figure}

The fact that different beam energies and phases of the LHC nominal cycle have been explored in this study justifies the need to include a brief description of the beam modes that are relevant to the next sections of this paper. Figure~\ref{fig:cycle} illustrates the operational cycle for a physics fill (Fill 7333). The different beam modes (gray) are presented along with the intensity evolution of Beam 1 (blue) and 2 (red) in the right axis and the beam energy (black) in the left axis.

In brief, the nominal LHC cycle is organized as follows. After injecting low-intensity single bunches for machine protection reasons, high-intensity batches (two or three trains of 48 or 72 bunches and a bunch spacing of 25~ns) are injected from the SPS to the LHC rings until the requested filling scheme is reached. The \textit{Injection} is performed in the Interaction Region (IR) 2 and 8 for Beam 1 and 2, respectively, with an energy per beam equal to 450~GeV. 

Then, during \textit{Ramp}, the current of the main dipoles and quadrupoles increases while the beams are accelerated. An intermediate squeeze of the $\beta$-functions at the IPs, $\beta^*$, is performed \cite{rampsqueeze}. 

At \textit{Flat Top}, each beam has reached the maximum total energy of 6.5~TeV (as compared to the nominal design total energy of 7~TeV). After a few minutes, the betatron tunes are trimmed from the injection $ (Q_x, Q_y) =\rm  (0.28, 0.31)$ to the collision $(Q_x, Q_y) = \rm (0.31, 0.32)$ values (magenta). With the Achromatic Telescopic Squeezing (ATS) optics \cite{ATS}, the beams are squeezed to $\beta^*=\rm 30~cm $ at the IPs of the two high luminosity experiments (ATLAS and CMS). 

During \textit{Adjust}, the separation bumps in the IRs collapse and the beams are brought to collision. At the end of this beam mode, the settings of the transverse damper are modified (cyan). 

The declaration of \textit{Stable Beams} signals the start of the data acquisition from the experiments. In this beam mode, luminosity optimization techniques are employed such as the crossing angle anti-leveling and the $\beta^*$-leveling \cite{levelling, antileveling}. Finally, the beams are extracted from the ring to the dump. 

\subsection{Beam measurements in normal operation}
\label{section_normal_operation}

\begin{figure}
\includegraphics[width = \columnwidth]{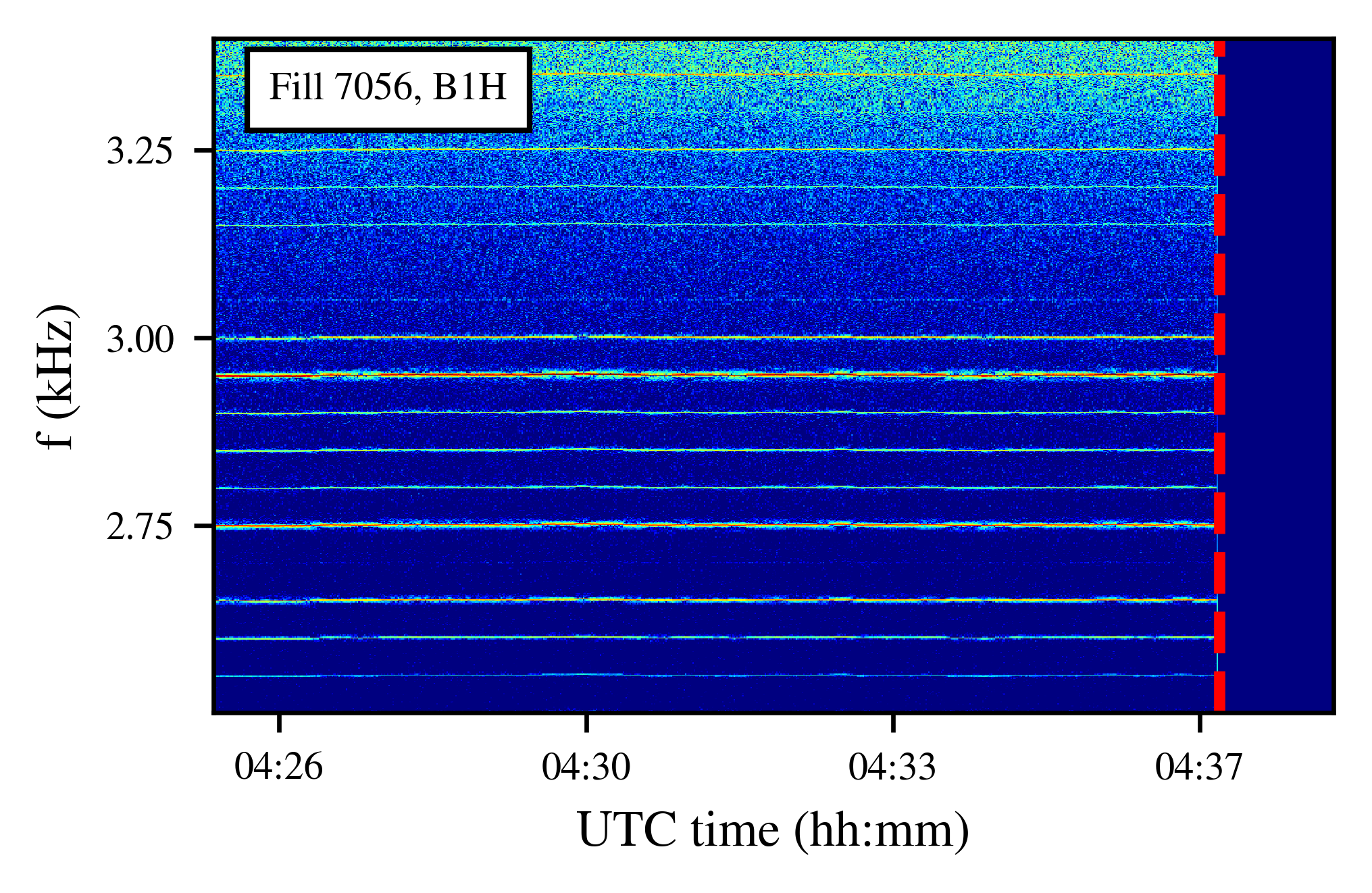} 
\caption{\label{fig:HSBBQ_spectrum_with_without_beam} Horizontal HS BBQ spectrogram of Beam 1 at the end of a physics fill (Fill 7056), centered around 3~kHz and color-coded with the PSD. The red dashed line indicates the end of the fill and the start of the dump of the beam.}
\end{figure}

The concept of the 50~Hz lines on the beam spectrum is illustrated using the the turn-by-turn data from the High Sensitivity Base-Band measurement system (HS BBQ) \cite{BBQ,jones2007lhc}. Figure~\ref{fig:HSBBQ_spectrum_with_without_beam} depicts the spectrogram of the horizontal plane of Beam 1 (Fill 7056) for the last few minutes of the fill, extending up to the first few minutes of the beam dump (red dashed line). The Fourier analysis for each time interval in the horizontal axis is performed with a window length of $\rm 2^{13}$  consecutive turns and an overlap of $\rm 2^{11}$ turns between windows. The frequency range is zoomed below the Beam 1 horizontal tune ($\approx$3.49~kHz) to observe the 50~Hz harmonics in its proximity. A color code is assigned to the Power Spectral Density (PSD) to distinguish the main spectral components (yellow and red) from the noise baseline (blue). 

The spectrum clearly shows that a series of 50~Hz harmonics are present in the beam signal. The fact that the lines appear as multiples of 50~Hz and not as sidebands around the betatron tune is one of the first indications among others (see Section~\ref{section_changing_operation}) that the nature of the power supply ripple is dipolar. 

Furthermore, the harmonics are visible only in the presence of the beam. All signals acquired after the end of the fill (red dashed line) are dominated by the noise of the instrument. A comparison between the signals prior and after the dump of the beam provides a first indication that the lines do not emerge as a by-product of instrumentation noise. 

To further exclude the factor of instrumental or environmental noise, the presence of these harmonics has been validated from several unrelated beam instruments. Position measurements from multiple pickups, located at different positions in the LHC ring, are collected. The main observables are the HS BBQ, the transverse damper Observation Box (ADTObsBox) \cite{ADT, ADT3, ADT4}, the Diode Orbit and Oscillation System (DOROS) \cite{DOROS, DOROS2} and the Multi-Band Instability Monitor (MIM) \cite{MIM1, MIM2}. Measurements from all the aforementioned instruments are available for the Machine Development (MD) Fill 7343, dedicated to studies concerning the 50~Hz harmonics \cite{Kostoglou:2703609}. 

\begin{figure}
\includegraphics[width = \columnwidth]{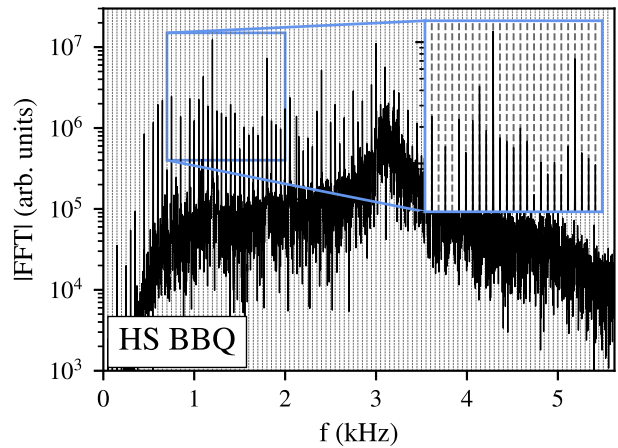}
\caption{\label{fig:multiple_instruments} The horizontal spectrum of Beam 1 at injection energy during the MD Fill 7343 from the HS BBQ and with a zoomed window (light blue). The vertical gray lines represent the multiples of 50~Hz.}
\end{figure}

Figure \ref{fig:multiple_instruments} shows the spectra for 
one of these instruments, the HS BBQ, for the horizontal plane of Beam 1, while the vertical gray lines represent the multiples of 50~Hz. To illustrate that the lines in the beam spectrum correspond to 50~Hz harmonics, a zoomed region of the spectrum is depicted (light blue). From the analysis of the various spectra, it is confirmed that a series of 50~Hz harmonics is visible across all unrelated instruments.

The turn-by-turn acquisitions, such as the ones from the HS BBQ and DOROS, allow accessing a frequency regime up to approximately 5.6~kHz, which is the Nyquist frequency assuming a single observation point along the accelerator (the sampling frequency is $f_s$=$f_{\rm rev}$ with $f_{\rm rev}$=11.245~kHz) \cite{nyquist1928certain}. If present in the signal, frequency components beyond this limit will be aliased in the spectrum.

On the contrary, the ADTObsBox and the MIM provide high sampling rate measurements. Specifically, the ADTObsBox instability buffer contains calibrated bunch-by-bunch position measurements for $\rm 2^{16}$ turns. Firstly, the fact that a calibrated metric is provided allows computing the offsets induced on the beam motion from the 50~Hz harmonics. Secondly, the bunch-by-bunch information is needed to study the evolution of the 50~Hz in the cycle and to compute a high bandwidth spectrum, in the presence of a regular filling scheme. 

As shown in Appendix~\ref{appendix:bbb_spectrum}, the noise floor of the single-bunch ADTObsBox spectrum exceeds the amplitude of the 50~Hz harmonics and therefore, a decrease of the noise baseline is necessary to study their evolution during the cycle. To overcome this problem, a method to combine the information from several bunches has been developed, taking into account the dephasing of the spectrum, due to the time delay, across the different bunches (Appendix~\ref{appendix:bbb_spectrum}). Assuming a regular filling scheme (equal spacing between bunches), this signal averaging algorithm not only provides a reduction of the noise floor but also extends the measurable frequency range of the beam spectrum, while suppressing the aliases and preserving the signal metric.

The horizontal spectrum of Beam 2 is computed for the physics Fill 7334 during collisions, using the bunch-by-bunch and turn-by-turn acquisitions from the Q7 pickup of the ADTObsBox. Although only the spectrum of Beam 2 is depicted, similar observations exist for both beams and planes.  

\begin{figure}
\includegraphics[width = \columnwidth]{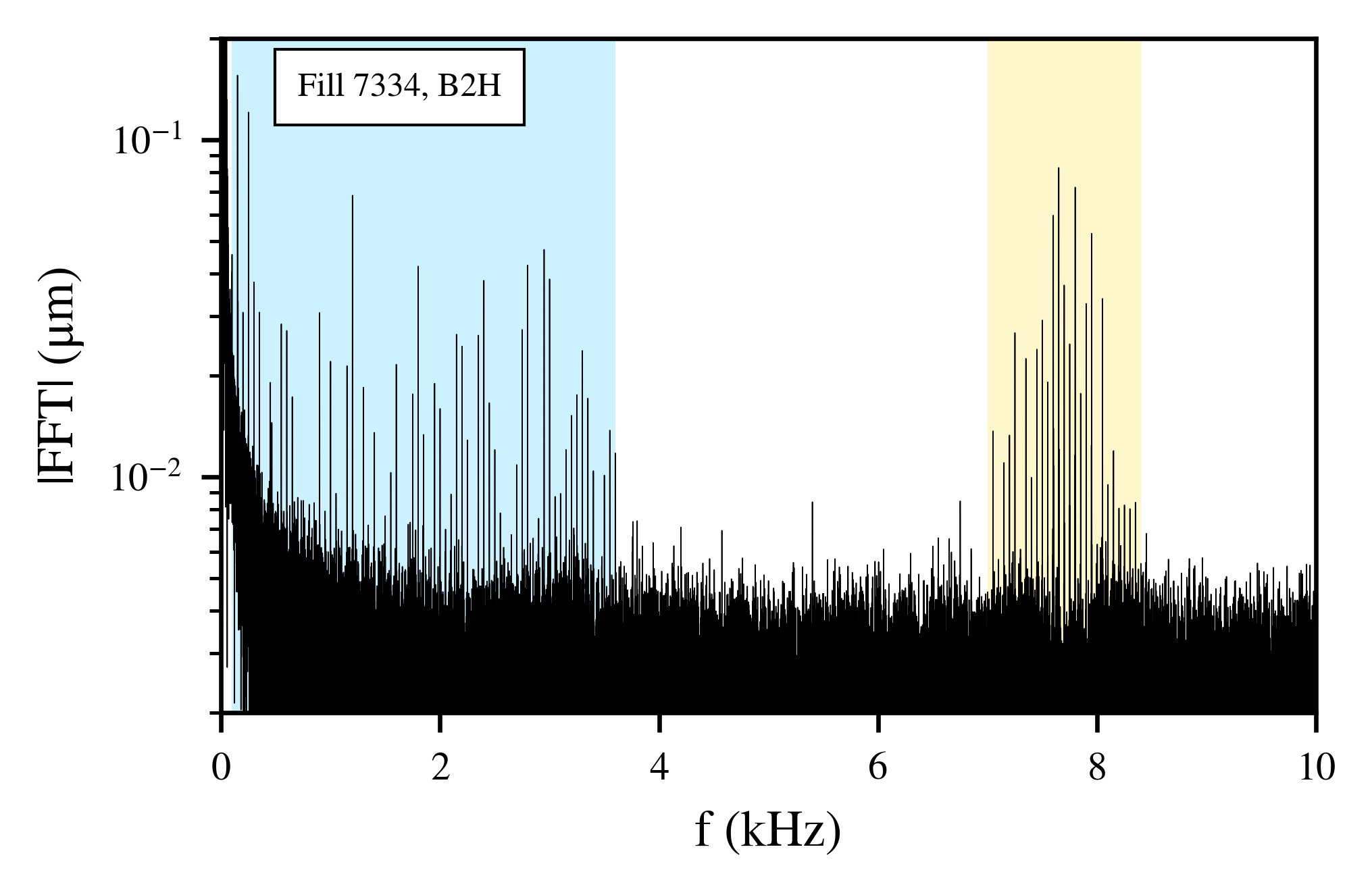} \\ 
\includegraphics[width = \columnwidth]{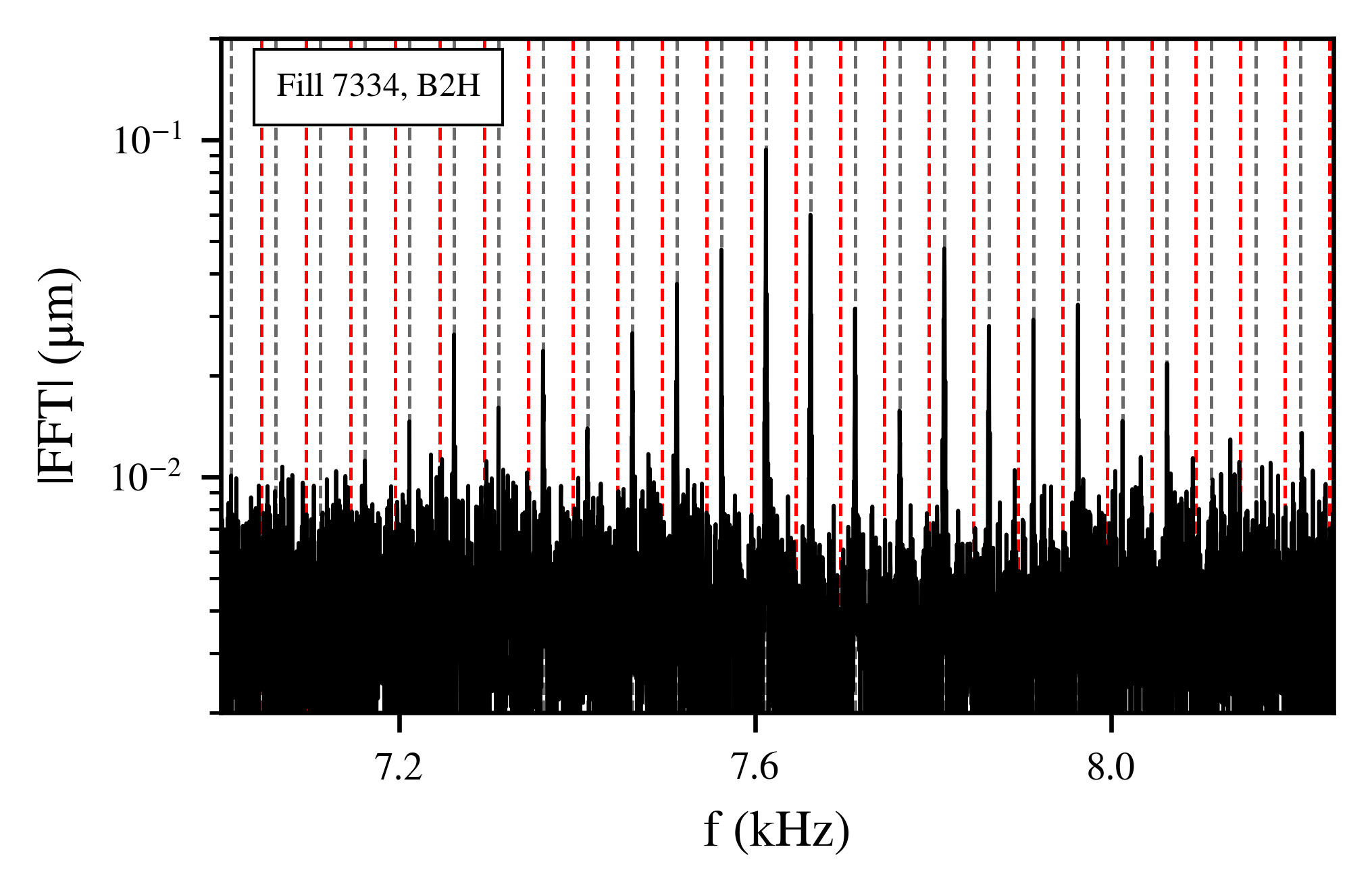}
\caption{\label{fig:ADTObs_spectrum_SB_high_cluster} The horizontal spectrum of Beam 2 at Flat Top for a frequency range up to 10~kHz (top), with the low and high-frequency cluster indicated by the blue and orange span, respectively, and centered around the high-frequency cluster (bottom). The red and gray dashed lines represent the expected position of aliased ($f_{\rm rev} - f_{50}$) and physical 50~Hz harmonics ($f_{50}$), respectively.}
\end{figure}

Figure~\ref{fig:ADTObs_spectrum_SB_high_cluster} illustrates the Fourier analysis, first, for a frequency range up to 10~kHz (Fig.~\ref{fig:ADTObs_spectrum_SB_high_cluster} top). From the review of the spectrum, two areas of particular interest are identified. The first regime (blue span) consists of 50~Hz harmonics extending up to 3.6~kHz. The second area (orange span) is a cluster of 50~Hz at 7-8 kHz. In particular, the cluster is centered at the frequency \(f_{\rm rev}-f_x\), where \(f_x\) is the horizontal betatron frequency, which is $\approx$3.15~kHz at injection and $\approx$3.49~kHz at collision energy(see also Section~\ref{section_changing_operation}). In the frequency interval between the two clusters, either no harmonics are present in the signal or their amplitude is below the noise threshold of the instrument. 

Throughout this paper, the two regimes of interest are referred to as the \textit{low-frequency cluster} and the  \textit{high-frequency cluster}, respectively. It must be noted that the lowest order harmonics are excluded from the analysis as their amplitude is affected by the noise of the instrument. Then, the calibrated spectrum indicates that the harmonics of the high-frequency cluster are more important in terms of amplitude.

As the high-frequency cluster is situated at \(f_{\rm rev}-f_x\), the question that naturally arises is whether these frequency components emerge from aliasing. In fact, even in the case of a physics fill, the sampling rate is only approximately uniform as not all trains are equally spaced. This error can give rise to the aliasing of the low-frequency cluster around the revolution frequency. It must be noted however, that the beam revolution frequency ($f_{\rm rev}$=11.245~kHz) is not a multiple of 50~Hz and therefore, the aliases can be distinguished from the excitations at 50~Hz.

Figure~\ref{fig:ADTObs_spectrum_SB_high_cluster} (bottom) presents the spectrum centered around the high-frequency cluster. The red dashed lines represent the expected position of aliased 50~Hz harmonics $f_{\rm rev} - f_{50}$, where \( f_{\rm 50}\) are the harmonics of the low-frequency cluster,  while the gray dashed lines illustrate the multiples of 50~Hz \( f_{\rm 50}=h \cdot 50\)~Hz, where $h$ is a positive integer. As the spectral components of the high-frequency cluster coincide with the 50~Hz multiples (gray), it is concluded that they are not aliased frequencies.

The time variation of the beam spectrum can reveal important information concerning the source of the perturbation. Due to the variation of the power grid load, the frequency of the mains power supply is not strictly 50~Hz. The study focuses on the impact of the aforementioned drift on the frequency evolution of the 50~Hz harmonics in order to illustrate their distinct signature in the frequency domain. 

\begin{figure}
\includegraphics[width = \columnwidth]{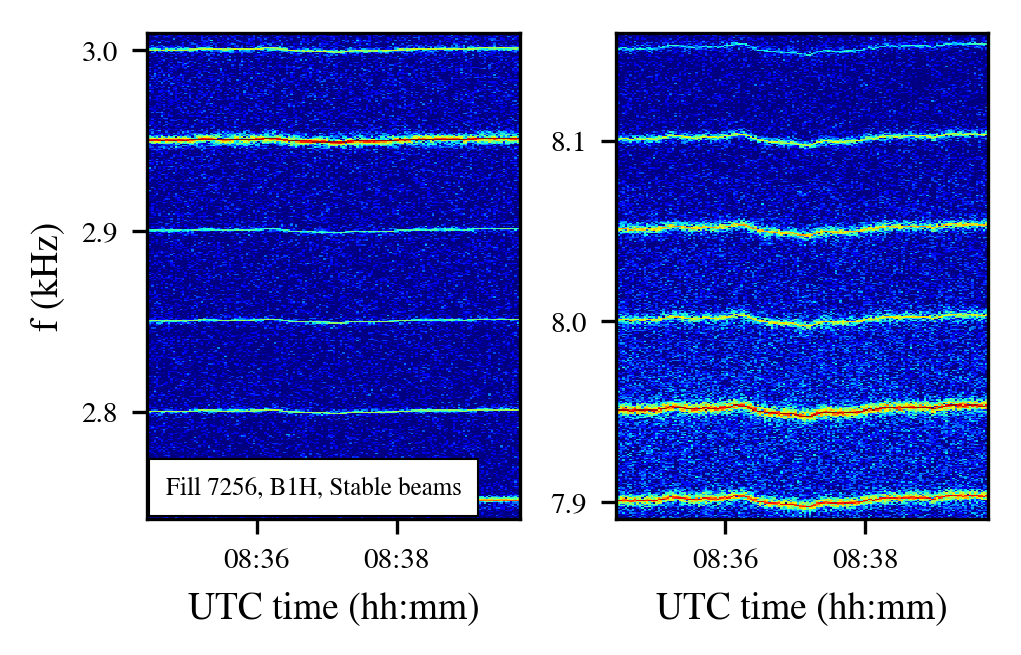} 
\caption{\label{fig:spectrogram_SB_MIM} The horizontal spectrogram of Beam 1 in a regime of the low (left) and high (right) frequency cluster.}
\end{figure}

The spectrogram of the horizontal position of Beam 1 is computed from the MIM data for a time interval at Stable Beams. In Fig.~\ref{fig:spectrogram_SB_MIM}, the horizontal axis represents the timestamp of each spectrum with a window length of $\rm 2^{14}$ turns, the vertical axis is centered around a value in the low (left) and high (right) frequency cluster and a color code is assigned to the PSD. 

An important finding is that, although the lines are harmonics of 50~Hz, a time variation of their frequency is observed. Specifically, all harmonics are affected by a similar frequency modulation, the amplitude of which is proportional to the order of the harmonic $h$. For this reason, the aforementioned effect is more pronounced in the harmonics of the high-frequency cluster, an observation which provides yet another indication that these components are not aliases.

To illustrate quantitatively that the harmonics experience a similar frequency modulation, the amplitude of which scales with the order of the harmonic $h$, an algorithm that can precisely follow their evolution has been implemented. The steps of the algorithm are the following: for each measured time interval the amplitude of the Fourier spectrum is computed. The algorithm focuses on a regime in the vicinity of a single harmonic and, by employing a maximization routine, an accurate determination of its frequency is achieved by detecting the local maximum. The algorithm returns the frequency and the amplitude of the harmonic at each time step. This procedure is repeated for all the time intervals in the spectrogram. 

\begin{figure}
\includegraphics[width = \columnwidth]{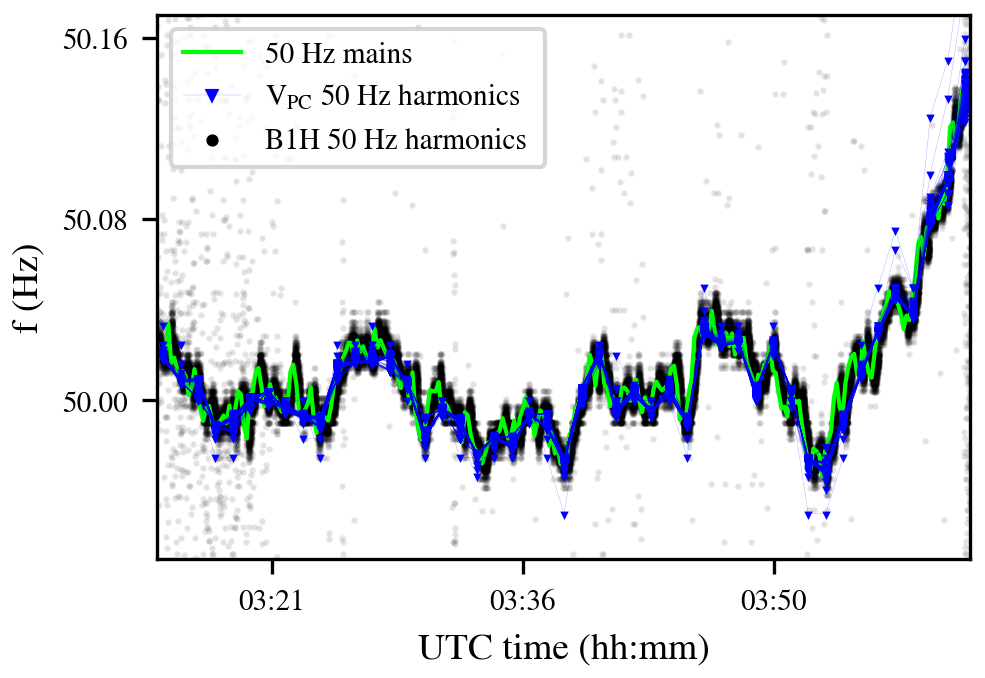} 
\caption{\label{fig:pc_B1_frequency_modulation} The mains power supply ripple frequency (green), the frequency modulation of the harmonics observed in the voltage spectrum of the main dipoles power supply of sector 1-2 (blue) and the ones of the beam spectrum (black), after normalizing with the order of each harmonic $h$.}
\end{figure}

Iterating over all the harmonics in the spectrum with the aforementioned algorithm validates the existence of a similar frequency modulation with an amplitude proportional to the order of the harmonic. Figure~\ref{fig:pc_B1_frequency_modulation} shows the frequency evolution of all the harmonics (black) after normalizing with the order of the harmonic $h$ and projecting to the fundamental frequency (50~Hz). The modulation is visible in both beams and planes, during all beams modes and across several unrelated instruments. The proportional relationship between the modulation amplitude and the harmonic order, observed both in the low and high-frequency cluster, suggests that they emerge from a common source (see Appendix~\ref{app:fm}). 

Figure~\ref{fig:pc_B1_frequency_modulation} also presents the evolution of the network frequency for the same time span (green) \cite{50Hz_evolution}. The origin of the frequency modulation in the harmonics of the beam is clearly related to the stability of the 50~Hz mains from the electrical network, which then propagates to all the harmonics. 

The beam spectrum is compared to the output voltage spectrum of the Main Bends power supply installed in sector 1-2. During the MD Fill 7343, voltage measurements of the power supply were collected every minute with a sampling rate of 50~kHz. The supply's voltage spectrum consists of 50~Hz harmonics, extending up to 10~kHz. Applying a similar analysis to the one used for the harmonics of the beam yields an identical frequency evolution of the 50~Hz components in the power supply. Figure~\ref{fig:pc_B1_frequency_modulation} presents the modulation of the lines in the power supply down-sampled to 50~Hz (blue). 

It is interesting to note that, at the end of the MD (6 am Central European Time or 4 am in Coordinated Universal Time), a frequency drift above the usual variation of the 50~Hz is observed. To validate that this effect is reproducible, fills for the same time and different days have been analyzed, yielding similar results. These variations appear to be the result of the changing load of the power grid at this time of the day.


The previous findings are not meant to establish a correlation between the dipole power supply in sector 1-2 and the beam. The importance of these observations resides on the fact that, if the 50~Hz harmonics are the results of a real beam excitation, their frequency domain signature points to a specific type of power supply as the source. 

The existence of multiple 50~Hz harmonics in combination with the frequency modulation induced by the mains suggests that the origin of these frequency lines are power supplies that are based on commutated Thyristor Technology. This can be understood with a frequency analysis of the variation of the magnetic field (B-Train) in two other machines of the accelerator complex: the Proton Synchrotron (PS) and the and SPS \cite{PS2,SPS_BTRAIN}. In the B-Train system, a pickup up coil is installed in the aperture of the reference dipole magnets. The measured signals correspond to the rate of change of the magnetic field. 

\begin{figure}
\includegraphics[width = \columnwidth]{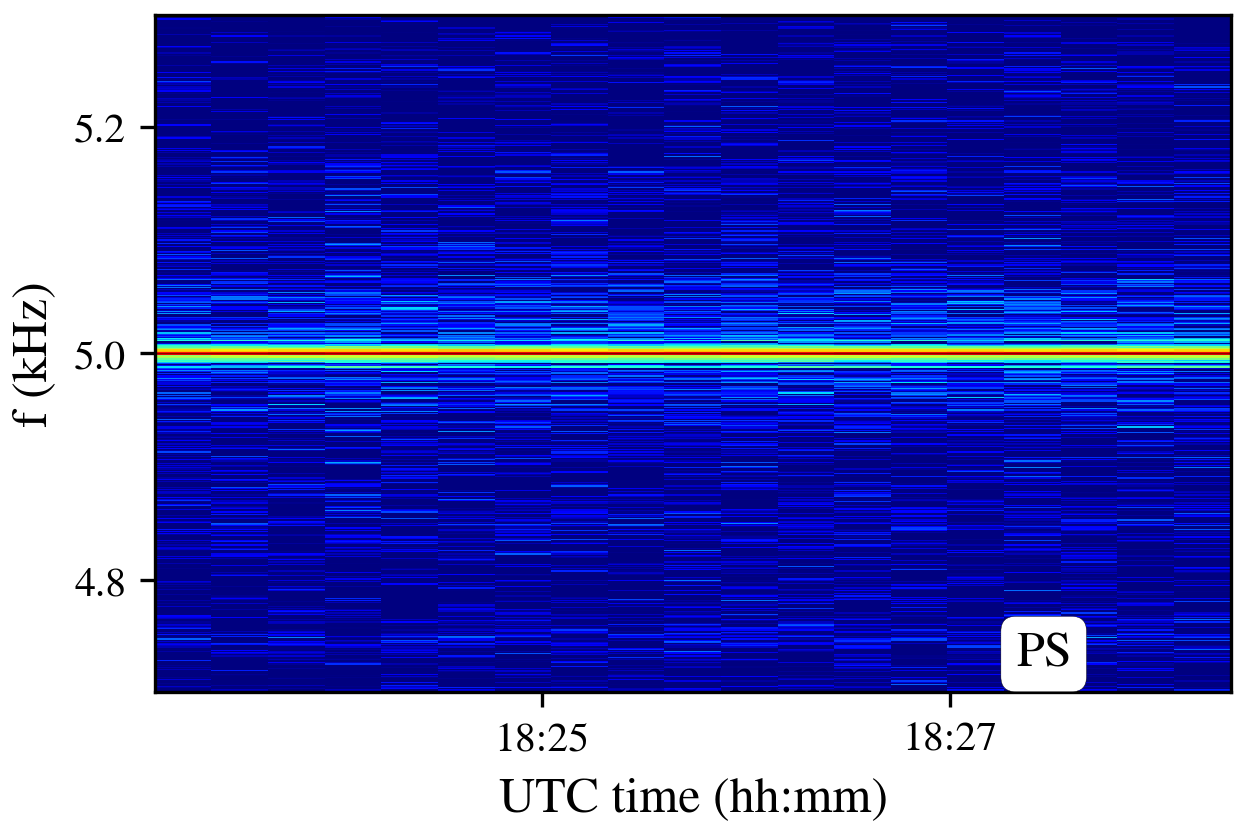} \\
\includegraphics[width = \columnwidth]{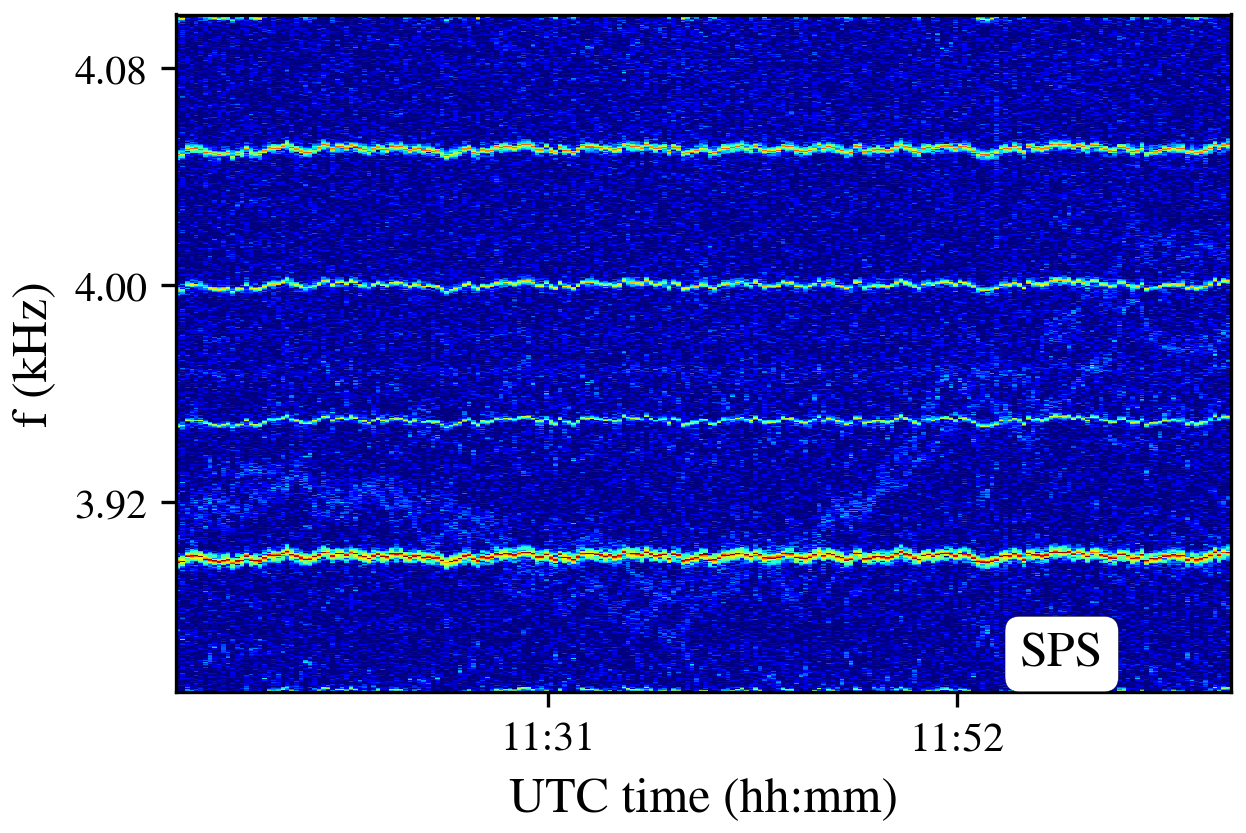} 
\caption{\label{fig:Spectrum_PS_SPS} The spectrum of the B-Train in PS (top) and SPS (bottom). The strong component at ~5 kHz in the PS spectrum is one of the switching frequencies of the switch-mode power supply. The SPS spectrum illustrates the signature of a SCR power supply, which is characterized by a series of 50~Hz harmonics with a frequency modulation induced by the mains.}
\end{figure}

Figure~\ref{fig:Spectrum_PS_SPS} shows the spectrogram of the magnetic measurements for the PS (top) and SPS (bottom). The PS spectrum reveals a strong component at 5~kHz, which is the frequency of some of its Switch-Mode power supplies \cite{PS}. The switching of this type of power supply is regulated by a clock. Subsequently, a negligible variation in the frequency evolution of the line is observed. As the switching frequencies are well defined, they can be easily identified and no 50 Hz harmonics are present in the signal. 

On the contrary, in the SPS case, the power supply is a Silicon Controlled Rectifier (SCR). Hence, the 50~Hz harmonics are visible on the signal and the stability of the mains has an impact on the output current of the power supplies. In the following studies, when the expected position of the 50~Hz harmonics is illustrated, the drift of the harmonics due to this modulation is taken into account.

Therefore, if environmental noise is excluded as the origin of the perturbation, the signature of the 50~Hz harmonics in the beam spectrum suggests that the possible sources are limited to magnets with SCR power supplies. The magnets powered by SCR in the LHC are presented in Table~\ref{tab:SCR} \cite{SCR, SCR2}. 
\begin{table}
\caption{\label{tab:SCR}%
The SCR power supplies in the LHC \cite{SCR, SCR2}.
}
\begin{ruledtabular}
\begin{tabular}{cc}
\textrm{Power supply type} &
\textrm{Use} \\ 
\colrule
 RPTE & Main dipoles \\ 
 RPTF & Warm quadrupoles (Q4, Q5) \\ 
 RPTG & Dogleg dipoles (D1-D4, spare)  \\ 
 RPTL & Alice compensator \\
 RPTM & Dump septa \\
 RPTI & Alice and LHCb dipoles \\
 RPTN & LHCb compensator \\
 RPTJ & CMS Solenoid \\
 RPTH & Alice Solenoid \\
 RPTK & RF Klystron \\
\end{tabular}
\end{ruledtabular}
\end{table}

 The phase evolution of the 50~Hz harmonics between two locations in the ring can clarify whether the power supply ripple lines are the result of a real beam excitation. To this end, their phase advance is measured between two closely separated pickups and compared to the betatron phase advance between the same pickups. For the validity of the comparison, the two observation points must be situated in a relatively close distance, so that the beam does not encounter a noise perturbation while crossing this path. 
 
 If the harmonics are a by-product of noise in the beam instrumentation then their phase advance is not expected to correspond to the betatronic one. Furthermore, an arbitrary dephasing between the low and high-frequency cluster should be observed. On the contrary, in the case of a real excitation, the power supply ripple phase advance must correspond to the betatronic one for all the harmonics present in the spectrum. 

Two pickups of the transverse damper, referred to as Q7 and Q9, are selected for the analysis. At collision energy, the betatron phase advance between the two observation points is defined by the optics and it is approximately equal to 110 degrees. The first step is to compute the complex Fourier spectra for a single pickup and for each bunch in a physics fill to observe the dephasing of the lines across the full machine. As previously reported, with the present noise floor, the evolution of the lines cannot be determined with a single bunch. For this reason, the average signal is computed from five consecutive LHC trains, each one of which consists of 48 bunches. Then, the phase evolution of each harmonic is computed across the accelerator.

\begin{figure}
\includegraphics[width = \columnwidth]{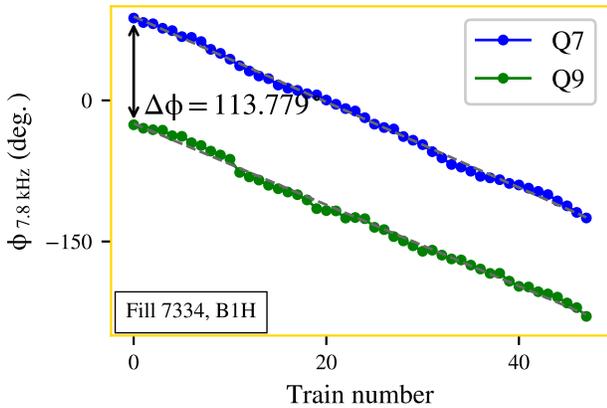}
\caption{\label{fig:Q7Q9} The phase evolution of the $h$=156 harmonic as a function of the train number for Q7 (blue) and Q9 (green). The gray lines illustrate the expected phase evolution of the harmonics in the accelerator (Appendix~\ref{appendix:bbb_spectrum}).}
\end{figure}

Figure~\ref{fig:Q7Q9} depicts the dephasing of a harmonic in the high-frequency cluster ($h$=156) as a function of the train number for Q7 (blue) and Q9 (green). The gray dashed lines illustrate the expected dephasing, which is proportional to the frequency and the time delay of the trailing trains from the first train in the machine (Appendix~\ref{appendix:bbb_spectrum}). It must be noted that by averaging over a few consecutive trains the signal is sub-sampled to \(f_{\rm rev}\), similarly to the single bunch case. The negative slope of the 7.8~kHz line shows that the phase evolution of the harmonic was computed through aliasing, i.e., by following the phase evolution of the reflected frequency component around the revolution frequency ($f_{\rm rev}\rm - 7.8~kHz$). The phase advance of each harmonic is the difference in the phase determination of the two pickups. A correspondence to the betatron phase advance is found, an observation that clearly proves, for the first time, that the two harmonics correspond to a real beam excitation. Similar results are obtained for all the harmonics present in the beam spectrum.

The filling scheme of the physics Fill 7334 is divided in three groups of consecutive trains. Each group corresponds to approximately one-third of the total beam. The average value and the standard deviation of the dephasing between Q7 and Q9 are computed from the three groups for all the harmonics above noise level. 

\begin{figure}
\includegraphics[width = \columnwidth]{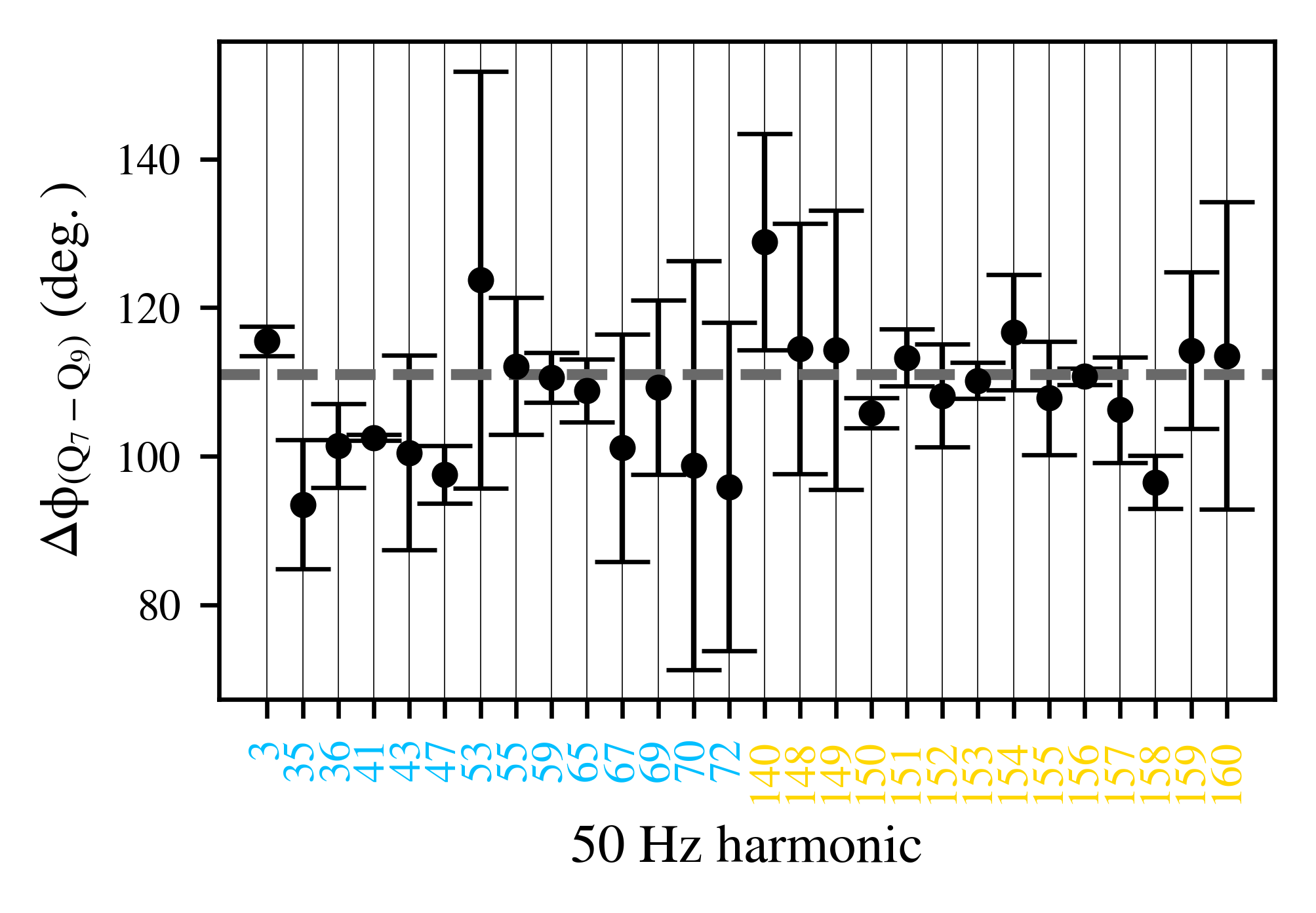} 
\caption{\label{fig:Q7Q9_all} The average phase advance from Q7 to Q9 for the harmonics in the low (blue) and high (orange) frequency cluster. The error bars represent one standard deviation and the gray dashed line illustrates the betatron phase advance (110 degrees).}
\end{figure}

Figure~\ref{fig:Q7Q9_all} demonstrates the average phase advance for the harmonics in the low (blue) and high (orange) frequency cluster. The error bars represent one standard deviation since following the frequency drift of lower-amplitude harmonics can introduce uncertainties. The gray dashed line indicates the betatron phase advance. The average value demonstrates that, within an uncertainty represented by the standard deviation, the phase advance for all the harmonics is the one of the beam, thus proving that the observations are not instrumental.

\subsection{Observations during changes in the beam and machine configuration}
\label{section_changing_operation}

\begin{figure}
\includegraphics[width = \columnwidth]{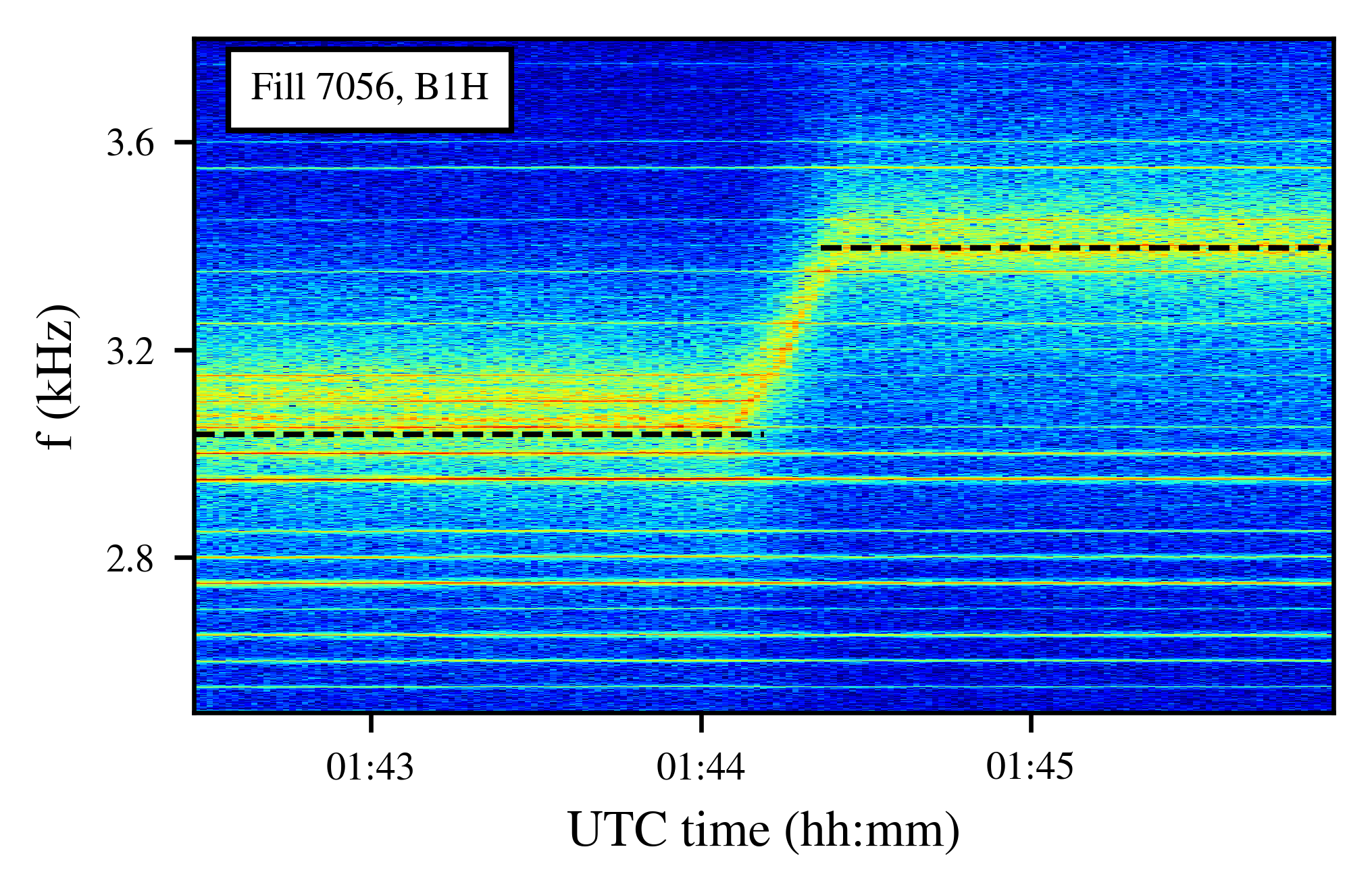} 
\caption{\label{fig:HS_BBQ_FT} The horizontal spectrogram of Beam 1 at Flat Top. The black dashed lines represent an approximation of the injection and the collision tune in the horizontal plane.}
\end{figure}

The response of the harmonics during a simple modification of the betatron motion such as the change of the tune at Flat Top is investigated. 
Figure~\ref{fig:HS_BBQ_FT} presents the HS BBQ spectrogram for the horizontal plane of Beam 1 in the physics Fill 7056. The spectrogram is centered around the betatron tune for the whole duration of Flat Top and a color code is assigned to the PSD. The black dashed line represents an approximation of the horizontal tune evolution. 

First, one must observe that the frequencies of the lines are not affected by the tune change. This fact proves that the harmonics are the result of a dipolar field error rather than a tune modulation \cite{kostoglou2020tune}. Second, a comparison prior and after the trim leads to the conclusion that the amplitude of the lines in the vicinity of the betatron tune is strongly enhanced. This resonant behavior is in agreement with a dipolar perturbation, with an excitation frequency that approaches the betatron tune \cite{previous}.

\begin{figure}
\includegraphics[width = \columnwidth]{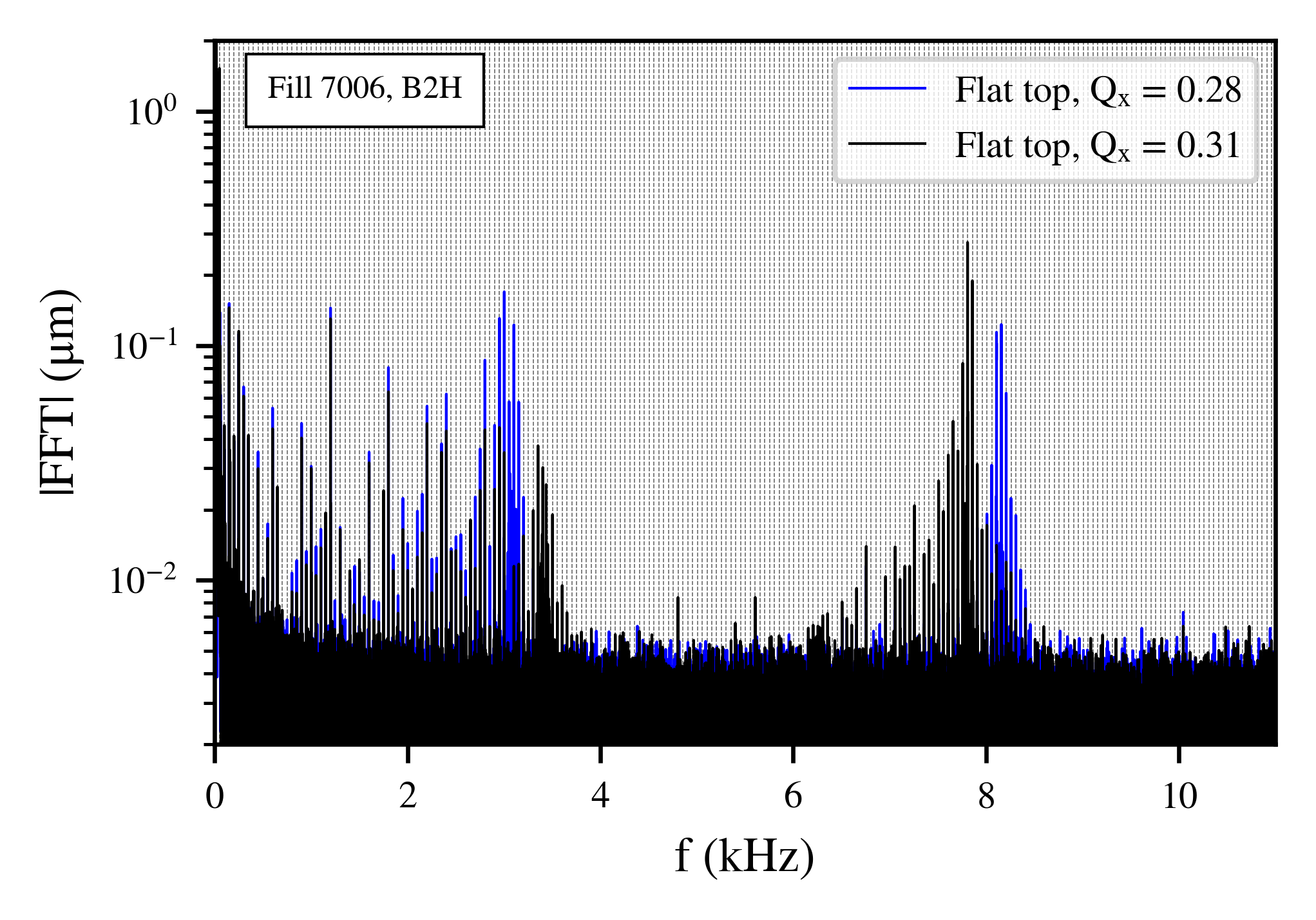} \\ 
\includegraphics[width = \columnwidth]{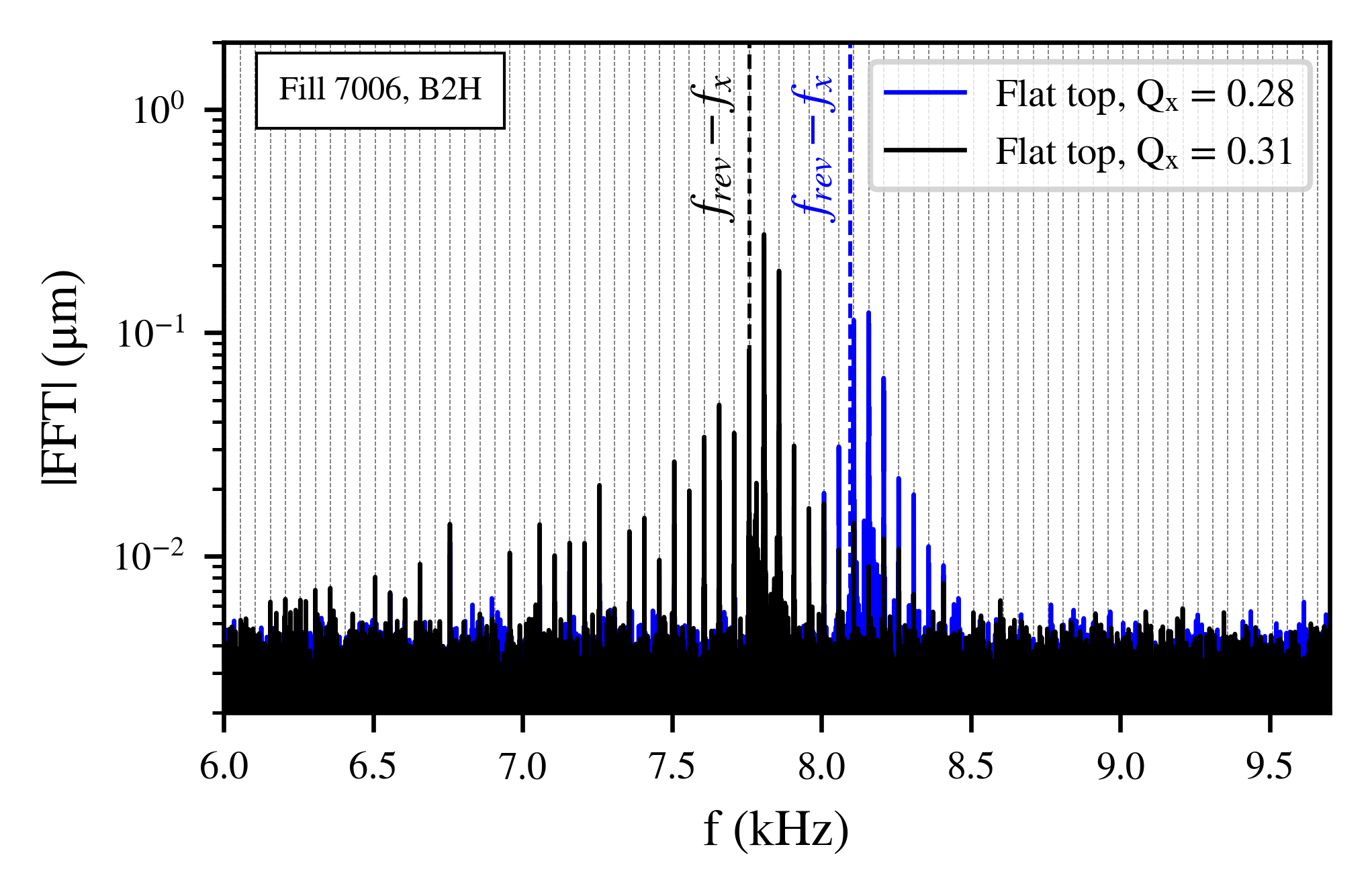} 
\caption{\label{fig:ADT_FT} The horizontal spectrum of Beam 2 prior (blue) and after (black) the tune change at Flat Top in a frequency range up to 10~kHz (top) and centered around the high-frequency cluster (bottom). The gray dashed lines represent the multiples of 50~Hz. The dashed blue and black vertical lines illustrate the location of $f_{\rm rev} - f_x$.}
\end{figure}

To investigate the impact of the tune change on the high-frequency cluster, the high bandwidth spectrum is computed from the ADTObsBox prior and after the tune trim. Figure~\ref{fig:ADT_FT} shows the horizontal spectrum of Beam 2 up to 10~kHz (top) at Flat Top with the injection (blue) and collision (black) tune. Similar observations exist for both beams and planes.

The closest harmonics to the tune of the low-frequency cluster are enhanced in terms of amplitude. A shift is also observed at the position of the high-frequency cluster. To further illustrate this effect, the spectrum is centered around the high-frequency cluster in Fig.~\ref{fig:ADT_FT} (bottom). Although the effect is dipolar in both cases (the harmonics coincide with the 50 Hz multiples indicated with the gray lines), this observation shows that the location of the cluster is at \(f_{\rm rev}-f_x\) and thus, depends on the betatron tune. The fact that the changes in the beam configuration affect the amplitude of the power supply ripple lines provides further proof that the harmonics are the result of a real beam excitation. %

The response of the harmonics is studied when another modification is applied to the betatron motion and specifically, to its phase advance, while the tune is constant. During the MD Fill 6984, the betatron phase advance between the Interaction Point (IP) 1 and 5 were modified \cite{IP15}. This was achieved through the incorporation of a set of optics knobs, which allow scanning the phase between the two IPs, based on the ATS scheme \cite{ATS}. The knobs lead to a trim in the current of the quadrupole families responsible for the control of the tune. Throughout these modifications, the betatron tune is constant.

\begin{figure}
\includegraphics[width = \columnwidth]{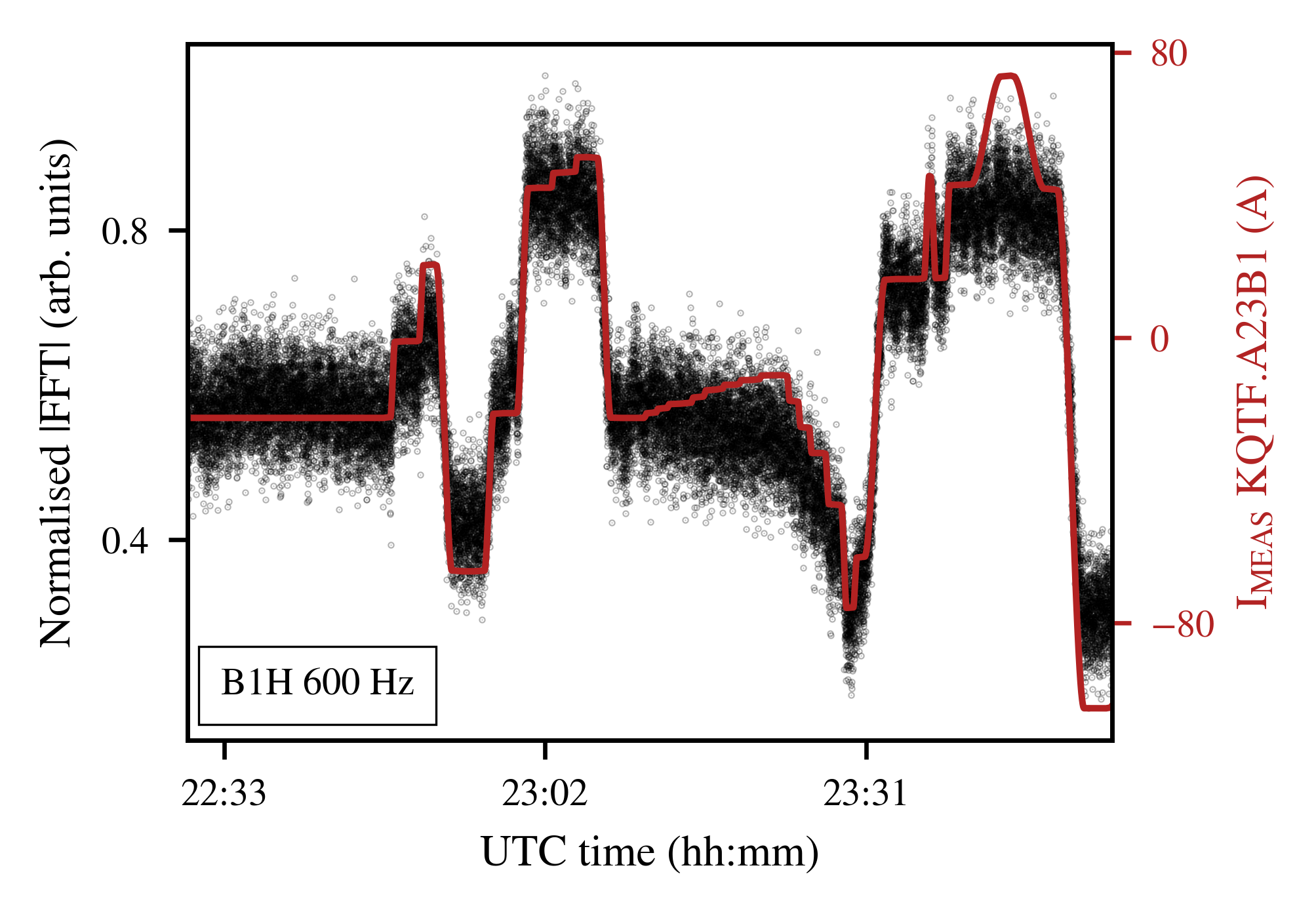}
\caption{\label{fig:HSBBQ_600Hz_IP15} The amplitude evolution of the 600~Hz lines (black) during the change of the phase advance between IP1 and IP5. while the tune is kept constant. The red line represents the current in the quadrupole trims employed for the phase scan.}
\end{figure}

Figure~\ref{fig:HSBBQ_600Hz_IP15} illustrates the supply current for a single quadrupole (red). The evolution of the current corresponds to a change of the phase advance within a range of \(\rm \pm\)20 degrees for the horizontal plane of Beam 1. During this time interval, the amplitude evolution of the harmonics is computed.

Figure~\ref{fig:HSBBQ_600Hz_IP15} also denotes the response of the $h$=12 harmonic (black curve). The amplitude evolution of the lines in the low-frequency cluster is clearly impacted by the variation of the betatron phase advance, an effect that provides definitive indications that they correspond to actual beam motion. As far as the high-frequency cluster is concerned, no impact is observed in their amplitude evolution throughout these tests, which is possibly due to the mitigation of these lines from the transverse damper as shown later in this paper.

Following the change of the betatron tune and phase advance, we explore the evolution of the spectrum across different beams modes and thereby, different energies and optics. Some of the magnets in Table~\ref{tab:SCR} can be excluded as potential sources by observing the evolution of the spectrum across the cycle.

First, due to the fact that the 50~Hz harmonics are systematically present in all beam modes and fills, the power supplies of the spare magnets and the septa are excluded. Specifically, the separation/recombination dipoles D1-D4 are currently powered by switch-mode power supplies, while the SCR power supplies of these dipoles are only used in case of failure. An analysis of the fills where the spare SCR power supplies were employed did not reveal any modification on the beam spectrum and thus, can be excluded. In addition, the magnetic field of the dump septa only affects the beam during extraction and not in operation.

Second, the amplitude of the power supply ripple lines does not significantly attenuate with increasing beam energy. Considering a non-ramping power supply as the source, a reduction of the angular deflection and thus, of the amplitude of the power supply ripple, should be observed with increasing beam rigidity. The absence of such an attenuation leads to the conclusion that the power supply ripple originates from a ramping power supply. Consequently, all non-ramping power supplies can also be excluded. 

Through this process of elimination, the remaining candidates of Table~\ref{tab:SCR} are the main dipoles and the warm quadrupoles. Combining this finding with previous indications of the dipolar nature of the source, the investigation focuses on the power supplies of the Main Bends. The main dipoles have undoubtedly the highest filling coefficient in the ring and, as previously mentioned, the studies conducted in other synchrotrons have proved that the arc circuit was systematically the dominant contributor.

Based on the previous findings, the pursued investigations focus on the main dipoles circuit. To establish a correlation between the harmonics of the beam and the ones in the output of their power supplies, a modification in the configuration of the latter is needed. An important observation was made when the status of the active filters of the Main Bends, which are installed for the attenuation of the 50~Hz ripple \cite{AF, AF2, burnet_active_filters}, was changed. During dedicated MD fills, the eight active filters were disabled sector-by-sector. 

\begin{figure}
\includegraphics[width =\linewidth]{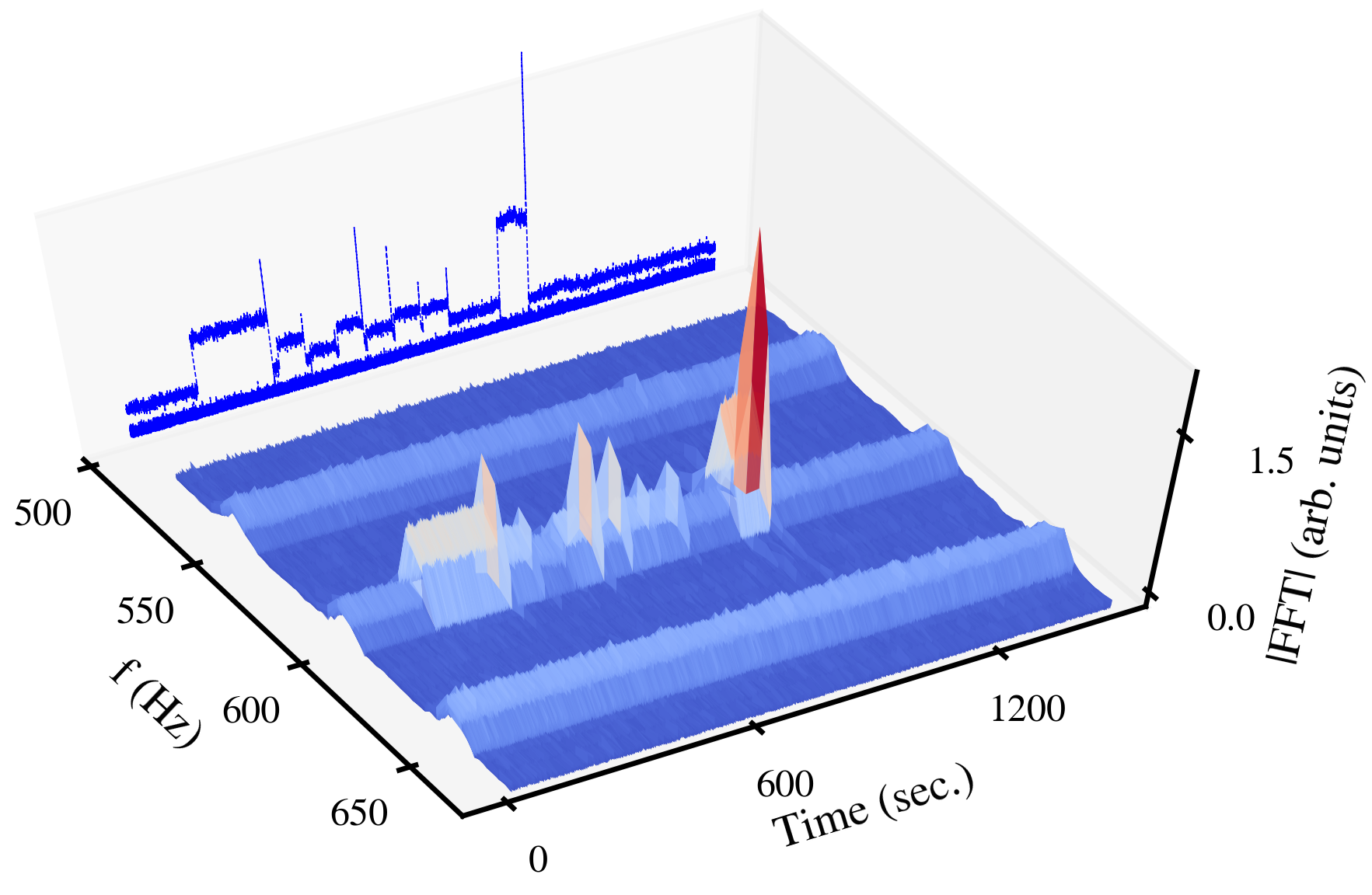} 
\caption{\label{fig:3D_spectrogram_AF} The horizontal spectrogram of Beam 1 during the tests of the active filters at injection centered around 600~Hz. The blue lines represent the amplitude evolution of the spectral components in this regime.}
\end{figure}

Figure~\ref{fig:3D_spectrogram_AF} depicts the 3D spectrogram for the horizontal plane of Beam 1, as acquired from the MIM, for the time interval of the tests conducted at injection (Fill 7343). For a first demonstration, the frequency range is limited around 600~Hz. The projection of the spectrogram, which represents the amplitude evolution of the $h$=12 harmonic, is shown with the blue curve. Disabling the eight filters leads to abrupt changes in its amplitude evolution.

\begin{figure}
\includegraphics[width = \columnwidth]{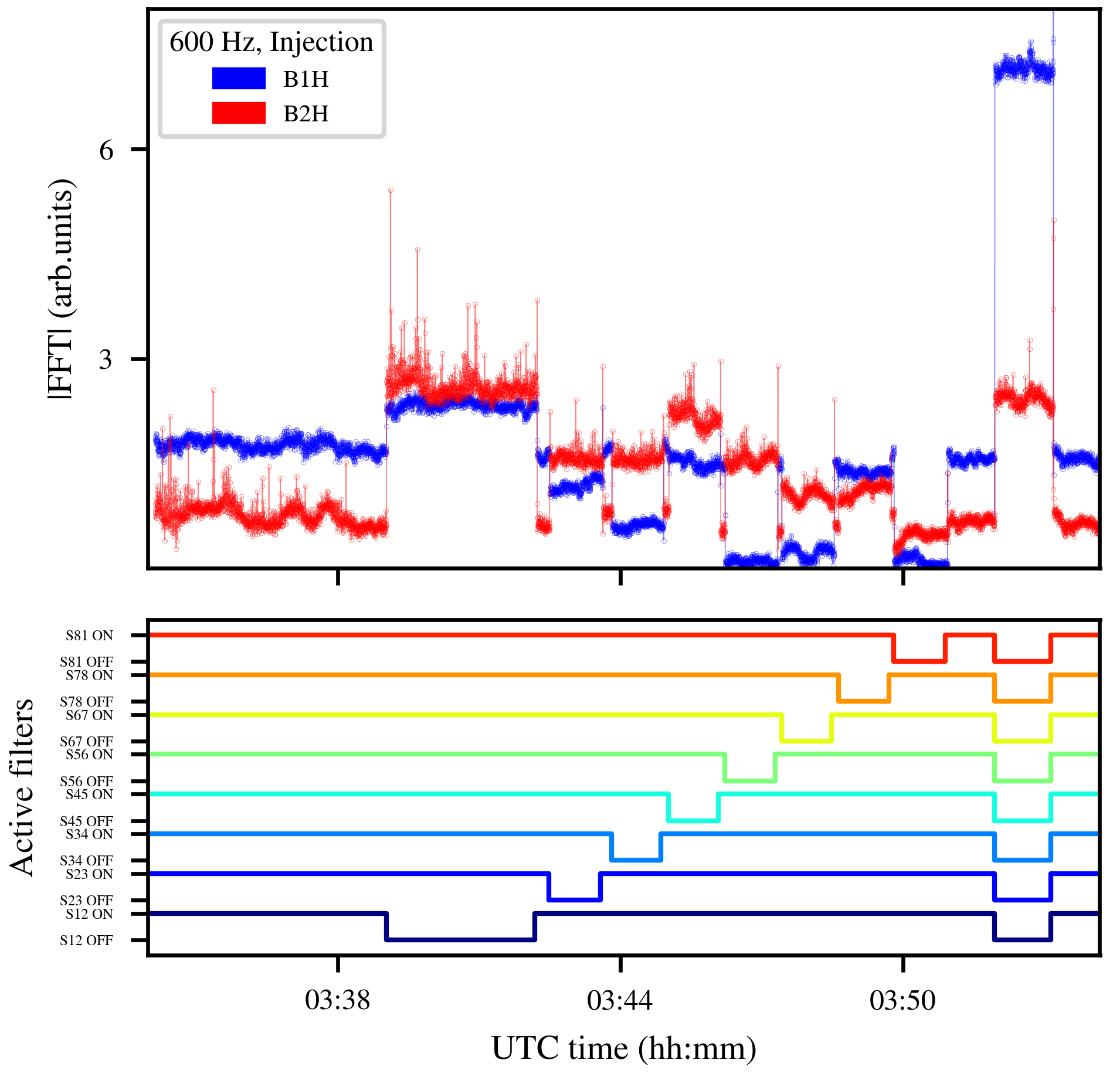}
\caption{\label{fig:HSBBQ_AF_600Hz} The response of the $h$=12 harmonic (top) during the tests with the active filters for Beam 1 (blue) and Beam 2 (red) at injection energy. The status of the active filters is color-coded with the sector number (bottom).}
\end{figure}

The amplitude evolution of the $h$=12 harmonic is extracted from the 3D spectrogram. Figure~\ref{fig:HSBBQ_AF_600Hz} (top) presents the response of the 600~Hz line in Beam 1 (blue) and 2 (red) at injection energy. The status of the eight active filters is presented for the same time span (bottom) and a color code is assigned to each sector. The distinct changes in the amplitude coincide with the disabling of the filter of each sector. As a last step, the filters were disabled simultaneously, which led to an important increase in the amplitude of the line. Similar observations are collected by conducting the same tests at top energy.

The observations on the $h$=12 harmonic provide evidence that all eight power supplies contribute to this effect. The question that arises is whether the most impacted sectors in terms of power supply ripple can be identified. Reviewing the results of Fig.~\ref{fig:HSBBQ_AF_600Hz} yields that the positive or negative impact of the filter compared to the baseline, which is defined as the amplitude of the harmonic prior to the test, depends on the sector. For instance, at injection in Beam 1, disabling the Filter of sector 1-2 leads to an increase of the ripple amplitude. Therefore, the filter, when active, suppresses the harmonic and its impact is characterized as positive. On the contrary, sector 5-6 has a negative contribution at injection. Then, comparing the same sector across the two beams reveals a different impact between the two (e.g., sector 5-6 at injection). This can be possibly attributed to the different phase advance of the two beams in the ring. Finally, a comparison between the results at injection and top energy reveal that the contribution for the same beam and sector also depends on the beam energy.

\begin{figure}
\includegraphics[width = \columnwidth]{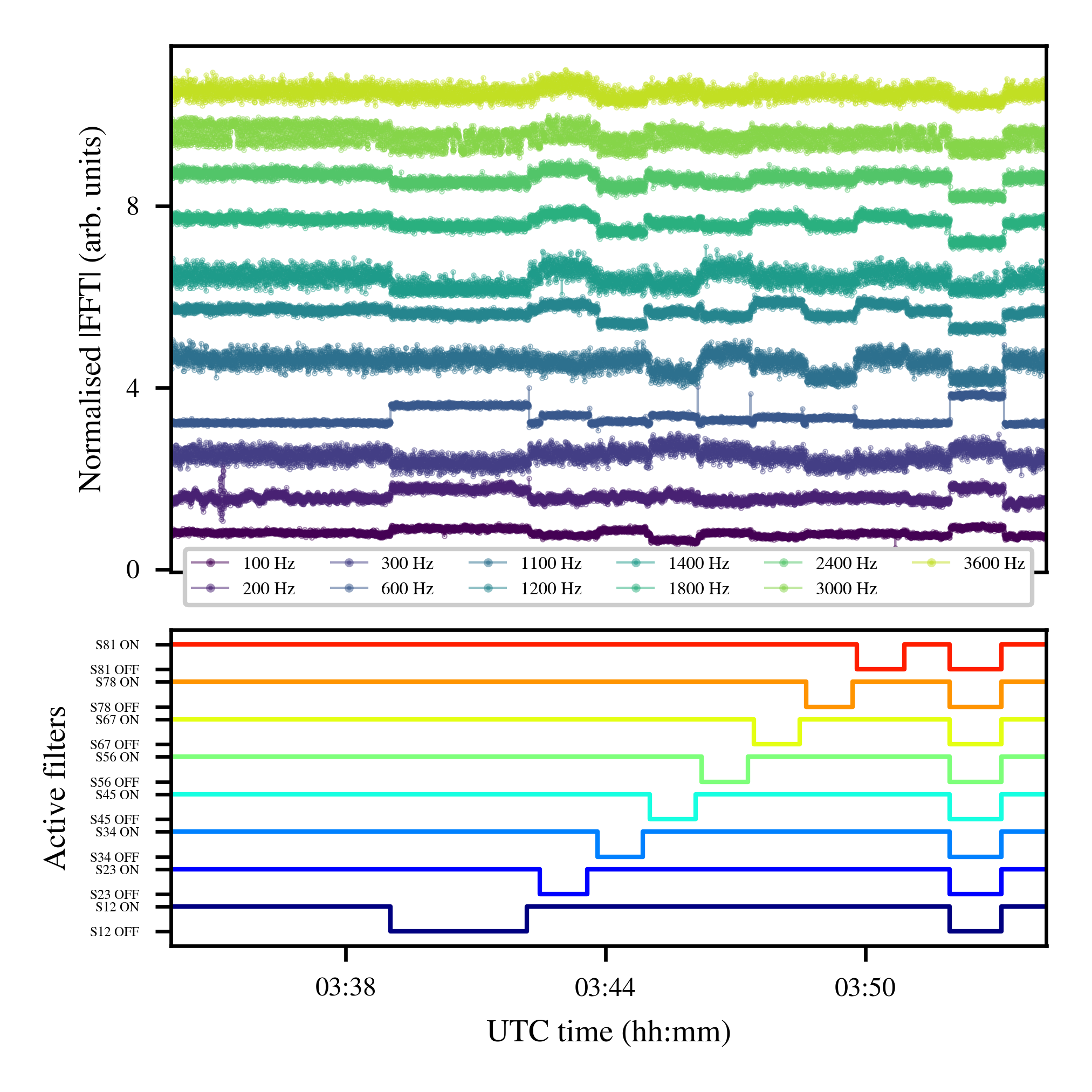} 
\caption{\label{fig:MIM_AF} The amplitude evolution of the harmonics in the low-frequency cluster (up to 3.6~kHz) during the tests with the active filters. The status of the active filters is color-coded with the sector number (bottom).}
\end{figure}

The correlation with the power supplies is not only valid for the 600~Hz line, but for most of the 50~Hz harmonics included in the low-frequency cluster. Figure~\ref{fig:MIM_AF} shows the amplitude evolution of various harmonics at injection, represented with a different color code. The abrupt changes in the amplitude when disabling the active filter of each sector are reproduced for harmonics up to 3.6~kHz. In addition to the observations at 600~Hz, the contribution of each sector also varies across the harmonics.

Applying a simple modification in the configuration of the dipole power supplies, such as modifying the status of the active filters, has a direct impact on the low-frequency cluster harmonics of the beam. These results provide evidence that the power supplies of the main dipoles are the source of the harmonics up to 3.6~kHz observed in the beam spectrum. It is the first time that such a correlation has been demonstrated in the LHC. As no modifications are envisaged for the power supplies of the main dipoles, this perturbation is expected to be present also in the future upgrade of the LHC, the High Luminosity-LHC (HL-LHC) \cite{Apollinari:2284929}. It must also be underlined that no change in the amplitude evolution of the harmonics in the high-frequency cluster is reported during these tests. 

\begin{figure}
\includegraphics[width = \columnwidth]{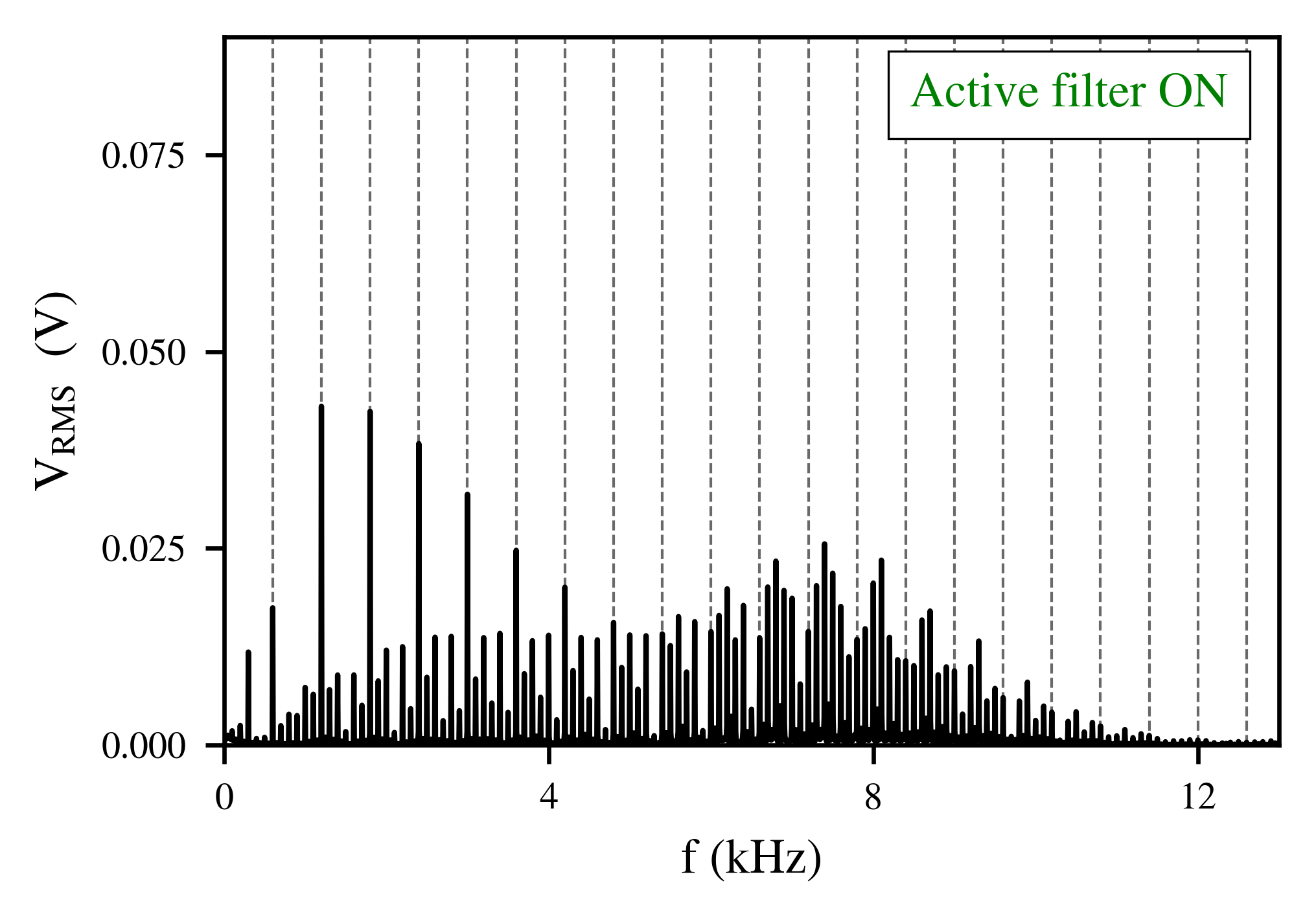} 
\includegraphics[width = \columnwidth]{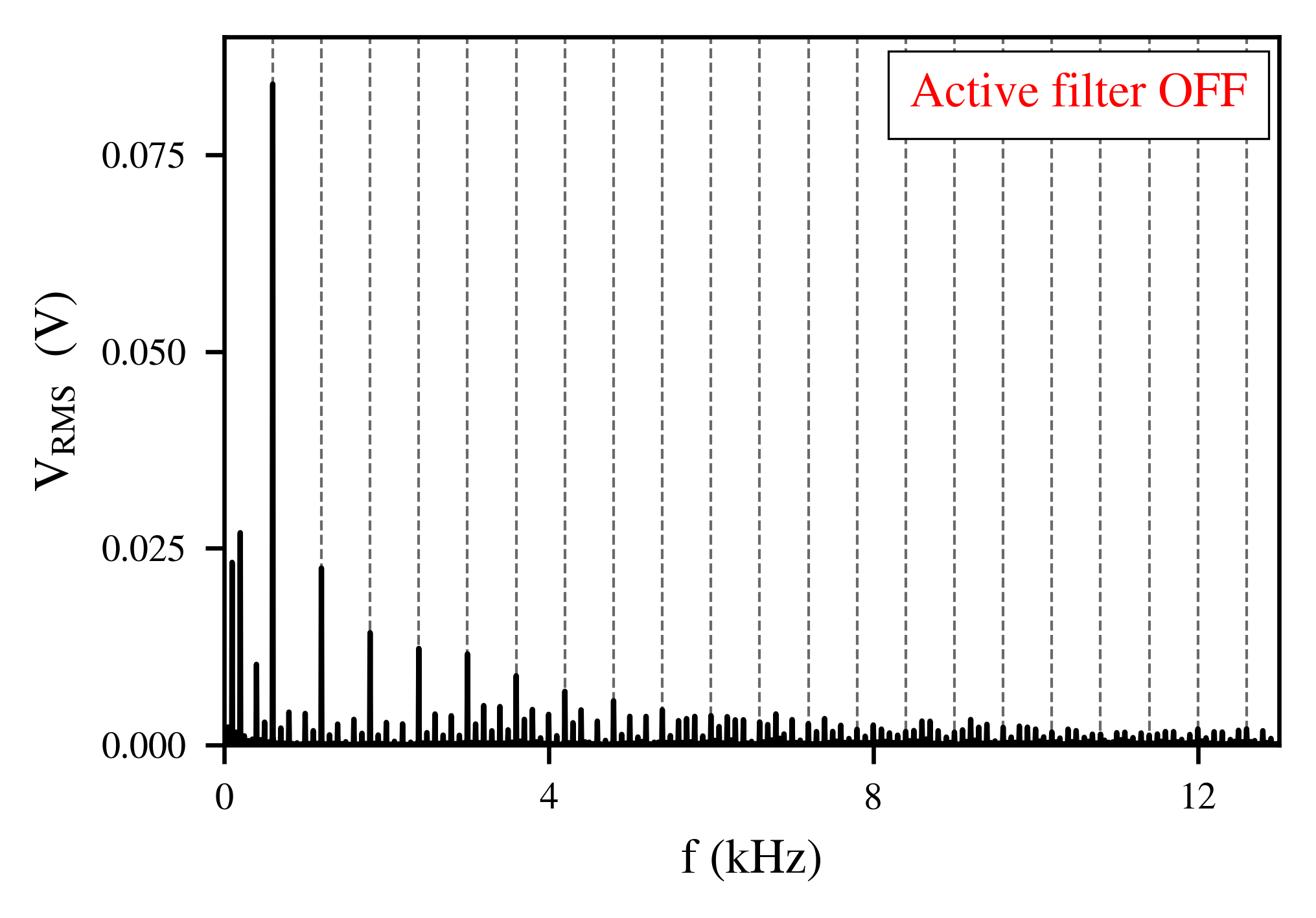} 
\caption{\label{fig:S12_AF_ON_OFF} Voltage spectrum of the power supply of the main dipoles in one of the LHC arcs (Sector 1-2) with the active filter enabled (top) and disabled (bottom). The vertical gray lines represent the multiples of 600~Hz.}
\end{figure}

Figure~\ref{fig:S12_AF_ON_OFF} demonstrates the voltage spectrum of the power supply in one of the LHC sectors, first, when the active filter is enabled (top) and, then, disabled (bottom). In this case, the vertical lines represent the multiples of 600~Hz. The comparison of the spectrum prior and after the modification shows that the active filter is suppressing some of the harmonics up to approximately 3~kHz, while it enhances the high-order harmonics \cite{ burnet_active_filters}. However, the amplitude of the high-frequency cluster in the beam spectrum did not increase when disabling the active filter possibly due to the interplay of this cluster with the transverse damper as explained later in the present paper.

The spectra of both beams and planes in a physics fill are measured and compared. As the pickups for the two beams and planes are located at different positions in the ring, the ADTObsBox calibrated spectra are normalized with the corresponding \(\beta\)-functions. The $\beta$-functions at the location of the ADTObsBox Q7 pickup for both beams and planes, at injection and top energy are listed in Table~\ref{tab:betas_Q7}.

\begin{table}
\caption{\label{tab:betas_Q7}%
The $\beta$-functions at the position of the Q7 pickup per beam and plane with injection and collision optics ($\beta^*=30 \ cm $).
}
\begin{ruledtabular}
\begin{tabular}{cccc}
\textrm{Beam} & \textrm{Plane} &
\textrm{$\beta$  at injection (m) } & \textrm{$\beta$  at collision (m) }\\ 
\colrule
1 & Horizontal & 130.9 & 100.7 \\
1 & Vertical & 173.6  & 99.6 \\ 
2 & Horizontal & 151.1 & 109.6 \\
2 & Vertical &  166.3 & 166.3 \\
\end{tabular}
\end{ruledtabular}
\end{table}

\begin{figure}
\includegraphics[width = \columnwidth]{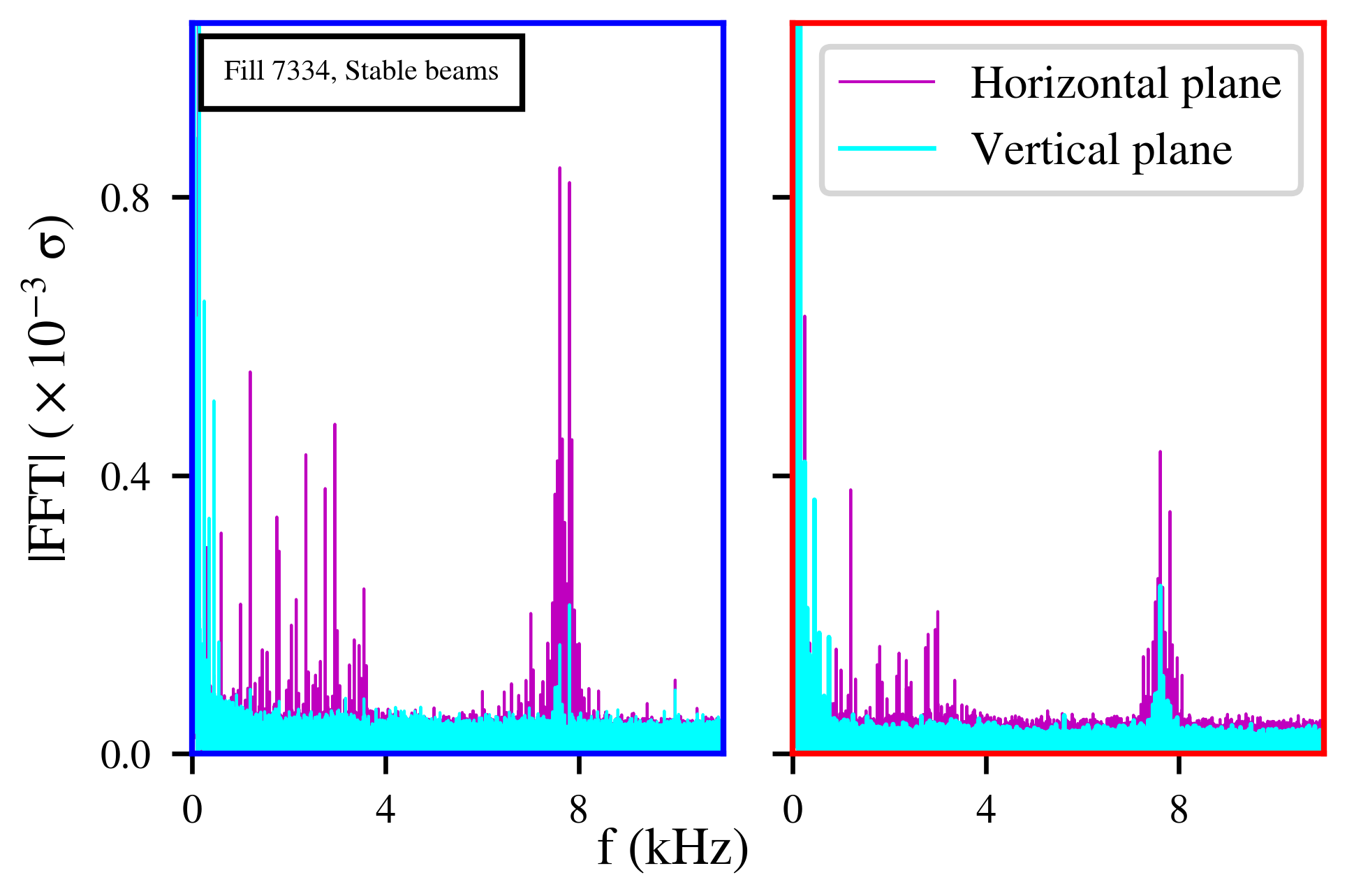} 
\caption{\label{fig:ADT_sectrum_horizontal_vertical}The spectrum of the horizontal (magenta) and vertical (cyan) plane of Beam 1 (left) and 2 (right) in Stable Beams, normalized with the corresponding $\beta$-functions.}
\end{figure}

Figure \ref{fig:ADT_sectrum_horizontal_vertical} shows the spectra for the horizontal (magenta) and vertical (cyan) plane for Beam 1 (left) and 2 (right). 
Comparing the amplitudes of the spectral lines yields that the perturbation is mainly affecting the horizontal plane, an effect compatible with a field error of the main dipoles. Due to the transverse coupling of the machine, an attenuated perturbation is also present in the vertical plane. To demonstrate that this effect results from the coupling, controlled excitations have been applied using the transverse damper during dedicated experiments. Although only the horizontal plane was excited, the oscillation was visible also in the vertical plane. 

The maximum offset observed in the horizontal spectrum of Beam 1 is approximately \(\rm 0.1 \ \mu m\), which corresponds to \(\rm 10^{-3} \ \sigma\). As shown in \cite{previous}, assuming a single dipolar perturbation, this value corresponds to a deflection of 0.09~nrad at a location with \(\beta=\rm 105 \ m\) for an excitation frequency in the vicinity of the tune (\( |Q-Q_p|=\rm 5\times 10^{-3}\), where $Q$ is the betatron tune and $Q_p$ is the ripple tune). Comparing the equivalent kick with the bending angle of a single dipole in the LHC (\(\rm \approx\)5~mrad) and neglecting additional effects (e.g. transverse damper, magnet's beam screen) yields a field stability of \(\rm 1.8\times 10^{-8}\), a value which is well within the power supply specifications.

Comparing the spectra of the two beams yields an asymmetry in the amplitude of the ripple between Beam 1 and 2 and that a more significant effect is visible in Beam 1. To verify the reproducibility of this observation, the spectra of both beams and planes are computed for all the proton physics fills of 2018. For each fill, the maximum offset induced by the 50~Hz harmonics is computed, which corresponds, in Stable Beams, to a frequency of 7.7~kHz. 

\begin{figure}
\includegraphics[width = \columnwidth]{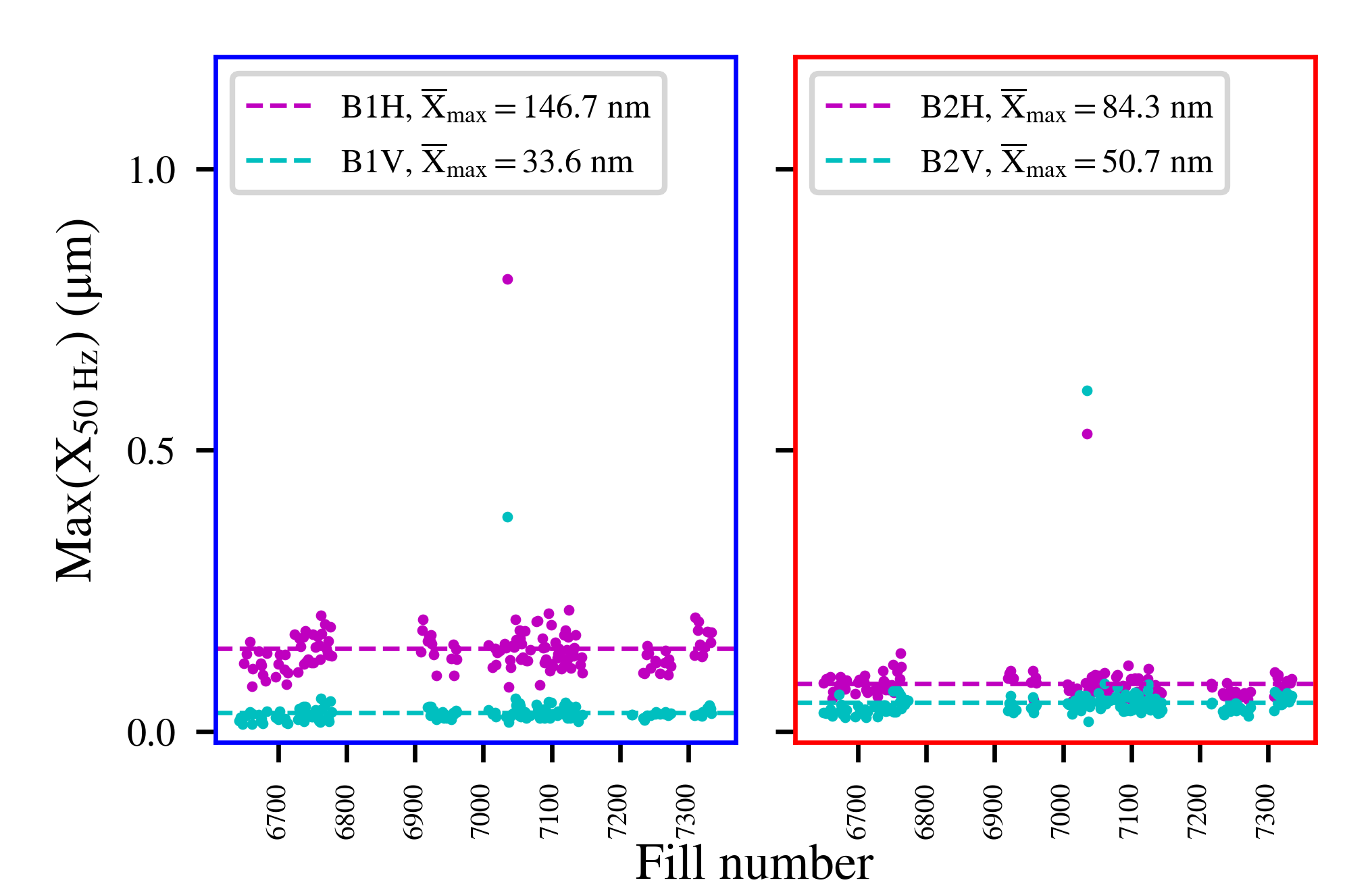} 
\caption{\label{fig:ADT_sectrum_horizontal_vertical_all} The maximum amplitude of the 50~Hz harmonics for the horizontal (magenta) and vertical plane (cyan) of Beam 1 (blue) and 2 (red) for all the proton physics fills of 2018.}
\end{figure}

Figure~\ref{fig:ADT_sectrum_horizontal_vertical_all} depicts the maximum amplitude observed in the spectrum as a function of the fill number for the horizontal (magenta) and vertical (cyan) plane in Beam 1 (blue) and 2 (red). The dashed lines represent the average offset over all the fills for each plane. These results confirm that the power supply ripple is systematically more pronounced in Beam 1 than Beam 2 by approximately a factor of two in the horizontal plane. Although the dipole power supplies are common for both beams, this discrepancy is possibly attributed to the different phase advances of the two beams in the accelerator.

The fill-by-fill analysis of the spectra in Fig.~\ref{fig:ADT_sectrum_horizontal_vertical_all} reveals an increase of the ripple amplitude in the physics Fill 7035. An additional parameter that has not been included in the analysis so far is the activity of the transverse damper and the interplay with the 50~Hz harmonics. 

In the nominal LHC cycle, the ADT settings are modified and in particular, the extended ADT bandwidth is changed to standard bandwidth at the end of Adjust \cite{Dubouchet:2012hzl, Komppula2019ADTAO}. An increase in the gain of the transverse damper is reported for the standard bandwidth compared to the extended one. In the Fill 7035, this modification was not applied and the extended bandwidth was used at Stable Beams. 

\begin{figure}
\includegraphics[width = \columnwidth]{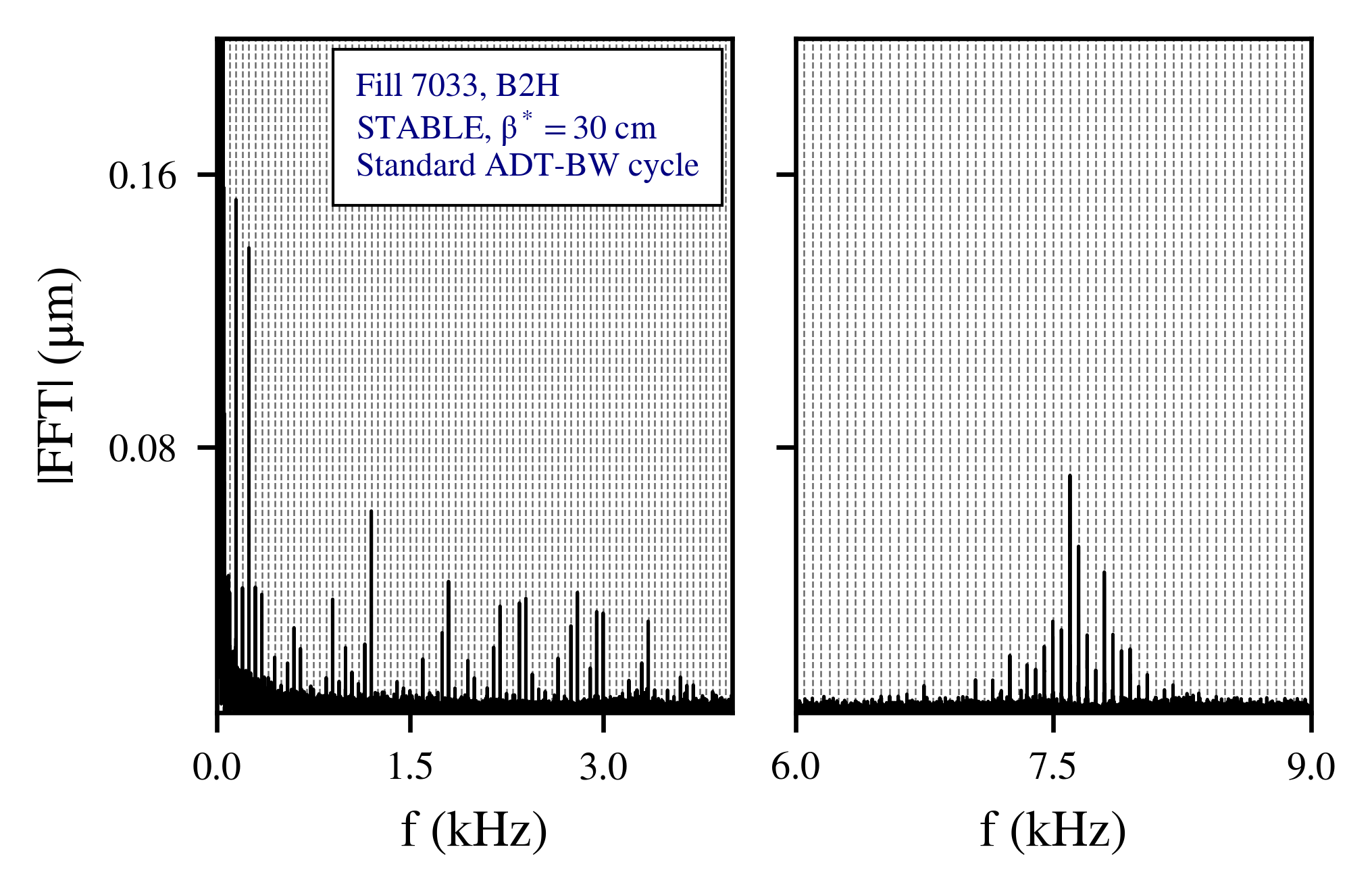} \\
\includegraphics[width = \columnwidth]{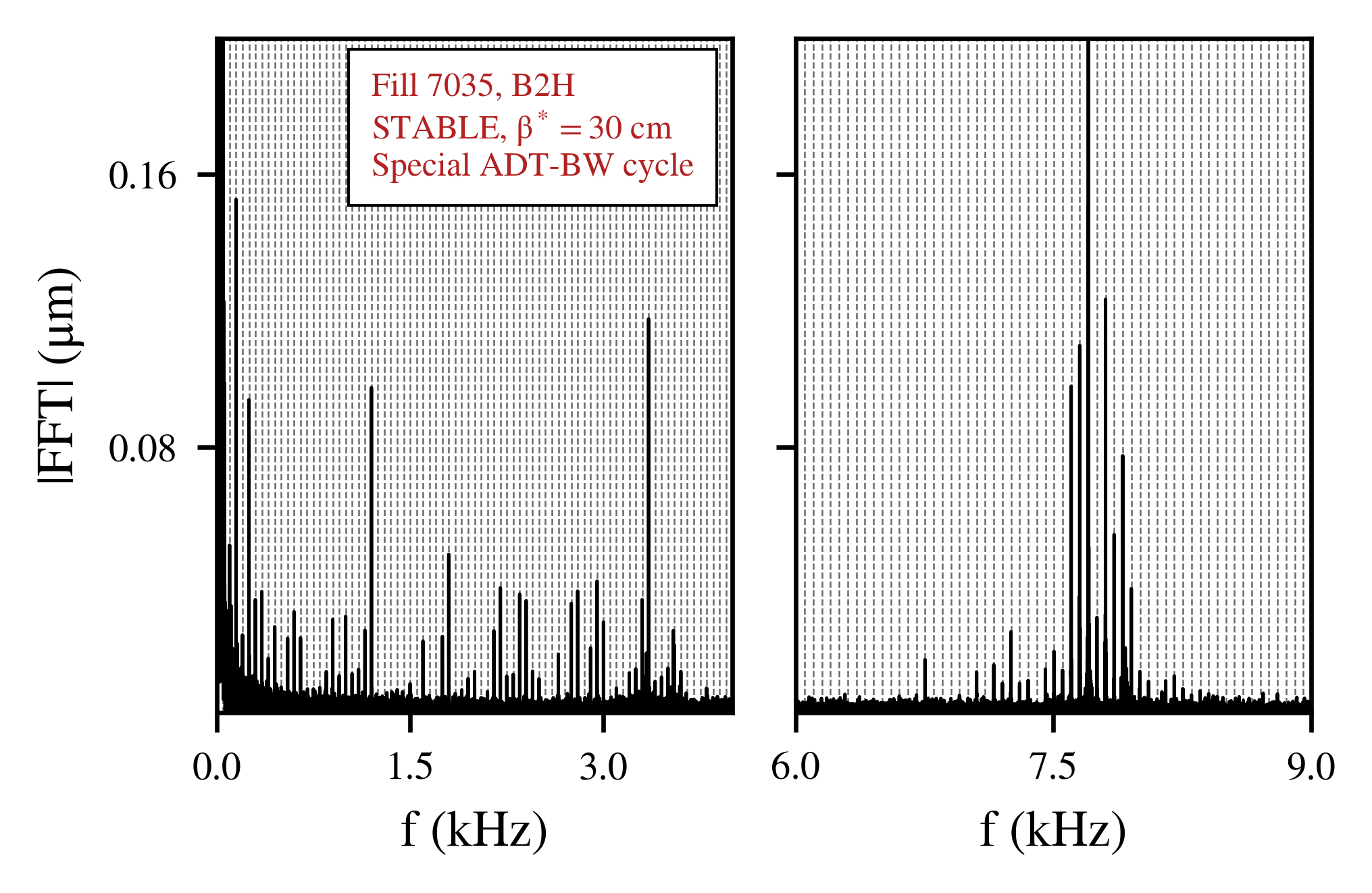} 
\caption{\label{fig:Fill_7035_spectra} The horizontal spectrum of Beam 2 at Stable Beams centered around the low (left) and high (right) frequency cluster for a fill with the standard (Fill 7033, top) and (b) extended (Fill 7035, bottom) ADT bandwidth.}
\end{figure}

Figure ~\ref{fig:Fill_7035_spectra} illustrates the horizontal spectrum of Beam 2 at Stable Beams centered around the low (left) and high (right) frequency cluster and for the Fills 7033 (top) and 7035 (bottom) with the standard ADT bandwidth and extended bandwidth, respectively. Comparing the two spectra yields an increase in the amplitude of the 50~Hz harmonics in the regime above 3~kHz, which is particularly important for the high-frequency cluster. 

This observation indicates that the amplitudes of the high-order harmonics are reduced by the damper in normal operation. This also explains why an amplitude increase of the high-frequency cluster was not observed when the corresponding harmonics in the power supply's voltage spectrum increased during the active filters tests. The impact of the ADT settings is also systematically observed in other beam modes of the machine cycle during which the bandwidth was modified such as the Adjust.

The importance of this finding resides on the fact that a strong asymmetry is present between the frequencies of the low and high cluster in terms of amplitude. In particular, these observations indicate that, lowering the transverse damper gain or in the absence of the damper, the amplitude of the harmonics in the high-frequency cluster is expected to be further enhanced compared to the values that have been observed experimentally. 

In contrast, Fig.~\ref{fig:S12_AF_ON_OFF} shows that the ripple in the power supply voltage spectrum attenuates with increasing frequency. Although the active filters enhance the high-order harmonics in the power supply voltage spectrum, their amplitude is still lower than the ones of the low-order harmonics. Furthermore, high-frequency perturbations such as the high-frequency cluster strongly exceed the cutoff frequency of the LHC main dipoles due to the shielding effect of the beam screen \cite{dipoles_transfer_function}. To this end, if the high-frequency cluster is driven by a direct excitation due to power supply ripple, a significant attenuation of its amplitude should be observed compared to the low-frequency cluster, while experimentally we observed the opposite.

Additionally, it should be mentioned that the increase of the power supply ripple by a factor of two in Fill 7035 did not lead to an increase of losses or emittance growth compared to the rest of the fills. However, as the duration of the fill was limited to 40 minutes, the impact of the power supply ripple lines on the beam lifetime cannot be excluded. 

To conclude, Table~\ref{tab:summary} presents a summary of the most important experimental observations for each cluster. Combining this information suggests that, rather than a direct excitation, the high-frequency cluster is the result of the interplay between ripple in the dipole power supplies and a mechanism originating from the beam. The transfer function from the power supply voltage to the magnetic field seen by the beam does not consider the beam's response. Therefore, the asymmetry between the two clusters can be explained if there is a higher sensitivity of the beam's response in the regime \(f_{\rm rev} - f_x\) compared to $f_x$, leading to important offsets from small power supply ripple perturbations at these frequencies. A potential candidate is the interplay of the beam with the machine transverse impedance, as the first unstable mode is at \( f_{\rm rev} - f_x\) \cite{Ruggiero:1995kv}. Other ripple source such as the Uninterruptible Power Supply (UPS) must be investigated \cite{UPS}. Further observations and experiments are necessary to identify the exact mechanism that is responsible for the high-frequency cluster. 

\begin{table}
\caption{\label{tab:summary}%
The summary of the observations for the low and high-frequency cluster.
}
\begin{ruledtabular}
\begin{tabular}{cc}
\textrm{low-frequency cluster\footnote{Extending up to 3.6~kHz.}}&
\textrm{high-frequency cluster\footnote{Located at 7-8~kHz, depending on the tune.}} \\
\colrule
\multicolumn{2}{c}{Presence of 50~Hz harmonics} \\
\multicolumn{2}{c}{Frequency modulation from the mains} \\
\multicolumn{2}{c}{Phase advance Q7-Q9 compatible with betatronic} \\
\multicolumn{2}{c}{Dipolar nature} \\
\multicolumn{2}{c}{Mainly in the horizontal plane} \\
\multicolumn{2}{c}{Larger amplitudes in Beam 1} \\
\thead{Impact from IP1/5 phase scan} & \thead{Impact from change of tune} \\
\thead{Impact from active filters} & \thead{Mitigation from transverse damper} \\
\end{tabular}
\end{ruledtabular}
\end{table}

\section{Summary and conclusions}
The purpose of the current study was to investigate the origin of the 50~Hz harmonics, an effect that has been observed in the beam signal since the start of the LHC operation. For this reason, a detailed review of the beam spectrum during several beam and machine configurations has been performed that revealed the existence of harmonics in two regimes in the frequency domain: the low-frequency cluster that extends up to 3.6~kHz and the high-frequency cluster at the location \( f_{\rm rev}-f_x\). The methodology presented in this paper allowed us to identify, for the first time in the LHC operation, the existence of the high-frequency cluster on the beam signal.  Although many similarities have been identified between the low and high-frequency cluster, the need to distinguish the two regimes is justified by their different response when modifications in the machine configuration are applied.

It is concluded that the two regimes are the result of a real beam excitation. A common signature between the two clusters has been identified; both regimes consist of a set of 50~Hz harmonics, that experience a frequency modulation induced by the mains. These findings indicate that the low and high-frequency clusters emerge from a common source. The signature of the harmonics in both regimes is compatible with a ramping SCR power supply. Based on the fact that the harmonics are multiples of 50~Hz rather than sidebands around the tune, and that the horizontal plane is mainly affected, it is concluded that the nature of the source is dipolar. 

In terms of amplitude, an asymmetry between the two clusters has been identified. In particular, more significant amplitudes of the beam oscillations are reported for the high-frequency cluster. During the proton run of 2018, the measured effect of the power supply ripple in the horizontal plane of Beam 1 was larger than the one of Beam 2 by a factor of two.

As far as the low-frequency cluster is concerned, a correlation with the eight thyristor, line-commutated power supplies of the Main Bends is established, through experiments with the active filters. It is concluded that the eight power supplies of the main dipoles are the source of the low-frequency cluster in the transverse beam spectrum. It is the first time that such a correlation has been demonstrated in the LHC operation.

The amplitude of the beam oscillations in the high-frequency cluster is larger if compared to the low-frequency cluster, hence the importance to identify its origin. If both clusters emerge from a common source, the question that arises is what is the mechanism that allows these high-frequency components to excite the beam. Oscillations at such high frequencies are expected to be significantly attenuated by the shielding effect of the beam screen in the dipole magnets. A review of the power supply's voltage spectrum reveals that there is a reduction of the ripple with increasing frequency. 

On the contrary, experimental observations indicate the presence of important spectral components in the high frequency cluster. More interestingly, the amplitude increase of the lines when the ADT settings are modified indicates that, in normal operation, a mitigation of the high-frequency cluster occurs due to the transverse damper. This fact underlines that, in the absence of the damper, the amplitude of the high-frequency cluster is expected to be further enhanced. The exact mechanism responsible for the appearance of the high-frequency cluster on the beam spectrum must be identified in the future operation of the accelerator.

Based on the source and the fact that no modifications are foreseen for the power supplies of the main dipoles, the 50~Hz harmonic are also expected to be present in the HL-LHC era. It has been demonstrated that the transverse damper can effectively reduce the amplitude of these harmonics and its capabilities can be employed in the future to mitigate this power supply ripple perturbation.

The analysis presented in this paper improves our understanding of the ripple effects that were present during the LHC operation. In the context of these studies, a general framework for the analysis of the experimental data has been developed, which can also be used to address other types of noise effects.  

\begin{acknowledgments}
The authors gratefully acknowledge H.~Bartosik, A.~Bland, O.~S.~Brüning, X.~Buffat, J.-P.~Burnet, L. R.~Carver, R.~De Maria, D.~Gamba, R.~T.~Garcia, M.~Martino, V.~Montabonnet, N.~Mounet, D.~Nisbet, H.~Thiesen, Y.~Thurel and J.~Wenninger for valuable suggestions and discussions on this work. We would like to thank D.~Valuch and M.~Soderen for all ADT related measurements and experiments, T.~Levens for the MIM measurements and discussions, M.~C.~Bastos and C.~Baccigalupi for the power supply acquisitions and discussions and J.~Olexa for the DOROS measurements. 
\end{acknowledgments}

\appendix

\section{Beam spectrum from bunch-by-bunch acquisitions}
\label{appendix:bbb_spectrum}
In the presence of a regular filling scheme, the bunch-by-bunch and turn-by-turn ADTObsBox data can be combined to increase the effective bandwidth of the instrument. Signal averaging is not only needed to access the high-frequency components of the signal without aliasing, but also to reduce the noise floor of the spectrum compared to the single bunch case. Averaging the signals of \(N_b\) bunches yields a \(\sqrt{N_b}\) increase in the signal to noise ratio, in the presence of random noise with zero mean that is uncorrelated with the signal \cite{snr}. 

\begin{figure}[htp]
\includegraphics[width = \columnwidth]{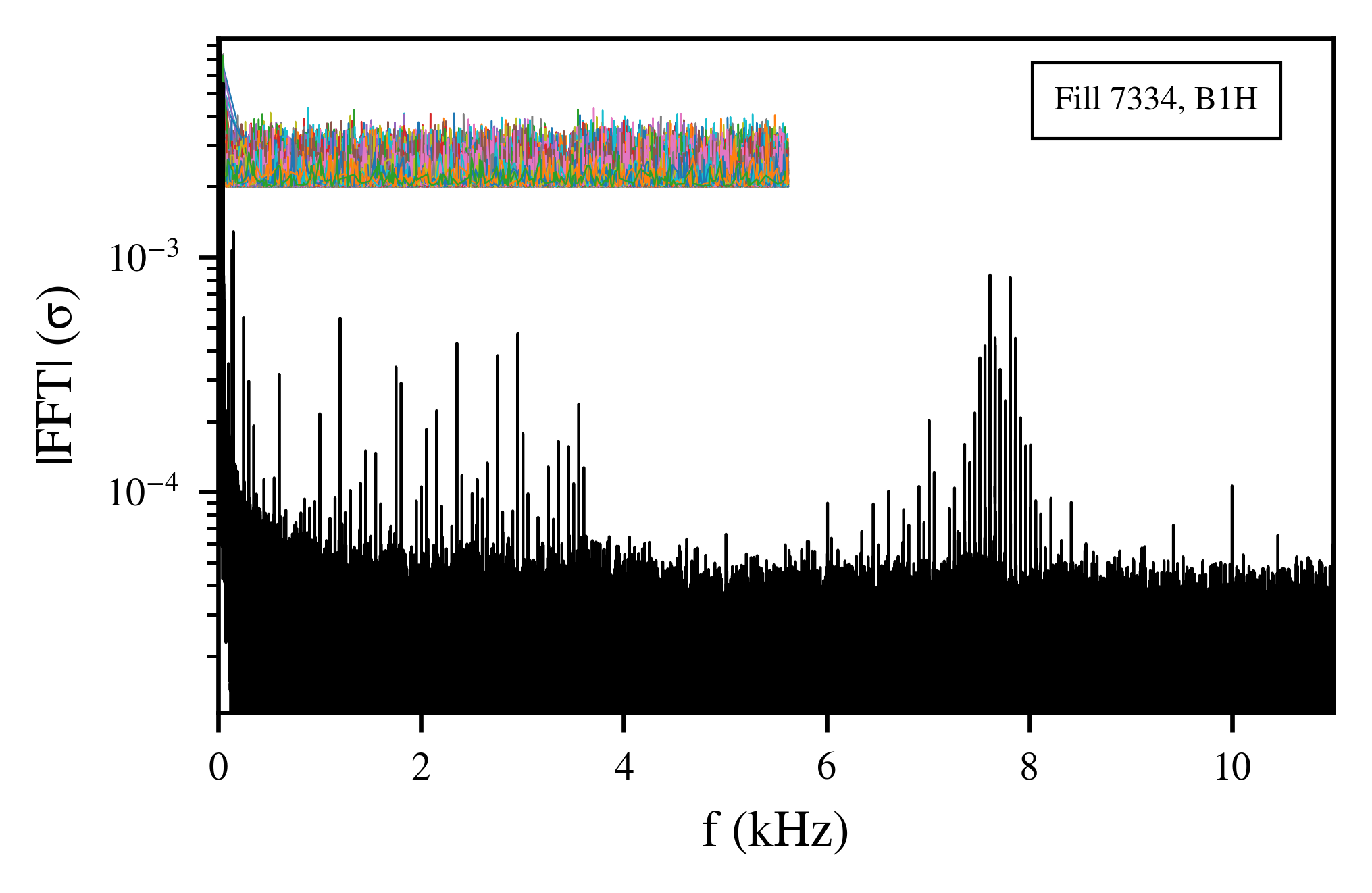} 
\caption{\label{app:fig:single_bunch_vs_average} The spectral envelope of several individual bunches (colored lines) and the spectrum after averaging over all the bunches in the machine (black).}
\end{figure}

The spectrum of individual bunches and after averaging over all the bunches in the machine is shown in Fig.~\ref{app:fig:single_bunch_vs_average} for the horizontal plane of Beam 1, for a physics fill and a window length of $4\times 10^4$ turns. The colored lines show the envelope of the spectra of several individual bunches, which is computed by setting a parametric peak threshold of $\rm 2\times10^{-3} \ \sigma$. The single bunch noise floor is approximately one order of magnitude higher than the 50~Hz harmonics and thus, signal averaging is necessary. 

The time delay \(\Delta t_i\) of a trailing bunch \(i\) with respect to the first bunch in the machine, considered as the reference, results in a phase angle:
\begin{equation}
\label{eq:dephasing}
\Delta \phi_i = 2\pi f \Delta t_i, 
\end{equation}
where \(f\) is the frequency under consideration. Consequently, the dephasing of the signals across the ring is proportional to the frequency and the longitudinal spacing of the bunches in the machine. 

To illustrate this effect, three trains of 48 bunches are considered in simulations with a dipolar excitation at 3~kHz. The bunch spacing is 25~ns and the trains are equally spaced in the LHC ring. The complex spectrum is computed for each bunch and the phase evolution of the 3~kHz line is extracted. 

\begin{figure}
\includegraphics[width = \columnwidth]{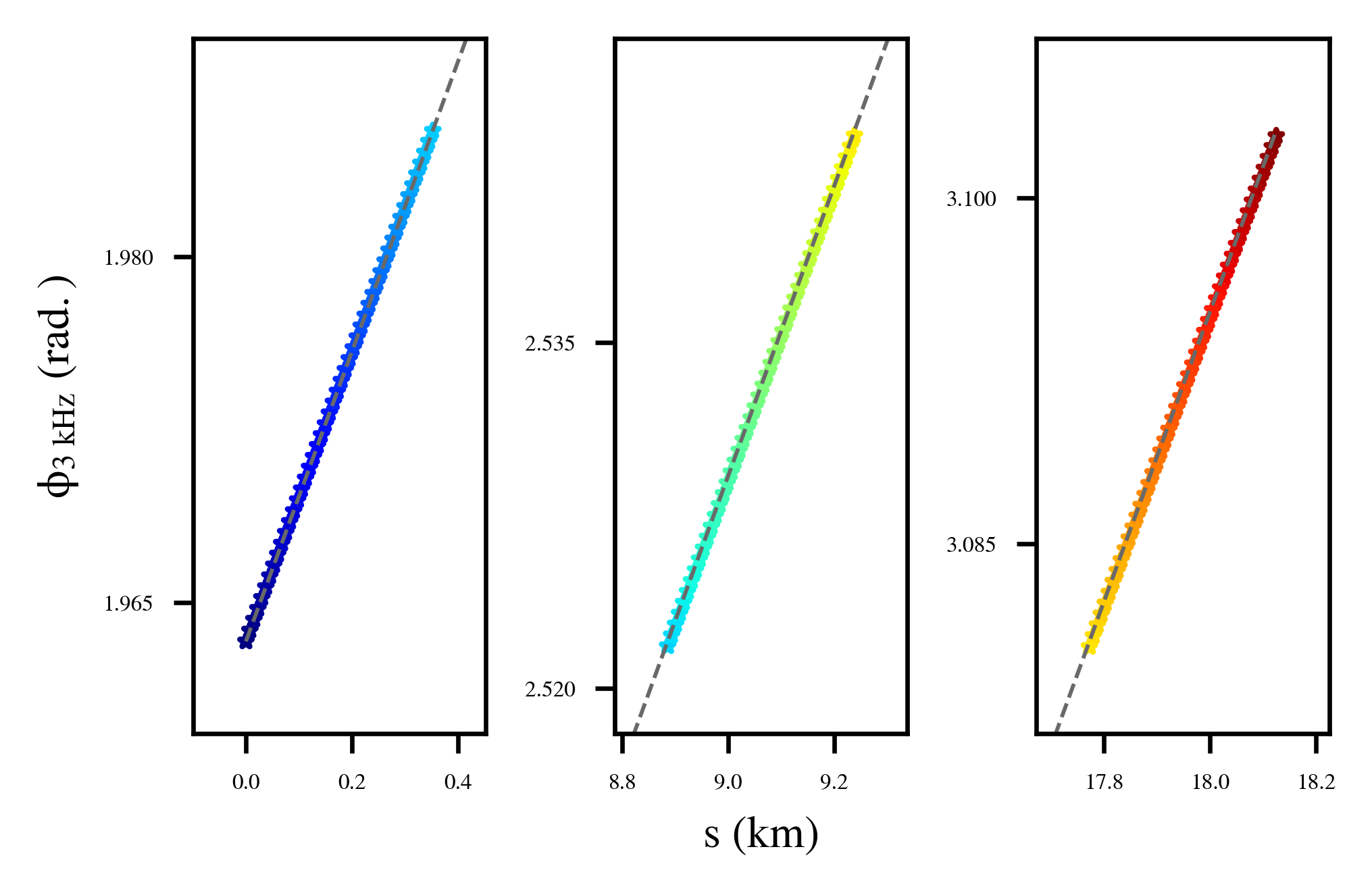}
\caption{\label{fig:bbb_dephasing} Phase evolution of the excitation at 3~kHz as a function of the bunch position for three trains of 48 bunches in the LHC ring. The dashed gray line represents the expected dephasing computed from Eq.~\eqref{eq:dephasing}.}
\end{figure}

Figure~\ref{fig:bbb_dephasing} depicts the phase evolution of the excitation for the three trains as a function of the bunch position in the ring. The color code represents the bunch number and the gray line is the expected phase evolution of Eq.~\eqref{eq:dephasing}, with respect to the first bunch that is considered as the reference. The linear phase evolution of an excitation across the trains in the machine has been experimentally verified by injecting power supply ripple with the transverse damper kicker.

For frequencies much lower than the sampling frequency \((f \ll f_{\rm rev} )\), the dephasing is negligible and the bunch-by-bunch data can be directly averaged in time domain. For frequencies comparable to the revolution frequency, such as the high-frequency cluster, the dephasing between the bunches cannot be neglected. In this case, simply averaging the bunch-by-bunch information will lead to an error in the resulting metric. To illustrate this effect, the first bunches of the three trains are selected.

\begin{figure}
\subfloat{\subfigimg[width = 0.98\columnwidth]{ \textbf{a)}}{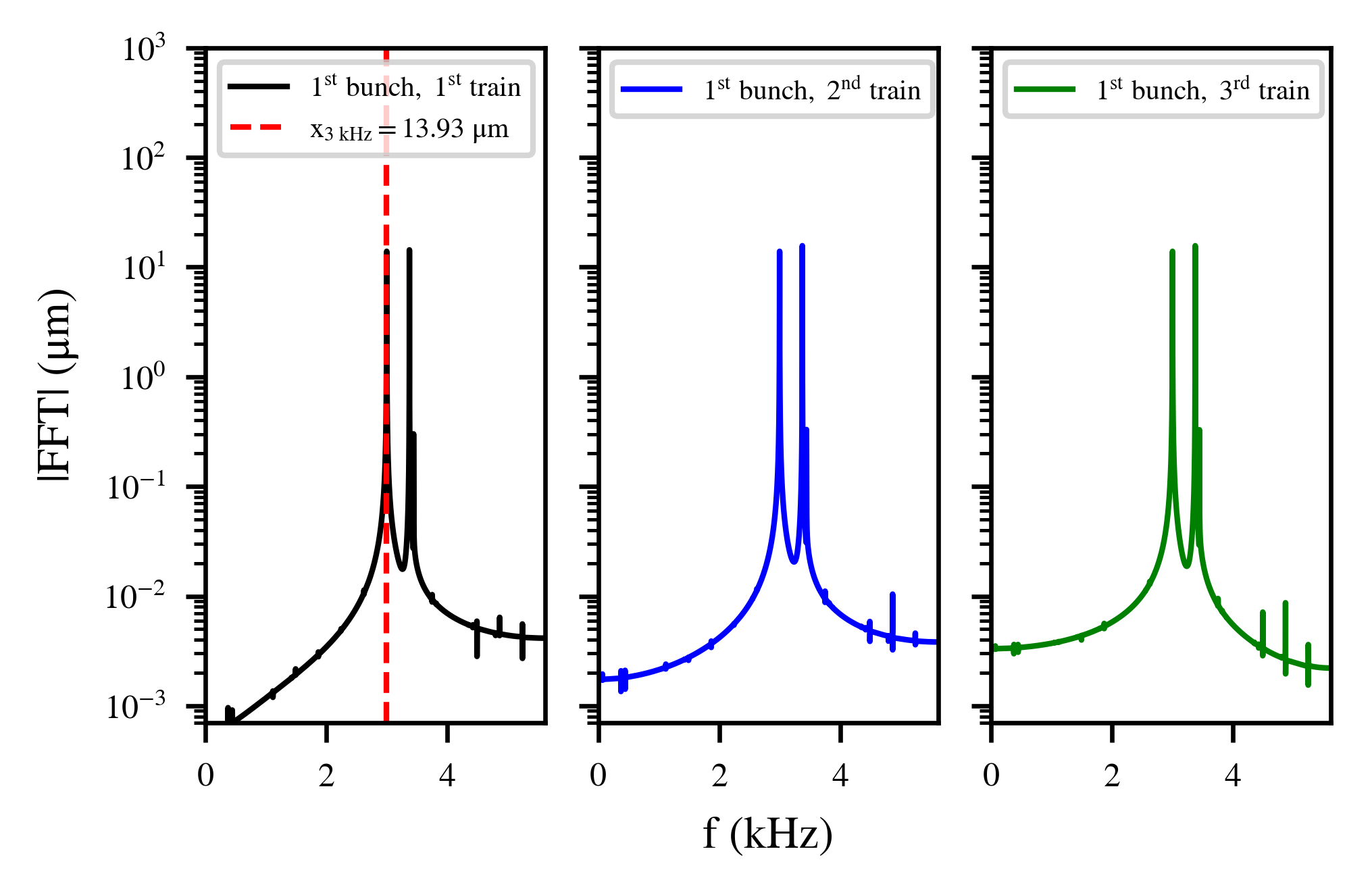} \label{subfig:bbb_app1}} \\
\subfloat{\subfigimg[width = 0.98\columnwidth]{\textbf{b)}}{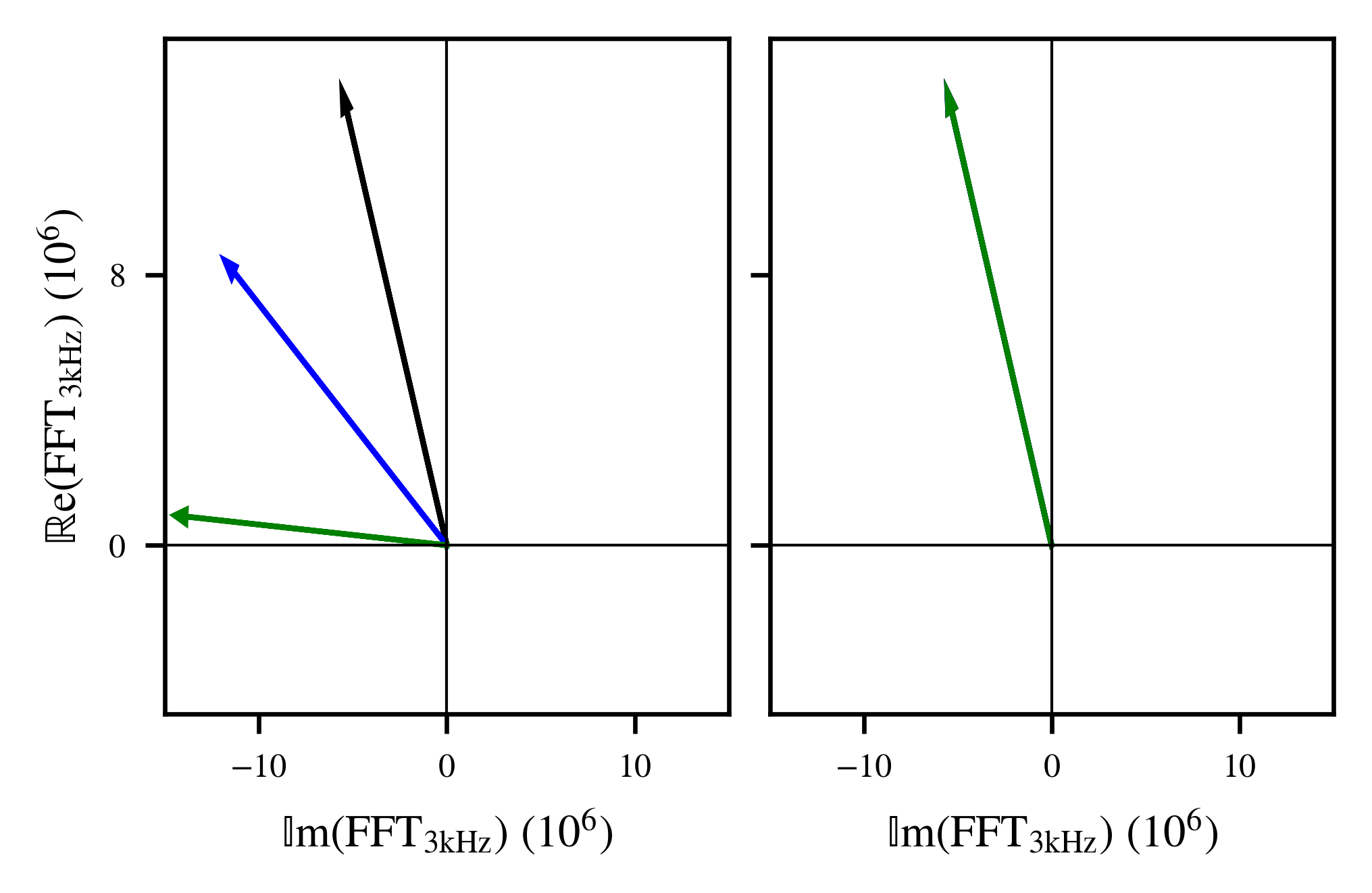} \label{subfig:bbb_app2}} 
\caption{\label{fig:3_bunches_excitation} (a) Spectrum of the first bunches of the first (black), second (blue) and third (green) train in the presence of a dipolar excitation at 3~kHz (red dashed line). (b) Phase of the excitation for the three bunches before (left) and after (right) the correction as computed from Eq.~\eqref{eq:correction}.}
\end{figure}

Figure~\ref{fig:3_bunches_excitation} illustrates the spectra for the first bunches (Fig.~\ref{subfig:bbb_app1}) of the first (black), second (blue) and third (green) train, respectively, in the presence of a dipolar excitation at 3~kHz. The excitation results in an offset of \(\rm 13.9~\mu m\) (red dashed line), while the second peak corresponds to the betatron tune.

Then, the complex Fourier coefficients at 3~kHz are computed. Figure~\ref{subfig:bbb_app2} presents the vector of the excitation in the spectrum, whose angle corresponds to the phase of the excitation, for each bunch (left). For a filling scheme consisting of three trains located in azimuthally symmetric locations in the ring, the dephasing at 3~kHz is important. Averaging over the three vectors without correcting for the dephasing will lead to an error in the offset of the final spectrum.

To this end, an algorithm that applies a phase correction has been implemented. The steps of the method are the following: first, the complex spectra \(F_i(\omega)\) are computed for each bunch, where \(\omega=2 \pi f\). Then, a rotation is applied to correct for the dephasing of Eq.~\eqref{eq:dephasing}. The impact of the rotation is depicted in the second plot of Fig.~\ref{subfig:bbb_app2}. Finally, the average over all bunches is computed. The procedure is described by the following expression:  
\begin{equation}
\label{eq:correction}
F(\omega) = \frac{1}{N_b}\sum_{i=1}^{N_b} F_i(\omega) \text{e}^{- j \omega \Delta t_i}.
\end{equation}

\section{Impact of a frequency modulation of the fundamental frequency on its harmonics}
\label{app:fm}
This section presents the impact of a frequency modulation on a harmonic dipolar excitation, similar to the one observed in the 50~Hz harmonics. To simulate this effect, a single particle is tracked in the LHC lattice using the single-particle tracking code SixTrack in the presence of a dipole field error \cite{sixtrack,sixtrack2}. 

The dipole strength is modulated with the absolute value of a sinusoidal function:
\begin{equation}
\label{eq:fm}
    \Delta k (t) = |A_r \cos{(2\pi f_m t )}|,
\end{equation}
where $t$ is the time, $A_r$ the amplitude and $f_m$ the frequency, which experiences a frequency modulation. The expression of $f_m$ as a function of the fundamental frequency $f_{0}$ is:
\begin{equation}
f_m = f_{0} + A_{\rm FM} \sin{(2\pi f_{\rm FM} t)},
\end{equation}
where $A_{\rm FM}=A_{\rm FM}'/(2\pi f_{\rm FM})$ is the amplitude and $f_{\rm FM}$ the frequency of the frequency modulation. This perturbation mimics a non-linear transfer function exciting all the even harmonics of the fundamental frequency $f_{0}$ that also experiences a low-frequency modulation at $f_{\rm FM}$. 

\begin{figure}
\includegraphics[width = \columnwidth]{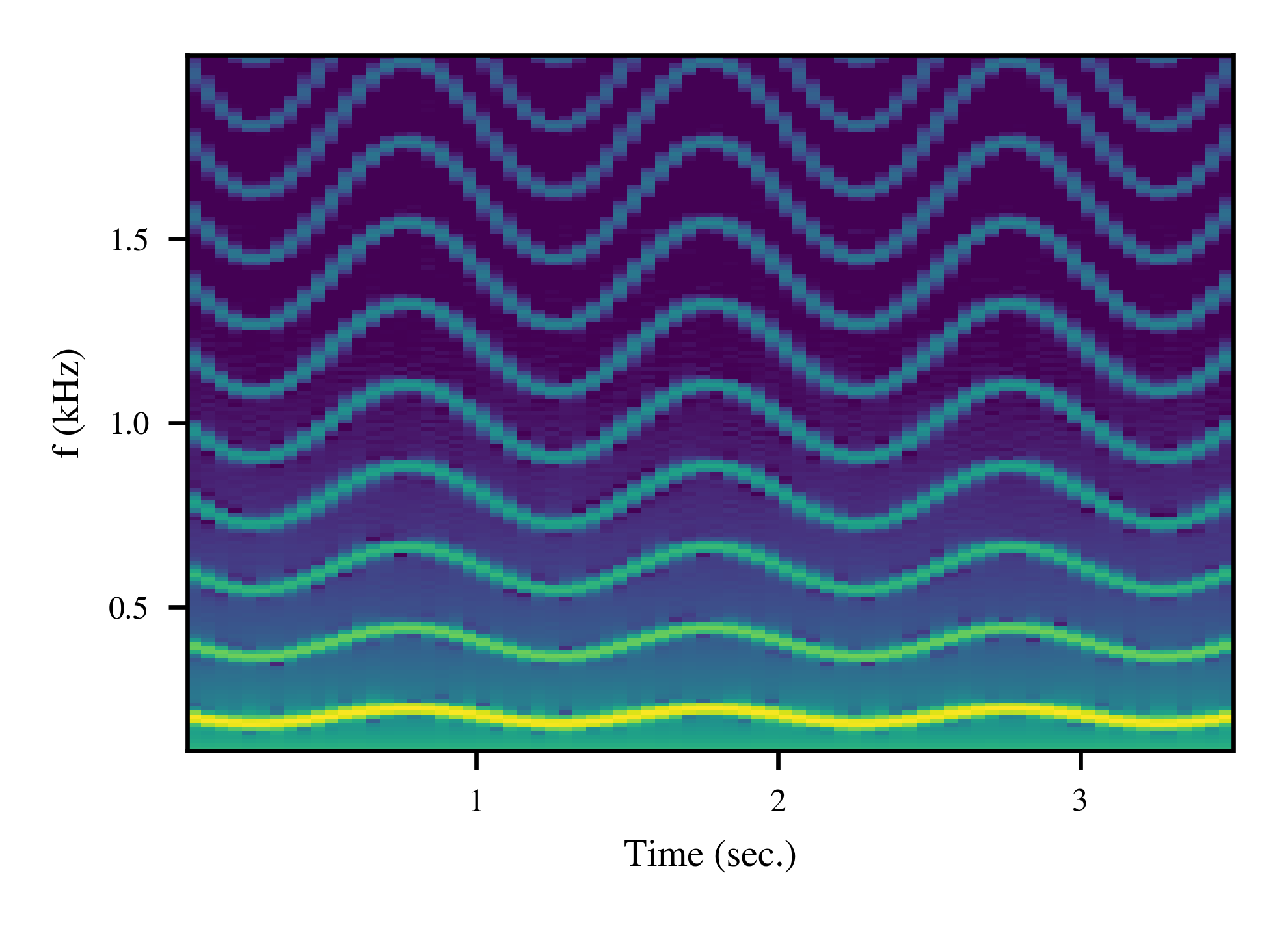}
\caption{\label{fig:FM_simple_signal} The impact of a frequency modulation on a harmonic dipolar excitation described in Eq.~\eqref{eq:fm} for $f_0$=100~Hz.}
\end{figure}

Figure~\ref{fig:FM_simple_signal} illustrates the spectrogram for a frequency range up to 1.8~kHz, color-coded with the PSD with $f_0$=100~Hz. All harmonics experience a similar frequency modulation with a peak-to-peak variation proportional to the order of the harmonic.

\FloatBarrier
\bibliography{bibliography}

\providecommand{\noopsort}[1]{}\providecommand{\singleletter}[1]{#1}%
\begin{thebibliography}{55}%
\makeatletter
\providecommand \@ifxundefined [1]{%
 \@ifx{#1\undefined}
}%
\providecommand \@ifnum [1]{%
 \ifnum #1\expandafter \@firstoftwo
 \else \expandafter \@secondoftwo
 \fi
}%
\providecommand \@ifx [1]{%
 \ifx #1\expandafter \@firstoftwo
 \else \expandafter \@secondoftwo
 \fi
}%
\providecommand \natexlab [1]{#1}%
\providecommand \enquote  [1]{``#1''}%
\providecommand \bibnamefont  [1]{#1}%
\providecommand \bibfnamefont [1]{#1}%
\providecommand \citenamefont [1]{#1}%
\providecommand \href@noop [0]{\@secondoftwo}%
\providecommand \href [0]{\begingroup \@sanitize@url \@href}%
\providecommand \@href[1]{\@@startlink{#1}\@@href}%
\providecommand \@@href[1]{\endgroup#1\@@endlink}%
\providecommand \@sanitize@url [0]{\catcode `\\12\catcode `\$12\catcode
  `\&12\catcode `\#12\catcode `\^12\catcode `\_12\catcode `\%12\relax}%
\providecommand \@@startlink[1]{}%
\providecommand \@@endlink[0]{}%
\providecommand \url  [0]{\begingroup\@sanitize@url \@url }%
\providecommand \@url [1]{\endgroup\@href {#1}{\urlprefix }}%
\providecommand \urlprefix  [0]{URL }%
\providecommand \Eprint [0]{\href }%
\providecommand \doibase [0]{https://doi.org/}%
\providecommand \selectlanguage [0]{\@gobble}%
\providecommand \bibinfo  [0]{\@secondoftwo}%
\providecommand \bibfield  [0]{\@secondoftwo}%
\providecommand \translation [1]{[#1]}%
\providecommand \BibitemOpen [0]{}%
\providecommand \bibitemStop [0]{}%
\providecommand \bibitemNoStop [0]{.\EOS\space}%
\providecommand \EOS [0]{\spacefactor3000\relax}%
\providecommand \BibitemShut  [1]{\csname bibitem#1\endcsname}%
\let\auto@bib@innerbib\@empty
\bibitem [{\citenamefont {Satogata}(1993)}]{intro1993_1}%
  \BibitemOpen
  \bibfield  {author} {\bibinfo {author} {\bibfnamefont {T.~J.}\ \bibnamefont
  {Satogata}},\ }\emph {\bibinfo {title} {{Nonlinear resonance islands and
  modulational effects in a proton synchrotron}}},\ \href
  {https://doi.org/10.2172/1372845} {Ph.D. thesis},\ \bibinfo  {school}
  {Northwestern U.} (\bibinfo {year} {1993})\BibitemShut {NoStop}%
\bibitem [{\citenamefont {Zimmermann}(1993)}]{intro1993_2}%
  \BibitemOpen
  \bibfield  {author} {\bibinfo {author} {\bibfnamefont {F.}~\bibnamefont
  {Zimmermann}},\ }\emph {\bibinfo {title} {{Emittance growth and proton beam
  lifetime in HERA}}},\ \href
  {http://lss.fnal.gov/archive/other/desy-93-059.pdf} {Ph.D. thesis},\ \bibinfo
   {school} {Hamburg U.} (\bibinfo {year} {1993})\BibitemShut {NoStop}%
\bibitem [{\citenamefont {{{O.~S.~Br{\"u}ning and
  F.~Willeke}}}(1994)}]{intro1994_1}%
  \BibitemOpen
  \bibfield  {author} {\bibinfo {author} {\bibnamefont {{{O.~S.~Br{\"u}ning and
  F.~Willeke}}}},\ }\bibfield  {title} {\bibinfo {title} {{Diffusion-like
  processes in proton storage rings due to the combined effect of non-linear
  fields and modulational effects with more than one frequency}},\ }in\ \href
  {https://cds.cern.ch/record/271555} {\emph {\bibinfo {booktitle} {European
  Particle Accelerator Conference (EPAC 1994)}}}\ (\bibinfo {year}
  {1994})\BibitemShut {NoStop}%
\bibitem [{\citenamefont {Br{\"u}ning}\ \emph
  {et~al.}(1994{\natexlab{a}})\citenamefont {Br{\"u}ning}, \citenamefont
  {Seidel}, \citenamefont {Mess},\ and\ \citenamefont {Willeke}}]{intro1994_2}%
  \BibitemOpen
  \bibfield  {author} {\bibinfo {author} {\bibfnamefont {O.~S.}\ \bibnamefont
  {Br{\"u}ning}}, \bibinfo {author} {\bibfnamefont {M.}~\bibnamefont {Seidel}},
  \bibinfo {author} {\bibfnamefont {K.}~\bibnamefont {Mess}}, and\ \bibinfo
  {author} {\bibfnamefont {F.}~\bibnamefont {Willeke}},\ }\href@noop {} {\emph
  {\bibinfo {title} {Measuring the effect of an external tune modulation on the
  particle diffusion in the proton storage ring of HERA}}},\ \bibinfo {type}
  {Tech. Rep.}\ (\bibinfo  {institution} {P00021147},\ \bibinfo {year}
  {1994})\BibitemShut {NoStop}%
\bibitem [{\citenamefont {Brüning}\ \emph
  {et~al.}(2004{\natexlab{a}})\citenamefont {Brüning}, \citenamefont
  {Collier}, \citenamefont {Lebrun}, \citenamefont {Myers}, \citenamefont
  {Ostojic}, \citenamefont {Poole},\ and\ \citenamefont
  {Proudlock}}]{Bruning:782076}%
  \BibitemOpen
  \bibfield  {author} {\bibinfo {author} {\bibfnamefont {O.~S.}\ \bibnamefont
  {Brüning}}, \bibinfo {author} {\bibfnamefont {P.}~\bibnamefont {Collier}},
  \bibinfo {author} {\bibfnamefont {P.}~\bibnamefont {Lebrun}}, \bibinfo
  {author} {\bibfnamefont {S.}~\bibnamefont {Myers}}, \bibinfo {author}
  {\bibfnamefont {R.}~\bibnamefont {Ostojic}}, \bibinfo {author} {\bibfnamefont
  {J.}~\bibnamefont {Poole}}, and\ \bibinfo {author} {\bibfnamefont
  {P.}~\bibnamefont {Proudlock}},\ }\href
  {https://doi.org/10.5170/CERN-2004-003-V-1} {\emph {\bibinfo {title} {{LHC
  Design Report}}}},\ CERN Yellow Reports: Monographs\ (\bibinfo  {publisher}
  {CERN},\ \bibinfo {address} {Geneva},\ \bibinfo {year} {2004})\BibitemShut
  {NoStop}%
\bibitem [{\citenamefont {Kuhn}\ \emph {et~al.}(2012)\citenamefont {Kuhn},
  \citenamefont {Arduini}, \citenamefont {Emery}, \citenamefont {Guerrero},
  \citenamefont {Hofle}, \citenamefont {Kain}, \citenamefont {Roncarolo},
  \citenamefont {Sapinski}, \citenamefont {Schaumann},\ and\ \citenamefont
  {Steinhagen}}]{50Hz_1}%
  \BibitemOpen
  \bibfield  {author} {\bibinfo {author} {\bibfnamefont {M.}~\bibnamefont
  {Kuhn}}, \bibinfo {author} {\bibfnamefont {G.}~\bibnamefont {Arduini}},
  \bibinfo {author} {\bibfnamefont {J.}~\bibnamefont {Emery}}, \bibinfo
  {author} {\bibfnamefont {A.}~\bibnamefont {Guerrero}}, \bibinfo {author}
  {\bibfnamefont {W.}~\bibnamefont {Hofle}}, \bibinfo {author} {\bibfnamefont
  {V.}~\bibnamefont {Kain}}, \bibinfo {author} {\bibfnamefont {F.}~\bibnamefont
  {Roncarolo}}, \bibinfo {author} {\bibfnamefont {M.}~\bibnamefont {Sapinski}},
  \bibinfo {author} {\bibfnamefont {M.}~\bibnamefont {Schaumann}}, and\
  \bibinfo {author} {\bibfnamefont {R.}~\bibnamefont {Steinhagen}},\ }\bibfield
   {title} {\bibinfo {title} {{{LHC} {Emittance} preservation during the 2012
  run.}},\ }in\ \href {https://cds.cern.ch/record/2302436} {\emph {\bibinfo
  {booktitle} {{Proceedings, 4th Evian Workshop on LHC beam operation: Evian
  Les Bains, France, December 17-20, 2012}}}}\ (\bibinfo {year} {2012})\ pp.\
  \bibinfo {pages} {161--170. 10 p}\BibitemShut {NoStop}%
\bibitem [{\citenamefont {Cettour~Cave}\ \emph {et~al.}(2013)\citenamefont
  {Cettour~Cave}, \citenamefont {De~Maria}, \citenamefont {Giovannozzi},
  \citenamefont {Ludwig}, \citenamefont {MacPherson}, \citenamefont {Redaelli},
  \citenamefont {Roncarolo}, \citenamefont {Solfaroli~Camillocci},\ and\
  \citenamefont {Venturini~Delsolaro}}]{50Hz_2}%
  \BibitemOpen
  \bibfield  {author} {\bibinfo {author} {\bibfnamefont {S.}~\bibnamefont
  {Cettour~Cave}}, \bibinfo {author} {\bibfnamefont {R.}~\bibnamefont
  {De~Maria}}, \bibinfo {author} {\bibfnamefont {M.}~\bibnamefont
  {Giovannozzi}}, \bibinfo {author} {\bibfnamefont {M.}~\bibnamefont {Ludwig}},
  \bibinfo {author} {\bibfnamefont {A.}~\bibnamefont {MacPherson}}, \bibinfo
  {author} {\bibfnamefont {S.}~\bibnamefont {Redaelli}}, \bibinfo {author}
  {\bibfnamefont {F.}~\bibnamefont {Roncarolo}}, \bibinfo {author}
  {\bibfnamefont {M.}~\bibnamefont {Solfaroli~Camillocci}}, and\ \bibinfo
  {author} {\bibfnamefont {W.}~\bibnamefont {Venturini~Delsolaro}},\ }\href
  {https://cds.cern.ch/record/1543434} {\bibinfo {title} {{Non-linear beam
  dynamics tests in the {LHC}: Measurement of intensity decay for probing
  dynamic aperture at injection}}} (\bibinfo {year} {2013})\BibitemShut
  {NoStop}%
\bibitem [{\citenamefont {{G.~Arduini}}(2015)}]{50Hz_3}%
  \BibitemOpen
  \bibfield  {author} {\bibinfo {author} {\bibnamefont {{G.~Arduini}}},\
  }\href@noop {} {\bibinfo {title} {50~{Hz} lines studies: First
  observations}},\ \bibinfo {howpublished}
  {\url{https://indico.cern.ch/event/436679/contributions/1085928}} (\bibinfo
  {year} {2015}),\ \bibinfo {note} {accessed: 2019-12-13}\BibitemShut {NoStop}%
\bibitem [{\citenamefont {{R.~De Maria}}(2012)}]{50Hz_4}%
  \BibitemOpen
  \bibfield  {author} {\bibinfo {author} {\bibnamefont {{R.~De Maria}}},\
  }\href@noop {} {\bibinfo {title} {Observation of 50~{Hz} lines in the {LHC
  BBQ} system}},\ \bibinfo {howpublished}
  {\url{https://lhc-beam-operation-committee.web.cern.ch/lhc-beam-operation-committee}}
  (\bibinfo {year} {2012}),\ \bibinfo {note} {accessed: 2019-12-14}\BibitemShut
  {NoStop}%
\bibitem [{\citenamefont {{{T.~Linnecar and W.~Scandale}}}(1986)}]{SPS}%
  \BibitemOpen
  \bibfield  {author} {\bibinfo {author} {\bibnamefont {{{T.~Linnecar and
  W.~Scandale}}}},\ }\href {http://cds.cern.ch/record/1214909} {\emph {\bibinfo
  {title} {{Phenomenology and causes of the 50 Hz spaced lines contaminating
  the {Schottky} signal}}}},\ \bibinfo {type} {Tech. Rep.}\ \bibinfo {number}
  {CERN-SPS-DI-MST-TL-WS-EEK. CERN-SPS-Improvement-Report-203}\ (\bibinfo
  {institution} {CERN},\ \bibinfo {address} {Geneva},\ \bibinfo {year}
  {1986})\BibitemShut {NoStop}%
\bibitem [{\citenamefont {Altuna}\ \emph {et~al.}(1992)\citenamefont {Altuna},
  \citenamefont {Arimatea}, \citenamefont {Bailey}, \citenamefont {Bohl},
  \citenamefont {Brandt}, \citenamefont {Cornelis}, \citenamefont {Depas},
  \citenamefont {Galluccio}, \citenamefont {Gareyte}, \citenamefont {Giachino},
  \citenamefont {Giovannozzi}, \citenamefont {Guo}, \citenamefont {Herr},
  \citenamefont {Hilaire}, \citenamefont {Lundberg}, \citenamefont {Miles},
  \citenamefont {Normann}, \citenamefont {Risselada}, \citenamefont {Scandale},
  \citenamefont {Schmidt}, \citenamefont {Spinks}, \citenamefont {Venturini},
  \citenamefont {Giovannozzi},\ and\ \citenamefont {Schmidt}}]{SPS1_tune}%
  \BibitemOpen
  \bibfield  {author} {\bibinfo {author} {\bibfnamefont {X.}~\bibnamefont
  {Altuna}}, \bibinfo {author} {\bibfnamefont {C.}~\bibnamefont {Arimatea}},
  \bibinfo {author} {\bibfnamefont {R.}~\bibnamefont {Bailey}}, \bibinfo
  {author} {\bibfnamefont {T.}~\bibnamefont {Bohl}}, \bibinfo {author}
  {\bibfnamefont {D.}~\bibnamefont {Brandt}}, \bibinfo {author} {\bibfnamefont
  {K.}~\bibnamefont {Cornelis}}, \bibinfo {author} {\bibfnamefont
  {C.}~\bibnamefont {Depas}}, \bibinfo {author} {\bibfnamefont
  {F.}~\bibnamefont {Galluccio}}, \bibinfo {author} {\bibfnamefont
  {J.}~\bibnamefont {Gareyte}}, \bibinfo {author} {\bibfnamefont
  {R.}~\bibnamefont {Giachino}}, \bibinfo {author} {\bibfnamefont
  {M.}~\bibnamefont {Giovannozzi}}, \bibinfo {author} {\bibfnamefont
  {Z.}~\bibnamefont {Guo}}, \bibinfo {author} {\bibfnamefont {W.}~\bibnamefont
  {Herr}}, \bibinfo {author} {\bibfnamefont {A.}~\bibnamefont {Hilaire}},
  \bibinfo {author} {\bibfnamefont {T.}~\bibnamefont {Lundberg}}, \bibinfo
  {author} {\bibfnamefont {J.}~\bibnamefont {Miles}}, \bibinfo {author}
  {\bibfnamefont {L.}~\bibnamefont {Normann}}, \bibinfo {author} {\bibfnamefont
  {T.}~\bibnamefont {Risselada}}, \bibinfo {author} {\bibfnamefont
  {W.}~\bibnamefont {Scandale}}, \bibinfo {author} {\bibfnamefont
  {F.}~\bibnamefont {Schmidt}}, \bibinfo {author} {\bibfnamefont
  {A.}~\bibnamefont {Spinks}}, \bibinfo {author} {\bibfnamefont
  {M.}~\bibnamefont {Venturini}}, \bibinfo {author} {\bibfnamefont
  {M.}~\bibnamefont {Giovannozzi}}, and\ \bibinfo {author} {\bibfnamefont
  {F.}~\bibnamefont {Schmidt}},\ }\bibfield  {title} {\bibinfo {title} {The
  1991 dynamic aperture experiment at the {CERN SPS}},\ }\href
  {https://doi.org/10.1063/1.42289} {\bibfield  {journal} {\bibinfo  {journal}
  {AIP Conference Proceedings}\ }\textbf {\bibinfo {volume} {255}},\ \bibinfo
  {pages} {355} (\bibinfo {year} {1992})},\ \Eprint
  {https://arxiv.org/abs/https://aip.scitation.org/doi/pdf/10.1063/1.42289}
  {https://aip.scitation.org/doi/pdf/10.1063/1.42289} \BibitemShut {NoStop}%
\bibitem [{\citenamefont {Brüning}(1992)}]{SPS2_tune}%
  \BibitemOpen
  \bibfield  {author} {\bibinfo {author} {\bibfnamefont {O.~S.}\ \bibnamefont
  {Brüning}},\ }\bibfield  {title} {\bibinfo {title} {Diffusion in a {FODO}
  cell due to modulation effects in the presence of nonlinear fields},\ }\href
  {https://cds.cern.ch/record/1108270} {\bibfield  {journal} {\bibinfo
  {journal} {Part. Accel.}\ }\textbf {\bibinfo {volume} {41}},\ \bibinfo
  {pages} {133} (\bibinfo {year} {1992})}\BibitemShut {NoStop}%
\bibitem [{\citenamefont {{W.~Fischer and M.~Giovannozzi}}(1997)}]{SPS3_tune}%
  \BibitemOpen
  \bibfield  {author} {\bibinfo {author} {\bibnamefont {{W.~Fischer and
  M.~Giovannozzi}}},\ }\bibfield  {title} {\bibinfo {title} {Dynamic aperture
  experiment at a synchrotron},\ }\href
  {https://doi.org/10.1103/PhysRevE.55.3507} {\bibfield  {journal} {\bibinfo
  {journal} {Physical Review E}\ }\textbf {\bibinfo {volume} {55}},\ \bibinfo
  {pages} {3507} (\bibinfo {year} {1997})}\BibitemShut {NoStop}%
\bibitem [{\citenamefont {{M.~Gasior and R.~Jones}}(2005)}]{SPS2}%
  \BibitemOpen
  \bibfield  {author} {\bibinfo {author} {\bibnamefont {{M.~Gasior and
  R.~Jones}}},\ }\href {http://cds.cern.ch/record/883298} {\emph {\bibinfo
  {title} {{The principle and first results of betatron tune measurement by
  direct diode detection}}}},\ \bibinfo {type} {Tech. Rep.}\ \bibinfo {number}
  {LHC-Project-Report-853. CERN-LHC-Project-Report-853}\ (\bibinfo
  {institution} {CERN},\ \bibinfo {address} {Geneva},\ \bibinfo {year} {2005})\
  \bibinfo {note} {revised version submitted on 2005-09-16
  09:23:15}\BibitemShut {NoStop}%
\bibitem [{\citenamefont {Br{\"u}ning}\ \emph
  {et~al.}(1994{\natexlab{b}})\citenamefont {Br{\"u}ning}, \citenamefont
  {Seidel}, \citenamefont {Mess},\ and\ \citenamefont {Willeke}}]{HERA1_tune}%
  \BibitemOpen
  \bibfield  {author} {\bibinfo {author} {\bibfnamefont {O.~S.}\ \bibnamefont
  {Br{\"u}ning}}, \bibinfo {author} {\bibfnamefont {M.}~\bibnamefont {Seidel}},
  \bibinfo {author} {\bibfnamefont {K.}~\bibnamefont {Mess}}, and\ \bibinfo
  {author} {\bibfnamefont {F.}~\bibnamefont {Willeke}},\ }\href@noop {} {\emph
  {\bibinfo {title} {Measuring the effect of an external tune modulation on the
  particle diffusion in the proton storage ring of HERA}}},\ \bibinfo {type}
  {Tech. Rep.}\ (\bibinfo  {institution} {P00021147},\ \bibinfo {year}
  {1994})\BibitemShut {NoStop}%
\bibitem [{\citenamefont {{O.~S.~Br{\"u}ning and
  F.~Willeke}}(1995)}]{HERA2_tune}%
  \BibitemOpen
  \bibfield  {author} {\bibinfo {author} {\bibnamefont {{O.~S.~Br{\"u}ning and
  F.~Willeke}}},\ }\bibfield  {title} {\bibinfo {title} {Reduction of particle
  losses in {HERA} by generating an additional harmonic tune modulation},\ }in\
  \href {https://doi.org/10.1109/PAC.1995.504677} {\emph {\bibinfo {booktitle}
  {Proceedings Particle Accelerator Conference}}},\ Vol.~\bibinfo {volume} {1}\
  (\bibinfo {year} {1995})\ pp.\ \bibinfo {pages} {420--422 vol.1}\BibitemShut
  {NoStop}%
\bibitem [{\citenamefont {Cameron}\ \emph {et~al.}(2005)\citenamefont
  {Cameron}, \citenamefont {Gasior}, \citenamefont {Jones},\ and\ \citenamefont
  {Tan}}]{RHIC1}%
  \BibitemOpen
  \bibfield  {author} {\bibinfo {author} {\bibfnamefont {P.}~\bibnamefont
  {Cameron}}, \bibinfo {author} {\bibfnamefont {M.}~\bibnamefont {Gasior}},
  \bibinfo {author} {\bibfnamefont {R.}~\bibnamefont {Jones}}, and\ \bibinfo
  {author} {\bibfnamefont {C.}~\bibnamefont {Tan}},\ }\href@noop {} {\emph
  {\bibinfo {title} {The effects and possible origins of mains ripple in the
  vicinity of the betatron spectrum.}}},\ \bibinfo {type} {Tech. Rep.}\
  (\bibinfo  {institution} {{Brookhaven National Lab.(BNL), Upton, NY (United
  States)}},\ \bibinfo {year} {2005})\BibitemShut {NoStop}%
\bibitem [{\citenamefont {Cameron}\ \emph {et~al.}(2006)\citenamefont
  {Cameron}, \citenamefont {Gasior}, \citenamefont {Jones},\ and\ \citenamefont
  {Tang}}]{RHIC2}%
  \BibitemOpen
  \bibfield  {author} {\bibinfo {author} {\bibfnamefont {P.}~\bibnamefont
  {Cameron}}, \bibinfo {author} {\bibfnamefont {M.}~\bibnamefont {Gasior}},
  \bibinfo {author} {\bibfnamefont {R.}~\bibnamefont {Jones}}, and\ \bibinfo
  {author} {\bibfnamefont {C.}~\bibnamefont {Tang}},\ }\href@noop {} {\emph
  {\bibinfo {title} {Observations of direct excitation of the betatron spectrum
  by mains harmonics in {RHIC}}}},\ \bibinfo {type} {Tech. Rep.}\ (\bibinfo
  {institution} {{Brookhaven National Laboratory (BNL) Relativistic Heavy Ion
  Collider}},\ \bibinfo {year} {2006})\BibitemShut {NoStop}%
\bibitem [{\citenamefont {Tan}(2005)}]{Tevatron1}%
  \BibitemOpen
  \bibfield  {author} {\bibinfo {author} {\bibfnamefont {C.-Y.}\ \bibnamefont
  {Tan}},\ }\bibfield  {title} {\bibinfo {title} {Novel tune diagnostics for
  the {Tevatron}},\ }in\ \href@noop {} {\emph {\bibinfo {booktitle}
  {Proceedings of the 2005 Particle Accelerator Conference}}}\ (\bibinfo
  {organization} {IEEE},\ \bibinfo {year} {2005})\ pp.\ \bibinfo {pages}
  {140--144}\BibitemShut {NoStop}%
\bibitem [{\citenamefont {Shiltsev}\ \emph {et~al.}(2011)\citenamefont
  {Shiltsev}, \citenamefont {Stancari},\ and\ \citenamefont
  {Valishev}}]{Tevatron2}%
  \BibitemOpen
  \bibfield  {author} {\bibinfo {author} {\bibfnamefont {V.}~\bibnamefont
  {Shiltsev}}, \bibinfo {author} {\bibfnamefont {G.}~\bibnamefont {Stancari}},
  and\ \bibinfo {author} {\bibfnamefont {A.}~\bibnamefont {Valishev}},\
  }\bibfield  {title} {\bibinfo {title} {Ambient betatron motion and its
  excitation by ``ghost lines'' in {Tevatron}},\ }\href@noop {} {\bibfield
  {journal} {\bibinfo  {journal} {Journal of Instrumentation}\ }\textbf
  {\bibinfo {volume} {6}}\bibinfo  {number} { (08)},\ \bibinfo {pages}
  {P08002}}\BibitemShut {NoStop}%
\bibitem [{\citenamefont {Solfaroli~Camillocci}\ \emph
  {et~al.}(2016)\citenamefont {Solfaroli~Camillocci}, \citenamefont {Redaelli},
  \citenamefont {Tomás},\ and\ \citenamefont {Wenninger}}]{rampsqueeze}%
  \BibitemOpen
\bibfield  {number} {  }\bibfield  {author} {\bibinfo {author} {\bibfnamefont
  {M.}~\bibnamefont {Solfaroli~Camillocci}}, \bibinfo {author} {\bibfnamefont
  {S.}~\bibnamefont {Redaelli}}, \bibinfo {author} {\bibfnamefont
  {R.}~\bibnamefont {Tomás}}, and\ \bibinfo {author} {\bibfnamefont
  {J.}~\bibnamefont {Wenninger}},\ }\bibfield  {title} {\bibinfo {title}
  {{Combined {Ramp} and {Squeeze} to 6.5~{TeV} in the {LHC}}},\ }in\ \href
  {https://doi.org/10.18429/JACoW-IPAC2016-TUPMW031} {\emph {\bibinfo
  {booktitle} {{Proceedings, 7th International Particle Accelerator Conference
  (IPAC 2016): Busan, Korea, May 8-13, 2016}}}}\ (\bibinfo {year} {2016})\ p.\
  \bibinfo {pages} {TUPMW031}\BibitemShut {NoStop}%
\bibitem [{\citenamefont {Fartoukh}(2013)}]{ATS}%
  \BibitemOpen
  \bibfield  {author} {\bibinfo {author} {\bibfnamefont {S.}~\bibnamefont
  {Fartoukh}},\ }\bibfield  {title} {\bibinfo {title} {Achromatic telescopic
  squeezing scheme and application to the {LHC} and its luminosity upgrade},\
  }\href {https://doi.org/10.1103/PhysRevSTAB.16.111002} {\bibfield  {journal}
  {\bibinfo  {journal} {Phys. Rev. ST Accel. Beams}\ }\textbf {\bibinfo
  {volume} {16}},\ \bibinfo {pages} {111002} (\bibinfo {year}
  {2013})}\BibitemShut {NoStop}%
\bibitem [{\citenamefont {{{B.~Muratori and T.~Pieloni}}}(2014)}]{levelling}%
  \BibitemOpen
  \bibfield  {author} {\bibinfo {author} {\bibnamefont {{{B.~Muratori and
  T.~Pieloni}}}},\ }\bibfield  {title} {\bibinfo {title} {{Luminosity levelling
  techniques for the {LHC}}}\ }(\bibinfo {year} {2014})\ pp.\ \bibinfo {pages}
  {177--181. 5 p},\ \bibinfo {note} {comments: 5 pages, contribution to the
  ICFA Mini-Workshop on Beam-Beam Effects in Hadron Colliders, CERN, Geneva,
  Switzerland, 18-22 Mar 2013}\BibitemShut {NoStop}%
\bibitem [{\citenamefont {Karastathis}\ \emph {et~al.}(2018)\citenamefont
  {Karastathis}, \citenamefont {Fuchsberger}, \citenamefont {Hostettler},
  \citenamefont {Papaphilippou},\ and\ \citenamefont
  {Pellegrini}}]{antileveling}%
  \BibitemOpen
  \bibfield  {author} {\bibinfo {author} {\bibfnamefont {N.}~\bibnamefont
  {Karastathis}}, \bibinfo {author} {\bibfnamefont {K.}~\bibnamefont
  {Fuchsberger}}, \bibinfo {author} {\bibfnamefont {M.}~\bibnamefont
  {Hostettler}}, \bibinfo {author} {\bibfnamefont {Y.}~\bibnamefont
  {Papaphilippou}}, and\ \bibinfo {author} {\bibfnamefont {D.}~\bibnamefont
  {Pellegrini}},\ }\bibfield  {title} {\bibinfo {title} {Crossing angle
  anti-leveling at the {LHC} in 2017},\ }in\ \href@noop {} {\emph {\bibinfo
  {booktitle} {Journal of Physics: Conference Series}}},\ Vol.\ \bibinfo
  {volume} {1067}\ (\bibinfo {organization} {IOP Publishing},\ \bibinfo {year}
  {2018})\ p.\ \bibinfo {pages} {022004}\BibitemShut {NoStop}%
\bibitem [{\citenamefont {Gasior}(2012)}]{BBQ}%
  \BibitemOpen
  \bibfield  {author} {\bibinfo {author} {\bibfnamefont {M.}~\bibnamefont
  {Gasior}},\ }\bibfield  {title} {\bibinfo {title} {{Faraday cup award: High
  sensitivity tune measurement using direct diode detection}},\ }\href
  {https://cds.cern.ch/record/1476069} {\bibfield  {journal} {\bibinfo
  {journal} {Conf. Proc.}\ }\textbf {\bibinfo {volume} {C1204151}},\ \bibinfo
  {pages} {MOAP02. 7 p} (\bibinfo {year} {2012})}\BibitemShut {NoStop}%
\bibitem [{\citenamefont {Jones}(2007)}]{jones2007lhc}%
  \BibitemOpen
  \bibfield  {author} {\bibinfo {author} {\bibfnamefont {O.}~\bibnamefont
  {Jones}},\ }\bibfield  {title} {\bibinfo {title} {{LHC} beam
  instrumentation},\ }in\ \href@noop {} {\emph {\bibinfo {booktitle} {2007 IEEE
  Particle Accelerator Conference (PAC)}}}\ (\bibinfo {organization} {IEEE},\
  \bibinfo {year} {2007})\ pp.\ \bibinfo {pages} {2630--2634}\BibitemShut
  {NoStop}%
\bibitem [{\citenamefont {Carver}\ \emph {et~al.}(2017)\citenamefont {Carver},
  \citenamefont {Buffat}, \citenamefont {Butterworth}, \citenamefont {Höfle},
  \citenamefont {Iadarola}, \citenamefont {Kotzian}, \citenamefont {Li},
  \citenamefont {Métral}, \citenamefont {Ojeda~Sandonís}, \citenamefont
  {Söderén},\ and\ \citenamefont {Valuch}}]{ADT}%
  \BibitemOpen
  \bibfield  {author} {\bibinfo {author} {\bibfnamefont {L.}~\bibnamefont
  {Carver}}, \bibinfo {author} {\bibfnamefont {X.}~\bibnamefont {Buffat}},
  \bibinfo {author} {\bibfnamefont {A.}~\bibnamefont {Butterworth}}, \bibinfo
  {author} {\bibfnamefont {W.}~\bibnamefont {Höfle}}, \bibinfo {author}
  {\bibfnamefont {G.}~\bibnamefont {Iadarola}}, \bibinfo {author}
  {\bibfnamefont {G.}~\bibnamefont {Kotzian}}, \bibinfo {author} {\bibfnamefont
  {K.}~\bibnamefont {Li}}, \bibinfo {author} {\bibfnamefont {E.}~\bibnamefont
  {Métral}}, \bibinfo {author} {\bibfnamefont {M.}~\bibnamefont
  {Ojeda~Sandonís}}, \bibinfo {author} {\bibfnamefont {M.}~\bibnamefont
  {Söderén}}, and\ \bibinfo {author} {\bibfnamefont {D.}~\bibnamefont
  {Valuch}},\ }\bibfield  {title} {\bibinfo {title} {{Usage of the transverse
  damper observation box for high sampling rate transverse position data in the
  {LHC}}},\ }in\ \href {https://doi.org/10.18429/JACoW-IPAC2017-MOPAB113}
  {\emph {\bibinfo {booktitle} {8th Int. Particle Accelerator Conf.
  (IPAC’17), Copenhagen, Denmark, May 2017}}},\ \bibinfo {series and number}
  {\bibinfo {number} {CERN-ACC-2017-117}}\ (\bibinfo {year} {2017})\ pp.\
  \bibinfo {pages} {389--392}\BibitemShut {NoStop}%
\bibitem [{\citenamefont {Ojeda~Sandonís}\ \emph {et~al.}(2015)\citenamefont
  {Ojeda~Sandonís}, \citenamefont {Baudrenghien}, \citenamefont {Butterworth},
  \citenamefont {Galindo}, \citenamefont {Höfle}, \citenamefont {Levens},
  \citenamefont {Molendijk}, \citenamefont {Vaga},\ and\ \citenamefont
  {Valuch}}]{ADT3}%
  \BibitemOpen
  \bibfield  {author} {\bibinfo {author} {\bibfnamefont {M.}~\bibnamefont
  {Ojeda~Sandonís}}, \bibinfo {author} {\bibfnamefont {P.}~\bibnamefont
  {Baudrenghien}}, \bibinfo {author} {\bibfnamefont {A.}~\bibnamefont
  {Butterworth}}, \bibinfo {author} {\bibfnamefont {J.}~\bibnamefont
  {Galindo}}, \bibinfo {author} {\bibfnamefont {W.}~\bibnamefont {Höfle}},
  \bibinfo {author} {\bibfnamefont {T.}~\bibnamefont {Levens}}, \bibinfo
  {author} {\bibfnamefont {J.}~\bibnamefont {Molendijk}}, \bibinfo {author}
  {\bibfnamefont {F.}~\bibnamefont {Vaga}}, and\ \bibinfo {author}
  {\bibfnamefont {D.}~\bibnamefont {Valuch}},\ }\bibfield  {title} {\bibinfo
  {title} {{Processing high-bandwidth bunch-by-bunch observation data from the
  {RF} and transverse damper systems of the {LHC}}},\ }in\ \href
  {https://doi.org/10.18429/JACoW-ICALEPCS2015-WEPGF062} {\emph {\bibinfo
  {booktitle} {{Proceedings, 15th International Conference on Accelerator and
  Large Experimental Physics Control Systems (ICALEPCS 2015): Melbourne,
  Australia, October 17-23, 2015}}}}\ (\bibinfo {year} {2015})\ p.\ \bibinfo
  {pages} {WEPGF062}\BibitemShut {NoStop}%
\bibitem [{\citenamefont {Söderén}\ \emph {et~al.}(2017)\citenamefont
  {Söderén}, \citenamefont {Kotzian}, \citenamefont {Ojeda~Sandonís},\ and\
  \citenamefont {Valuch}}]{ADT4}%
  \BibitemOpen
  \bibfield  {author} {\bibinfo {author} {\bibfnamefont {M.}~\bibnamefont
  {Söderén}}, \bibinfo {author} {\bibfnamefont {G.}~\bibnamefont {Kotzian}},
  \bibinfo {author} {\bibfnamefont {M.}~\bibnamefont {Ojeda~Sandonís}}, and\
  \bibinfo {author} {\bibfnamefont {D.}~\bibnamefont {Valuch}},\ }\bibfield
  {title} {\bibinfo {title} {{Online bunch by bunch transverse instability
  detection in {LHC}}},\ }in\ \href
  {https://doi.org/10.18429/JACoW-IPAC2017-MOPAB117} {\emph {\bibinfo
  {booktitle} {{Proceedings, 8th International Particle Accelerator Conference
  (IPAC 2017): Copenhagen, Denmark, May 14-19, 2017}}}}\ (\bibinfo {year}
  {2017})\ p.\ \bibinfo {pages} {MOPAB117}\BibitemShut {NoStop}%
\bibitem [{\citenamefont {Persson}\ \emph {et~al.}(2016)\citenamefont
  {Persson}, \citenamefont {Coello~de Portugal}, \citenamefont
  {Garcia-Tabares}, \citenamefont {Gąsior}, \citenamefont {Langner},
  \citenamefont {Lefèvre}, \citenamefont {Maclean}, \citenamefont {Malina},
  \citenamefont {Olexa}, \citenamefont {Skowroński},\ and\ \citenamefont
  {Tomás}}]{DOROS}%
  \BibitemOpen
  \bibfield  {author} {\bibinfo {author} {\bibfnamefont {T.}~\bibnamefont
  {Persson}}, \bibinfo {author} {\bibfnamefont {J.~M.}\ \bibnamefont {Coello~de
  Portugal}}, \bibinfo {author} {\bibfnamefont {A.}~\bibnamefont
  {Garcia-Tabares}}, \bibinfo {author} {\bibfnamefont {M.}~\bibnamefont
  {Gąsior}}, \bibinfo {author} {\bibfnamefont {A.}~\bibnamefont {Langner}},
  \bibinfo {author} {\bibfnamefont {T.}~\bibnamefont {Lefèvre}}, \bibinfo
  {author} {\bibfnamefont {E.}~\bibnamefont {Maclean}}, \bibinfo {author}
  {\bibfnamefont {L.}~\bibnamefont {Malina}}, \bibinfo {author} {\bibfnamefont
  {J.}~\bibnamefont {Olexa}}, \bibinfo {author} {\bibfnamefont
  {P.}~\bibnamefont {Skowroński}}, and\ \bibinfo {author} {\bibfnamefont
  {R.}~\bibnamefont {Tomás}},\ }\bibfield  {title} {\bibinfo {title}
  {Experience with {DOROS BPMs} for coupling measurement and correction},\ }in\
  \href@noop {} {\emph {\bibinfo {booktitle} {7th Int. Particle Accelerator
  Conf.(IPAC'16), Busan, Korea, May 8-13, 2016}}}\ (\bibinfo {organization}
  {JACOW, Geneva, Switzerland},\ \bibinfo {year} {2016})\ pp.\ \bibinfo {pages}
  {303--305}\BibitemShut {NoStop}%
\bibitem [{\citenamefont {Olexa}\ \emph {et~al.}(2013)\citenamefont {Olexa},
  \citenamefont {Ondracek}, \citenamefont {Brezovic},\ and\ \citenamefont
  {Gasior}}]{DOROS2}%
  \BibitemOpen
  \bibfield  {author} {\bibinfo {author} {\bibfnamefont {J.}~\bibnamefont
  {Olexa}}, \bibinfo {author} {\bibfnamefont {O.}~\bibnamefont {Ondracek}},
  \bibinfo {author} {\bibfnamefont {Z.}~\bibnamefont {Brezovic}}, and\ \bibinfo
  {author} {\bibfnamefont {M.}~\bibnamefont {Gasior}},\ }\href
  {https://cds.cern.ch/record/1546401} {\emph {\bibinfo {title} {{Prototype
  system for phase advance measurements of {LHC} small beam oscillations}}}},\
  \bibinfo {type} {Tech. Rep.}\ \bibinfo {number} {CERN-ATS-2013-038}\
  (\bibinfo  {institution} {CERN},\ \bibinfo {address} {Geneva},\ \bibinfo
  {year} {2013})\BibitemShut {NoStop}%
\bibitem [{\citenamefont {Steinhagen}\ \emph {et~al.}(2013)\citenamefont
  {Steinhagen}, \citenamefont {Lucas},\ and\ \citenamefont {Boland}}]{MIM1}%
  \BibitemOpen
  \bibfield  {author} {\bibinfo {author} {\bibfnamefont {R.}~\bibnamefont
  {Steinhagen}}, \bibinfo {author} {\bibfnamefont {T.}~\bibnamefont {Lucas}},
  and\ \bibinfo {author} {\bibfnamefont {M.}~\bibnamefont {Boland}},\
  }\bibfield  {title} {\bibinfo {title} {A {Multiband-Instability-Monitor} for
  high-frequency intra-bunch beam diagnostics}\ }(\bibinfo {year}
  {2013})\BibitemShut {NoStop}%
\bibitem [{\citenamefont {Levens}\ \emph {et~al.}(2019)\citenamefont {Levens},
  \citenamefont {Lef{\`e}vre},\ and\ \citenamefont {Valuch}}]{MIM2}%
  \BibitemOpen
  \bibfield  {author} {\bibinfo {author} {\bibfnamefont {T.}~\bibnamefont
  {Levens}}, \bibinfo {author} {\bibfnamefont {T.}~\bibnamefont {Lef{\`e}vre}},
  and\ \bibinfo {author} {\bibfnamefont {D.}~\bibnamefont {Valuch}},\
  }\bibfield  {title} {\bibinfo {title} {Initial results from the {LHC}
  {Multi-Band Instability Monitor}},\ }in\ \href@noop {} {\emph {\bibinfo
  {booktitle} {Int. Beam Instrumentation Conf.(IBIC'18), Shanghai, China, 09-13
  September 2018}}}\ (\bibinfo {organization} {JACOW Publishing, Geneva,
  Switzerland},\ \bibinfo {year} {2019})\ pp.\ \bibinfo {pages}
  {314--318}\BibitemShut {NoStop}%
\bibitem [{\citenamefont {Kostoglou}\ \emph {et~al.}(2019)\citenamefont
  {Kostoglou}, \citenamefont {Arduini}, \citenamefont {Baccigalupi},
  \citenamefont {Bartosik}, \citenamefont {Buffat}, \citenamefont {Burnet},
  \citenamefont {Carver}, \citenamefont {Cerqueira~Bastos}, \citenamefont
  {De~Maria}, \citenamefont {Fartoukh}, \citenamefont {Tomas~Garcia},
  \citenamefont {Iadarola}, \citenamefont {Intelisano}, \citenamefont {Levens},
  \citenamefont {Louro~Alves}, \citenamefont {Michels}, \citenamefont
  {Montabonnet}, \citenamefont {Nisbet}, \citenamefont {Olexa}, \citenamefont
  {Papaphilippou}, \citenamefont {Pojer}, \citenamefont {Poyet}, \citenamefont
  {Soderen}, \citenamefont {Solfaroli~Camillocci}, \citenamefont {Sterbini},
  \citenamefont {Thiesen}, \citenamefont {Trad}, \citenamefont {Triantafyllou},
  \citenamefont {Valuch},\ and\ \citenamefont {Wenninger}}]{Kostoglou:2703609}%
  \BibitemOpen
  \bibfield  {author} {\bibinfo {author} {\bibfnamefont {S.}~\bibnamefont
  {Kostoglou}}, \bibinfo {author} {\bibfnamefont {G.}~\bibnamefont {Arduini}},
  \bibinfo {author} {\bibfnamefont {C.}~\bibnamefont {Baccigalupi}}, \bibinfo
  {author} {\bibfnamefont {H.}~\bibnamefont {Bartosik}}, \bibinfo {author}
  {\bibfnamefont {X.}~\bibnamefont {Buffat}}, \bibinfo {author} {\bibfnamefont
  {J.-P.}\ \bibnamefont {Burnet}}, \bibinfo {author} {\bibfnamefont {L.~R.}\
  \bibnamefont {Carver}}, \bibinfo {author} {\bibfnamefont {M.}~\bibnamefont
  {Cerqueira~Bastos}}, \bibinfo {author} {\bibfnamefont {R.}~\bibnamefont
  {De~Maria}}, \bibinfo {author} {\bibfnamefont {S.}~\bibnamefont {Fartoukh}},
  \bibinfo {author} {\bibfnamefont {R.}~\bibnamefont {Tomas~Garcia}}, \bibinfo
  {author} {\bibfnamefont {G.}~\bibnamefont {Iadarola}}, \bibinfo {author}
  {\bibfnamefont {L.}~\bibnamefont {Intelisano}}, \bibinfo {author}
  {\bibfnamefont {T.}~\bibnamefont {Levens}}, \bibinfo {author} {\bibfnamefont
  {D.~M.}\ \bibnamefont {Louro~Alves}}, \bibinfo {author} {\bibfnamefont
  {O.}~\bibnamefont {Michels}}, \bibinfo {author} {\bibfnamefont
  {V.}~\bibnamefont {Montabonnet}}, \bibinfo {author} {\bibfnamefont
  {D.}~\bibnamefont {Nisbet}}, \bibinfo {author} {\bibfnamefont
  {J.}~\bibnamefont {Olexa}}, \bibinfo {author} {\bibfnamefont
  {Y.}~\bibnamefont {Papaphilippou}}, \bibinfo {author} {\bibfnamefont
  {M.}~\bibnamefont {Pojer}}, \bibinfo {author} {\bibfnamefont
  {A.}~\bibnamefont {Poyet}}, \bibinfo {author} {\bibfnamefont
  {M.}~\bibnamefont {Soderen}}, \bibinfo {author} {\bibfnamefont
  {M.}~\bibnamefont {Solfaroli~Camillocci}}, \bibinfo {author} {\bibfnamefont
  {G.}~\bibnamefont {Sterbini}}, \bibinfo {author} {\bibfnamefont
  {H.}~\bibnamefont {Thiesen}}, \bibinfo {author} {\bibfnamefont
  {G.}~\bibnamefont {Trad}}, \bibinfo {author} {\bibfnamefont {N.}~\bibnamefont
  {Triantafyllou}}, \bibinfo {author} {\bibfnamefont {D.}~\bibnamefont
  {Valuch}}, and\ \bibinfo {author} {\bibfnamefont {J.}~\bibnamefont
  {Wenninger}},\ }\href {https://cds.cern.ch/record/2703609} {\emph {\bibinfo
  {title} {{{MD4147: 50 Hz harmonics perturbation}}}}},\ \bibinfo {type} {Tech.
  Rep.}\ (\bibinfo  {institution} {CERN},\ \bibinfo {year} {2019})\BibitemShut
  {NoStop}%
\bibitem [{\citenamefont {Nyquist}(1928)}]{nyquist1928certain}%
  \BibitemOpen
  \bibfield  {author} {\bibinfo {author} {\bibfnamefont {H.}~\bibnamefont
  {Nyquist}},\ }\bibfield  {title} {\bibinfo {title} {Certain topics in
  telegraph transmission theory},\ }\href@noop {} {\bibfield  {journal}
  {\bibinfo  {journal} {Transactions of the American Institute of Electrical
  Engineers}\ }\textbf {\bibinfo {volume} {47}},\ \bibinfo {pages} {617}
  (\bibinfo {year} {1928})}\BibitemShut {NoStop}%
\bibitem [{50H(2020)}]{50Hz_evolution}%
  \BibitemOpen
  \href@noop {} {\bibinfo {title} {Network frequency}},\ \bibinfo
  {howpublished} {\url{https://clients.rte-france.com}} (\bibinfo {year}
  {2020}),\ \bibinfo {note} {accessed: 2020-06-03}\BibitemShut {NoStop}%
\bibitem [{\citenamefont {Buzio}\ \emph {et~al.}(2010)\citenamefont {Buzio},
  \citenamefont {Galbraith}, \citenamefont {Gilardoni}, \citenamefont
  {Giloteaux}, \citenamefont {Golluccio}, \citenamefont {Petrone},
  \citenamefont {Walckiers},\ and\ \citenamefont {Beaumont}}]{PS2}%
  \BibitemOpen
  \bibfield  {author} {\bibinfo {author} {\bibfnamefont {M.}~\bibnamefont
  {Buzio}}, \bibinfo {author} {\bibfnamefont {P.}~\bibnamefont {Galbraith}},
  \bibinfo {author} {\bibfnamefont {S.}~\bibnamefont {Gilardoni}}, \bibinfo
  {author} {\bibfnamefont {D.}~\bibnamefont {Giloteaux}}, \bibinfo {author}
  {\bibfnamefont {G.}~\bibnamefont {Golluccio}}, \bibinfo {author}
  {\bibfnamefont {C.}~\bibnamefont {Petrone}}, \bibinfo {author} {\bibfnamefont
  {L.}~\bibnamefont {Walckiers}}, and\ \bibinfo {author} {\bibfnamefont
  {A.}~\bibnamefont {Beaumont}},\ }\bibfield  {title} {\bibinfo {title}
  {{Development of upgraded magnetic instrumentation for {CERN's} real-time
  reference field measurement systems}},\ }\bibfield  {booktitle} {\emph
  {\bibinfo {booktitle} {{Proceedings, 1st International Particle Accelerator
  Conference (IPAC'10): Kyoto, Japan, May 23-28, 2010}}},\ }\href@noop {}
  {\bibfield  {journal} {\bibinfo  {journal} {Conf. Proc.}\ }\textbf {\bibinfo
  {volume} {C100523}},\ \bibinfo {pages} {MOPEB016} (\bibinfo {year}
  {2010})}\BibitemShut {NoStop}%
\bibitem [{\citenamefont {Bohl}(2016)}]{SPS_BTRAIN}%
  \BibitemOpen
  \bibfield  {author} {\bibinfo {author} {\bibfnamefont {T.}~\bibnamefont
  {Bohl}},\ }\bibfield  {title} {\bibinfo {title} {Functional specification for
  upgrade of {SPS B-train}},\ }\href@noop {} {\bibfield  {journal} {\bibinfo
  {journal} {CERN, Geneva, Switzerland, Rep. EDMS}\ }\textbf {\bibinfo {volume}
  {1689759}} (\bibinfo {year} {2016})}\BibitemShut {NoStop}%
\bibitem [{\citenamefont {Huschauer}\ \emph {et~al.}(2017)\citenamefont
  {Huschauer}, \citenamefont {Blas}, \citenamefont {Borburgh}, \citenamefont
  {Damjanovic}, \citenamefont {Gilardoni}, \citenamefont {Giovannozzi},
  \citenamefont {Hourican}, \citenamefont {Kahle}, \citenamefont {Le~Godec},
  \citenamefont {Michels}, \citenamefont {Sterbini},\ and\ \citenamefont
  {Hernalsteens}}]{PS}%
  \BibitemOpen
  \bibfield  {author} {\bibinfo {author} {\bibfnamefont {A.}~\bibnamefont
  {Huschauer}}, \bibinfo {author} {\bibfnamefont {A.}~\bibnamefont {Blas}},
  \bibinfo {author} {\bibfnamefont {J.}~\bibnamefont {Borburgh}}, \bibinfo
  {author} {\bibfnamefont {S.}~\bibnamefont {Damjanovic}}, \bibinfo {author}
  {\bibfnamefont {S.}~\bibnamefont {Gilardoni}}, \bibinfo {author}
  {\bibfnamefont {M.}~\bibnamefont {Giovannozzi}}, \bibinfo {author}
  {\bibfnamefont {M.}~\bibnamefont {Hourican}}, \bibinfo {author}
  {\bibfnamefont {K.}~\bibnamefont {Kahle}}, \bibinfo {author} {\bibfnamefont
  {G.}~\bibnamefont {Le~Godec}}, \bibinfo {author} {\bibfnamefont
  {O.}~\bibnamefont {Michels}}, \bibinfo {author} {\bibfnamefont
  {G.}~\bibnamefont {Sterbini}}, and\ \bibinfo {author} {\bibfnamefont
  {C.}~\bibnamefont {Hernalsteens}},\ }\bibfield  {title} {\bibinfo {title}
  {Transverse beam splitting made operational: Key features of the multiturn
  extraction at the {CERN Proton Synchrotron}},\ }\href
  {https://doi.org/10.1103/PhysRevAccelBeams.20.061001} {\bibfield  {journal}
  {\bibinfo  {journal} {Phys. Rev. Accel. Beams}\ }\textbf {\bibinfo {volume}
  {20}},\ \bibinfo {pages} {061001} (\bibinfo {year} {2017})}\BibitemShut
  {NoStop}%
\bibitem [{\citenamefont {{H.~Thiesen and D.~Nisbet}}(2008)}]{SCR}%
  \BibitemOpen
  \bibfield  {author} {\bibinfo {author} {\bibnamefont {{H.~Thiesen and
  D.~Nisbet}}},\ }\bibfield  {title} {\bibinfo {title} {{Review of the initial
  phases of the {LHC} power converter commissioning}},\ }\bibfield  {booktitle}
  {\emph {\bibinfo {booktitle} {{Particle accelerator. Proceedings, 11th
  European Conference, EPAC 2008, Genoa, Italy, June 23-27, 2008}}},\
  }\href@noop {} {\bibfield  {journal} {\bibinfo  {journal} {Conf. Proc.}\
  }\textbf {\bibinfo {volume} {C0806233}},\ \bibinfo {pages} {THPP132}
  (\bibinfo {year} {2008})}\BibitemShut {NoStop}%
\bibitem [{\citenamefont {Brüning}\ \emph
  {et~al.}(2004{\natexlab{b}})\citenamefont {Brüning}, \citenamefont
  {Collier}, \citenamefont {Lebrun}, \citenamefont {Myers}, \citenamefont
  {Ostojic}, \citenamefont {Poole},\ and\ \citenamefont {Proudlock}}]{SCR2}%
  \BibitemOpen
  \bibfield  {author} {\bibinfo {author} {\bibfnamefont {O.~S.}\ \bibnamefont
  {Brüning}}, \bibinfo {author} {\bibfnamefont {P.}~\bibnamefont {Collier}},
  \bibinfo {author} {\bibfnamefont {P.}~\bibnamefont {Lebrun}}, \bibinfo
  {author} {\bibfnamefont {S.}~\bibnamefont {Myers}}, \bibinfo {author}
  {\bibfnamefont {R.}~\bibnamefont {Ostojic}}, \bibinfo {author} {\bibfnamefont
  {J.}~\bibnamefont {Poole}}, and\ \bibinfo {author} {\bibfnamefont
  {P.}~\bibnamefont {Proudlock}},\ }\href
  {https://doi.org/10.5170/CERN-2004-003-V-1} {\emph {\bibinfo {title} {{{LHC}
  Design Report}}}},\ CERN Yellow Reports: Monographs\ (\bibinfo  {publisher}
  {CERN},\ \bibinfo {address} {Geneva},\ \bibinfo {year} {2004})\ Chap.\
  \bibinfo {chapter} {Power converter system}\BibitemShut {NoStop}%
\bibitem [{\citenamefont {Kostoglou}\ \emph {et~al.}(2020)\citenamefont
  {Kostoglou}, \citenamefont {Bartosik}, \citenamefont {Papaphilippou},
  \citenamefont {Sterbini},\ and\ \citenamefont
  {Triantafyllou}}]{kostoglou2020tune}%
  \BibitemOpen
  \bibfield  {author} {\bibinfo {author} {\bibfnamefont {S.}~\bibnamefont
  {Kostoglou}}, \bibinfo {author} {\bibfnamefont {H.}~\bibnamefont {Bartosik}},
  \bibinfo {author} {\bibfnamefont {Y.}~\bibnamefont {Papaphilippou}}, \bibinfo
  {author} {\bibfnamefont {G.}~\bibnamefont {Sterbini}}, and\ \bibinfo {author}
  {\bibfnamefont {N.}~\bibnamefont {Triantafyllou}},\ }\href@noop {} {\bibinfo
  {title} {Tune modulation effects in the high luminosity large hadron
  collider}} (\bibinfo {year} {2020}),\ \Eprint
  {https://arxiv.org/abs/2003.00960} {arXiv:2003.00960 [physics.acc-ph]}
  \BibitemShut {NoStop}%
\bibitem [{\citenamefont {Karastathis}\ \emph {et~al.}(2020)\citenamefont
  {Karastathis}, \citenamefont {Fartoukh}, \citenamefont {Kostoglou},
  \citenamefont {Papaphilippou}, \citenamefont {Pojer}, \citenamefont {Poyet},
  \citenamefont {Solfaroli~Camillocci},\ and\ \citenamefont {Sterbini}}]{IP15}%
  \BibitemOpen
  \bibfield  {author} {\bibinfo {author} {\bibfnamefont {N.}~\bibnamefont
  {Karastathis}}, \bibinfo {author} {\bibfnamefont {S.}~\bibnamefont
  {Fartoukh}}, \bibinfo {author} {\bibfnamefont {S.}~\bibnamefont {Kostoglou}},
  \bibinfo {author} {\bibfnamefont {Y.}~\bibnamefont {Papaphilippou}}, \bibinfo
  {author} {\bibfnamefont {M.}~\bibnamefont {Pojer}}, \bibinfo {author}
  {\bibfnamefont {A.}~\bibnamefont {Poyet}}, \bibinfo {author} {\bibfnamefont
  {M.}~\bibnamefont {Solfaroli~Camillocci}}, and\ \bibinfo {author}
  {\bibfnamefont {G.}~\bibnamefont {Sterbini}},\ }\href
  {http://cds.cern.ch/record/2707081} {\emph {\bibinfo {title} {{{MD} 3584:
  Long-Range Beam-Beam 2018}}}},\ \bibinfo {type} {Tech. Rep.}\ (\bibinfo
  {institution} {CERN},\ \bibinfo {year} {2020})\BibitemShut {NoStop}%
\bibitem [{\citenamefont {Verweij}\ \emph {et~al.}(2008)\citenamefont
  {Verweij}, \citenamefont {Baggiolini}, \citenamefont {Ballarino},
  \citenamefont {Bellesia}, \citenamefont {Bordry}, \citenamefont {Cantone},
  \citenamefont {Casas~Lino}, \citenamefont {Serra}, \citenamefont {Trello},
  \citenamefont {Catalan~Lasheras}, \citenamefont {Charifoulline},
  \citenamefont {Coelingh}, \citenamefont {Dahlerup-Petersen}, \citenamefont
  {D'Angelo}, \citenamefont {Denz}, \citenamefont {Feher}, \citenamefont
  {Flora}, \citenamefont {Gruwé}, \citenamefont {Kain},\ and\ \citenamefont
  {Zerlauth}}]{AF}%
  \BibitemOpen
  \bibfield  {author} {\bibinfo {author} {\bibfnamefont {A.}~\bibnamefont
  {Verweij}}, \bibinfo {author} {\bibfnamefont {V.}~\bibnamefont {Baggiolini}},
  \bibinfo {author} {\bibfnamefont {A.}~\bibnamefont {Ballarino}}, \bibinfo
  {author} {\bibfnamefont {B.}~\bibnamefont {Bellesia}}, \bibinfo {author}
  {\bibfnamefont {F.}~\bibnamefont {Bordry}}, \bibinfo {author} {\bibfnamefont
  {A.}~\bibnamefont {Cantone}}, \bibinfo {author} {\bibfnamefont
  {M.}~\bibnamefont {Casas~Lino}}, \bibinfo {author} {\bibfnamefont
  {A.}~\bibnamefont {Serra}}, \bibinfo {author} {\bibfnamefont
  {C.}~\bibnamefont {Trello}}, \bibinfo {author} {\bibfnamefont
  {N.}~\bibnamefont {Catalan~Lasheras}}, \bibinfo {author} {\bibfnamefont
  {Z.}~\bibnamefont {Charifoulline}}, \bibinfo {author} {\bibfnamefont
  {G.}~\bibnamefont {Coelingh}}, \bibinfo {author} {\bibfnamefont
  {K.}~\bibnamefont {Dahlerup-Petersen}}, \bibinfo {author} {\bibfnamefont
  {G.}~\bibnamefont {D'Angelo}}, \bibinfo {author} {\bibfnamefont
  {R.}~\bibnamefont {Denz}}, \bibinfo {author} {\bibfnamefont {S.}~\bibnamefont
  {Feher}}, \bibinfo {author} {\bibfnamefont {R.}~\bibnamefont {Flora}},
  \bibinfo {author} {\bibfnamefont {M.}~\bibnamefont {Gruwé}}, \bibinfo
  {author} {\bibfnamefont {V.}~\bibnamefont {Kain}}, and\ \bibinfo {author}
  {\bibfnamefont {M.}~\bibnamefont {Zerlauth}},\ }\bibfield  {title} {\bibinfo
  {title} {Performance of the main dipole magnet circuits of the {LHC} during
  commissioning},\ }\href@noop {} {\bibfield  {journal} {\bibinfo  {journal}
  {EPAC 2008 - Contributions to the Proceedings}\ } (\bibinfo {year}
  {2008})}\BibitemShut {NoStop}%
\bibitem [{\citenamefont {Brüning}\ \emph
  {et~al.}(2004{\natexlab{c}})\citenamefont {Brüning}, \citenamefont
  {Collier}, \citenamefont {Lebrun}, \citenamefont {Myers}, \citenamefont
  {Ostojic}, \citenamefont {Poole},\ and\ \citenamefont {Proudlock}}]{AF2}%
  \BibitemOpen
  \bibfield  {author} {\bibinfo {author} {\bibfnamefont {O.~S.}\ \bibnamefont
  {Brüning}}, \bibinfo {author} {\bibfnamefont {P.}~\bibnamefont {Collier}},
  \bibinfo {author} {\bibfnamefont {P.}~\bibnamefont {Lebrun}}, \bibinfo
  {author} {\bibfnamefont {S.}~\bibnamefont {Myers}}, \bibinfo {author}
  {\bibfnamefont {R.}~\bibnamefont {Ostojic}}, \bibinfo {author} {\bibfnamefont
  {J.}~\bibnamefont {Poole}}, and\ \bibinfo {author} {\bibfnamefont
  {P.}~\bibnamefont {Proudlock}},\ }\bibfield  {title} {\bibinfo {title}
  {{Power converter system}},\ }in\ \href
  {https://doi.org/10.5170/CERN-2004-003-V-1} {\emph {\bibinfo {booktitle}
  {{LHC Design Report}}}},\ \bibinfo {series and number} {CERN Yellow Reports:
  Monographs}\ (\bibinfo  {publisher} {CERN},\ \bibinfo {address} {Geneva},\
  \bibinfo {year} {2004})\ Chap.~\bibinfo {chapter} {10}, pp.\ \bibinfo {pages}
  {275--307}\BibitemShut {NoStop}%
\bibitem [{\citenamefont {Burnet}(2019)}]{burnet_active_filters}%
  \BibitemOpen
  \bibfield  {author} {\bibinfo {author} {\bibfnamefont {J.-P.}\ \bibnamefont
  {Burnet}},\ }\bibfield  {title} {\bibinfo {title} {Test results of {RPTE LHC}
  thyristor rectifier with active filter}} (\bibinfo {year} {2019}),\ \bibinfo
  {note} {unpublished}\BibitemShut {NoStop}%
\bibitem [{\citenamefont {Apollinari}\ \emph {et~al.}(2017)\citenamefont
  {Apollinari}, \citenamefont {Béjar~Alonso}, \citenamefont {Brüning},
  \citenamefont {Fessia}, \citenamefont {Lamont}, \citenamefont {Rossi},\ and\
  \citenamefont {Tavian}}]{Apollinari:2284929}%
  \BibitemOpen
  \bibfield  {author} {\bibinfo {author} {\bibfnamefont {G.}~\bibnamefont
  {Apollinari}}, \bibinfo {author} {\bibfnamefont {I.}~\bibnamefont
  {Béjar~Alonso}}, \bibinfo {author} {\bibfnamefont {O.}~\bibnamefont
  {Brüning}}, \bibinfo {author} {\bibfnamefont {P.}~\bibnamefont {Fessia}},
  \bibinfo {author} {\bibfnamefont {M.}~\bibnamefont {Lamont}}, \bibinfo
  {author} {\bibfnamefont {L.}~\bibnamefont {Rossi}}, and\ \bibinfo {author}
  {\bibfnamefont {L.}~\bibnamefont {Tavian}},\ }\bibfield  {title} {\bibinfo
  {title} {{High-Luminosity Large Hadron Collider (HL-LHC)}},\ }\href
  {https://doi.org/10.23731/CYRM-2017-004} {\bibfield  {journal} {\bibinfo
  {journal} {CERN Yellow Rep. Monogr.}\ }\textbf {\bibinfo {volume} {4}},\
  \bibinfo {pages} {1} (\bibinfo {year} {2017})}\BibitemShut {NoStop}%
\bibitem [{\citenamefont {Dubouchet}\ \emph {et~al.}(2012)\citenamefont
  {Dubouchet}, \citenamefont {Hofle}, \citenamefont {Kotzian},\ and\
  \citenamefont {Valuch}}]{Dubouchet:2012hzl}%
  \BibitemOpen
  \bibfield  {author} {\bibinfo {author} {\bibfnamefont {F.}~\bibnamefont
  {Dubouchet}}, \bibinfo {author} {\bibfnamefont {W.}~\bibnamefont {Hofle}},
  \bibinfo {author} {\bibfnamefont {G.}~\bibnamefont {Kotzian}}, and\ \bibinfo
  {author} {\bibfnamefont {D.}~\bibnamefont {Valuch}},\ }\bibfield  {title}
  {\bibinfo {title} {{"What you get" - {Transverse} damper}},\ }in\ \href@noop
  {} {\emph {\bibinfo {booktitle} {{Proceedings, 4th Evian Workshop on LHC beam
  operation: Evian Les Bains, France, December 17-20, 2012}}}},\ \bibinfo
  {organization} {CERN}\ (\bibinfo  {publisher} {CERN},\ \bibinfo {address}
  {Geneva},\ \bibinfo {year} {2012})\ pp.\ \bibinfo {pages}
  {73--78}\BibitemShut {NoStop}%
\bibitem [{\citenamefont {Komppula}\ \emph {et~al.}(2019)\citenamefont
  {Komppula}, \citenamefont {Kotzian}, \citenamefont {Rains}, \citenamefont
  {S{\"o}der{\'e}n},\ and\ \citenamefont {Valuch}}]{Komppula2019ADTAO}%
  \BibitemOpen
  \bibfield  {author} {\bibinfo {author} {\bibfnamefont {J.~P.~O.}\
  \bibnamefont {Komppula}}, \bibinfo {author} {\bibfnamefont {G.}~\bibnamefont
  {Kotzian}}, \bibinfo {author} {\bibfnamefont {S.}~\bibnamefont {Rains}},
  \bibinfo {author} {\bibfnamefont {M.}~\bibnamefont {S{\"o}der{\'e}n}}, and\
  \bibinfo {author} {\bibfnamefont {D.}~\bibnamefont {Valuch}},\ }\bibfield
  {title} {\bibinfo {title} {{ADT} and {ObsBox} in {LHC} {Run} 2, plans for
  {LS2}},\ }in\ \href@noop {} {\emph {\bibinfo {booktitle} {{Proceedings, 9th
  Evian Workshop on LHC beam operation: Evian Les Bains, France, January,
  2019}}}}\ (\bibinfo {year} {2019})\BibitemShut {NoStop}%
\bibitem [{\citenamefont {Morrone}\ \emph {et~al.}(2019)\citenamefont
  {Morrone}, \citenamefont {Martino}, \citenamefont {De~Maria}, \citenamefont
  {Fitterer},\ and\ \citenamefont {Garion}}]{dipoles_transfer_function}%
  \BibitemOpen
  \bibfield  {author} {\bibinfo {author} {\bibfnamefont {M.}~\bibnamefont
  {Morrone}}, \bibinfo {author} {\bibfnamefont {M.}~\bibnamefont {Martino}},
  \bibinfo {author} {\bibfnamefont {R.}~\bibnamefont {De~Maria}}, \bibinfo
  {author} {\bibfnamefont {M.}~\bibnamefont {Fitterer}}, and\ \bibinfo {author}
  {\bibfnamefont {C.}~\bibnamefont {Garion}},\ }\bibfield  {title} {\bibinfo
  {title} {Magnetic frequency response of {High-Luminosity Large Hadron
  Collider} beam screens},\ }\href
  {https://doi.org/10.1103/PhysRevAccelBeams.22.013501} {\bibfield  {journal}
  {\bibinfo  {journal} {Phys. Rev. Accel. Beams}\ }\textbf {\bibinfo {volume}
  {22}},\ \bibinfo {pages} {013501} (\bibinfo {year} {2019})}\BibitemShut
  {NoStop}%
\bibitem [{\citenamefont {Ruggiero}(1995)}]{Ruggiero:1995kv}%
  \BibitemOpen
  \bibfield  {author} {\bibinfo {author} {\bibfnamefont {F.}~\bibnamefont
  {Ruggiero}},\ }\bibfield  {title} {\bibinfo {title} {{Single beam collective
  effects in the {LHC}}},\ }\bibfield  {booktitle} {\emph {\bibinfo {booktitle}
  {{Collective effects in large hadron colliders. Proceedings, International
  Workshop, Montreux, Switzerland, October 17-22, 1994}}},\ }\href@noop {}
  {\bibfield  {journal} {\bibinfo  {journal} {Part. Accel.}\ }\textbf {\bibinfo
  {volume} {50}},\ \bibinfo {pages} {83} (\bibinfo {year} {1995})}\BibitemShut
  {NoStop}%
\bibitem [{\citenamefont {{V.~Chareyre}}(2015)}]{UPS}%
  \BibitemOpen
  \bibfield  {author} {\bibinfo {author} {\bibnamefont {{V.~Chareyre}}},\
  }\href {https://edms.cern.ch/document/1539776/} {\emph {\bibinfo {title}
  {{{Assessment of the high frequency noise produced by the UPS systems in the
  LHC Machine}}}}},\ \bibinfo {type} {Tech. Rep.}\ (\bibinfo  {institution}
  {CERN},\ \bibinfo {year} {2015})\BibitemShut {NoStop}%
\bibitem [{\citenamefont {{U.~Hassan and S.~Anwar}}(2010)}]{snr}%
  \BibitemOpen
  \bibfield  {author} {\bibinfo {author} {\bibnamefont {{U.~Hassan and
  S.~Anwar}}},\ }\bibfield  {title} {\bibinfo {title} {Reducing noise by
  repetition: Introduction to signal averaging},\ }\href
  {https://doi.org/10.1088/0143-0807/31/3/003} {\bibfield  {journal} {\bibinfo
  {journal} {European Journal of Physics}\ }\textbf {\bibinfo {volume} {31}},\
  \bibinfo {pages} {453} (\bibinfo {year} {2010})}\BibitemShut {NoStop}%
\bibitem [{six(2019)}]{sixtrack}%
  \BibitemOpen
  \href@noop {} {\bibinfo {title} {Six{T}rack}},\ \bibinfo {howpublished}
  {\url{http://sixtrack.web.cern.ch/SixTrack/}} (\bibinfo {year} {2019}),\
  \bibinfo {note} {accessed: 2019-11-26}\BibitemShut {NoStop}%
\bibitem [{\citenamefont {Maria}\ \emph {et~al.}(2019)\citenamefont {Maria},
  \citenamefont {Andersson}, \citenamefont {Olsen}, \citenamefont {Field},
  \citenamefont {Giovannozzi}, \citenamefont {Hermes}, \citenamefont
  {H{\o}imyr}, \citenamefont {Kostoglou}, \citenamefont {Iadarola},
  \citenamefont {Mcintosh}, \citenamefont {Mereghetti}, \citenamefont {Molson},
  \citenamefont {Pellegrini}, \citenamefont {Persson}, \citenamefont
  {Schwinzerl}, \citenamefont {Maclean}, \citenamefont {Sjobak}, \citenamefont
  {Zacharov},\ and\ \citenamefont {Singh}}]{sixtrack2}%
  \BibitemOpen
  \bibfield  {author} {\bibinfo {author} {\bibfnamefont {R.~D.}\ \bibnamefont
  {Maria}}, \bibinfo {author} {\bibfnamefont {J.}~\bibnamefont {Andersson}},
  \bibinfo {author} {\bibfnamefont {V.~K.~B.}\ \bibnamefont {Olsen}}, \bibinfo
  {author} {\bibfnamefont {L.}~\bibnamefont {Field}}, \bibinfo {author}
  {\bibfnamefont {M.}~\bibnamefont {Giovannozzi}}, \bibinfo {author}
  {\bibfnamefont {P.~D.}\ \bibnamefont {Hermes}}, \bibinfo {author}
  {\bibfnamefont {N.}~\bibnamefont {H{\o}imyr}}, \bibinfo {author}
  {\bibfnamefont {S.}~\bibnamefont {Kostoglou}}, \bibinfo {author}
  {\bibfnamefont {G.}~\bibnamefont {Iadarola}}, \bibinfo {author}
  {\bibfnamefont {E.}~\bibnamefont {Mcintosh}}, \bibinfo {author}
  {\bibfnamefont {A.}~\bibnamefont {Mereghetti}}, \bibinfo {author}
  {\bibfnamefont {J.~W.}\ \bibnamefont {Molson}}, \bibinfo {author}
  {\bibfnamefont {D.}~\bibnamefont {Pellegrini}}, \bibinfo {author}
  {\bibfnamefont {T.}~\bibnamefont {Persson}}, \bibinfo {author} {\bibfnamefont
  {M.}~\bibnamefont {Schwinzerl}}, \bibinfo {author} {\bibfnamefont {E.~H.}\
  \bibnamefont {Maclean}}, \bibinfo {author} {\bibfnamefont {K.}~\bibnamefont
  {Sjobak}}, \bibinfo {author} {\bibfnamefont {I.}~\bibnamefont {Zacharov}},
  and\ \bibinfo {author} {\bibfnamefont {S.}~\bibnamefont {Singh}},\ }\bibfield
   {title} {\bibinfo {title} {{{SixTrack} project: Status, runtime environment,
  and new developments}},\ }in\ \href
  {https://doi.org/10.18429/JACoW-ICAP2018-TUPAF02} {\emph {\bibinfo
  {booktitle} {{Proceedings, 13th International Computational Accelerator
  Physics Conference, ICAP2018: Key West, FL, USA, 20-24 October 2018}}}}\
  (\bibinfo {year} {2019})\ p.\ \bibinfo {pages} {TUPAF02}\BibitemShut
  {NoStop}%
\end{thebibliography}%


\providecommand{\noopsort}[1]{}\providecommand{\singleletter}[1]{#1}%
%
\end{document}